\documentclass[aps, prb, nosuperscriptaddress, nofootinbib, reprint, longbibliography, twocolumn]{revtex4}
\usepackage{graphicx}
\usepackage{mathrsfs}
\usepackage{amsmath}
\usepackage{subfigure}
\usepackage{bm}
\usepackage{verbatim}
\usepackage[colorlinks=true,letterpaper=true, pdfstartview=FitV,urlcolor=blue,linkcolor=red,citecolor=cyan]{hyperref}
\usepackage{color}
\usepackage{xcolor}
\usepackage{hyperref}
\usepackage{amssymb}
\usepackage{bbm}
\usepackage{siunitx}

\newcommand{\bq}{\bm{q}}

\newcommand{\bra}[1]{\langle #1\rvert}
\newcommand{\ket}[1]{\lvert #1\rangle}

\newcommand{\br}{\bm{r}}

\usepackage{siunitx}

\newcommand{\TwoDMatrix}[4]{\begin{pmatrix} #1 & #2\\#3 & #4\end{pmatrix}}

\newcommand{\kfa}{k_{F,\alpha}}
\newcommand{\kfb}{k_{F,\beta}}
\newcommand{\abs}[1]{\left|#1\right|}
\newcommand{\qtf}{q_\text{TF}}

\begin{document}
\title{
Weyl fermions with arbitrary monopoles in magnetic fields: Landau levels, longitudinal magnetotransport, and density-wave ordering 
}
\author{Xiao Li}
\email{lixiao@umd.edu}
\affiliation{Condensed Matter Theory Center and Joint Quantum Institute, University of Maryland, College Park, MD 20742, USA}

\author{Bitan Roy}
\affiliation{Condensed Matter Theory Center and Joint Quantum Institute, University of Maryland, College Park, MD 20742, USA}

\author{S. Das Sarma}
\affiliation{Condensed Matter Theory Center and Joint Quantum Institute, University of Maryland, College Park, MD 20742, USA}

\date{\today}

\begin{abstract}
We theoretically address the effects of strong magnetic fields in three-dimensional Weyl semimetals (WSMs) built out of Weyl nodes with a monopole charge $n$. For $n=1$, $2$, and $3$ we realize single, double, and triple WSM, respectively, and the monopole charge $n$ determines the integer topological invariant of the WSM. 
Within the linearized continuum description, the quasiparticle spectrum is then composed of Landau levels (LLs), containing exactly $n$ number of chiral zeroth Landau levels (ZLLs), irrespective of the orientation of the magnetic field. 
In the presence of strong backscattering, for example, (due to quenched disorder associated with random impurities), these systems generically give rise to longitudinal magnetotransport. Restricting ourselves to the quantum limit (and assuming only the subspace of the ZLLs to be partially filled) and mainly accounting for Gaussian impurities, we show that the longitudinal magnetoconductivity (LMC) in all members of the Weyl family displays a positive linear-$B$ scaling, when the field is applied along the axis that separates the Weyl nodes. But, in double and triple WSM, LMC displays a smooth crossover to a nonlinear $B$-dependence as the field is tilted away from such a high-symmetry direction. In addition, due to the enhanced density of states, the LL quantization can trigger instabilities toward the formation of translational symmetry breaking density-wave orderings for sufficiently weak interaction (BCS instability), which gaps out the ZLLs. Concomitantly as the temperature (magnetic field) is gradually decreased (increased) the LMC becomes negative. Thus WSMs with arbitrary monopole charge ($n$) can host an intriguing interplay of LL quantization, longitudinal magnetotransport (a possible manifestation of one-dimensional chiral or axial anomaly), and density-wave ordering, when placed in a strong magnetic field.
\end{abstract}

\maketitle

\section{Introduction}

Strong spin-orbit coupling is the fundamental origin of several topological phases of matter, such as topological insulators and superconductors in two and three spatial dimensions~\cite{kane-hasan-RMP, qi-zhang-RMP, sau-prl, hasan-neupane,fu-ando, galitski-coleman}. The salient features of these systems are (i) an insulating bulk (electrical or thermal), and (ii) topologically protected metallic surface states (which could be superconducting depending on the situation). 
Among such gapless phases, the ones (so-called Weyl materials) formed by chiral Weyl fermions have attracted considerable recent attention and may serve as an experimental platform where various exotic phenomena (such as axionic electrodynamics and chiral anomaly), which were originally proposed in the context of high energy physics and quantum field theory, may be observed in the context of solid state physics~\cite{volovik, burkov-review, rao-review}. The current work is a comprehensive theoretical study of the strong-field magnetotransport properties of generalized three-dimensional Weyl systems (with arbitrary monopole charges) in the quantum limit.

\begin{figure}[!]
\includegraphics[width=8cm, height=6.0cm]{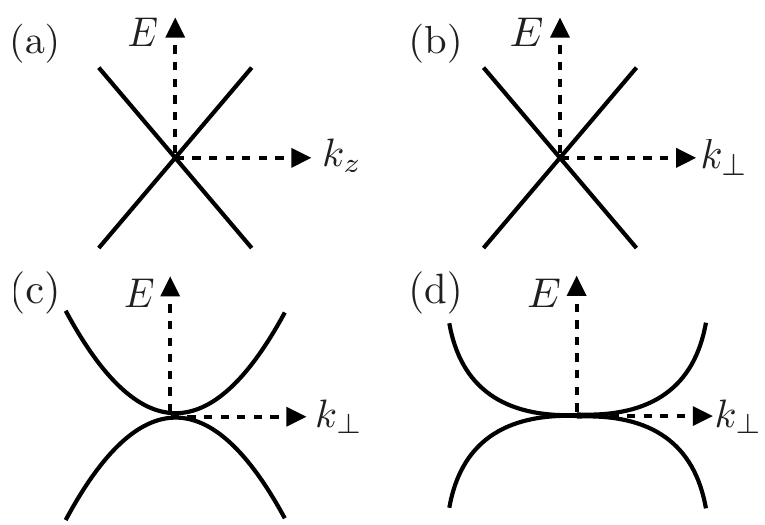}
\caption{(a) Spectrum of Weyl fermions along the $k_z$ direction in any Weyl semimetal. Dispersion in the $xy$ plane respectively for (b) single, (c) double, and (d) triple Weyl semimetals. Notice that $E \sim |\boldsymbol{k}_\perp|^{n}$, where $n$ is the monopole charge and ${\boldsymbol k}_\perp=\left( k_x, k_y \right)$, but $E \sim |k_z|$ in all Weyl semimetals.}\label{dispersion}
\end{figure}

Weyl fermions may result from complex band structures in strong spin-orbit coupled semiconductors~\cite{TaAs1, TaAs2, TaAs4, Shekhar, Borisenko, Liu-Mao, Chang-Hasan, TaAs3, Ding-PRX, Hasan-NaturePhysics, Xu-Shi, yazdani-1, yazdani-2, TaAs5, Wang-Xu}, multilayer heterostructures~\cite{Burkov1, Burkov2, Zyuzin1, Das, Cava2, babaganesh}, and they can also be found inside a broken symmetry phase in strongly correlated materials, such as $227$ pyrochlore iridates, as emergent quasiparticles~\cite{vishwanath, balents, sunbin, goswami-roy-dassarma}. 
The Weyl semimetals (WSMs) can be classified into two broad categories: (i) Inversion ($\mathcal P$) symmetry-breaking WSMs that are commonly found in weakly correlated semiconductors, and (ii) time-reversal ($\mathcal T$) odd WSMs that can be found in strongly correlated materials with comparable strength of spin-orbit coupling and electronic interaction. However, irrespective of the microscopic origin, WSMs are composed of Weyl nodes in the reciprocal space, where Kramers non-degenerate valence and conduction bands touch each other at the so-called diabolic points in momentum space~\cite{herring}. 
The Weyl nodes act as the source (monopole) and sink (anti-monopole) of Abelian Berry flux, and in its close vicinity, Weyl fermions can be identified as left (right) chiral fermions. A \emph{no-go theorem} guarantees the existence of an equal number of left and right chiral fermions in the system~\cite{nielsen-nogo}. The monopole charge ($n$) also dictates the amount of Berry flux enclosed by a plane perpendicular to the line joining these two points and in turn defines the \emph{integer topological invariant} of the system. Thus monopole charge permits a topological classification of Weyl semimetals, and for $n=1$, $2$, and $3$ we can call them single, double and triple WSM, respectively. 
The most common Weyl system has just $n=1$, whereas the system with arbitrary $n$ can be regarded as a generalized Weyl system.  (In the current work we consider magnetotransport in single, double, and triple WSMs although our theory is generalizable to even higher values of $n$, but it is well-known that Weyl systems with monopole charge $n>3$ is not allowed in three-dimensional lattice systems~\cite{Fang2012:Weyl,nagaosa-yang}, and therefore our work applies to all possible physical Weyl materials that can be studied in the laboratory.) 
The energy-momentum dispersion of Weyl quasiparticles in these systems along various high-symmetry directions is shown in Fig.~\ref{dispersion}. Although the Weyl nodes are often accompanied by \emph{unit} monopole charge, it is nonetheless conceivable to realize Weyl nodes with a higher integer charge and the system with arbitrary $n$ can be regarded as a generalized Weyl system, such as the proposed double WSM in HgCr$_2$Se$_4$ and SrSi$_2$~\cite{Fang2012:Weyl, nagaosa-yang, hasan-DW}. 
Even though the material existence of a triple WSM has remained elusive so far, with the anticipated progress of materials science the discovery of such a generalized triple WSM is certainly possible. 
Most, if essentially not all, of the substantial theoretical literature on WSMs have focused on the single WSM because the existing experimental materials are all WSMs with $n=1$. 
On the other hand, there are significant differences in the properties of generalized WSMs depending on their monopole charges, necessitating theoretical studies of multi-WSMs with higher ($n>1$) values of monopole charges, which is what we do in the current work considering $n=2$ and $3$, and contrasting their properties with $n=1$ WSM (and with each other).

The integer topological invariant also determines various thermodynamic, transport, and topological properties of a WSM. For example, one of the hallmark signatures of a WSM is the presence of topologically protected \emph{Fermi arc}~\cite{vishwanath} that has recently been seen through angle resolved photoemission spectroscopy (ARPES), scanning tunneling microscopy (STM), and quantum oscillation measurements~\cite{TaAs3, Ding-PRX, Hasan-NaturePhysics, Xu-Shi, yazdani-1, yazdani-2,Potter2014,Zhang2016}. In a multi-WSM the Fermi arc bears an additional $n$-fold degeneracy, where $n$ is the monopole charge of Weyl nodes. 
Therefore, in double and triple WSMs, the Fermi arc respectively possesses two- and three fold orbital degeneracies, which, for example, can be detected in ARPES and STM. 
We here explore the ramifications of the underlying topological invariant of a pristine WSM, when the system is placed in a strong magnetic field. Our theoretical findings are the following:  

\begin{enumerate}

\item{When placed in a strong magnetic field, WSMs undergo Landau level (LL) quantization, supporting \emph{exactly} $n$ number of zeroth Landau levels (ZLLs), irrespective of the field orientation (within continuum approximation). 
All LLs, including the zeroth ones, disperse along the direction of the applied field. Thus, the orbital degeneracy of ZLL ($n$) gets tied with the monopole charge or topological invariant of the Weyl nodes.}

\item{Even though $n$ dispersive ZLLs always go through zero energy (in the continuum limit), they are in general non-degenerate, unless the field is applied along the axis on which Weyl nodes reside in the absence of the field [see Figs.~\ref{LL_DW_general} and \ref{LL_TW_general}].}

\item{Density of states (DOS) displays nontrivial dependence (such as its periodicity) on the tilting angle of the magnetic field away from a high-symmetry direction, separating the Weyl nodes [see Figs.~\ref{Fig:DOS_DW} and \ref{Fig:DOS_TW}]. Such intriguing features can be observed in \emph{angle resolved} quantum oscillation measurements~\cite{Pippard, behnia3}. }

\end{enumerate}

A strong magnetic field causes an \emph{effective} reduction of the dimensionality of the system. In particular, each branch of ZLLs cuts the zero energy (where the Fermi energy is pinned) at two isolated points in the Brillouin zone, yielding two field-induced Weyl nodes, around which the quasiparticles are once again described as (one-dimensional) chiral (left and right) fermions (We note that the effective reduction of a generic three-dimensional electron system to an apparent one-dimensional electron gas along the magnetic field direction is of course a well-known effect of the strong field limit where LL coupling can be ignored---for WSMs this dimensional reduction leads to effectively chiral one-dimensional fermion systems.). 
If an electric field ($E$) is now applied in parallel to the magnetic field, it can produce charge transfer from left to right Weyl nodes, causing \emph{violation of separate conservation laws for left and right chiral fermions}, captured by the following anomaly equation~\cite{peskin, fujikawa} 
\begin{align}~\label{anomaly-1D}
\partial_{\mu} \left( j_{\mu, R}-j_{\mu, L}\right)= N \: \frac{e E}{\hbar \pi}.
\end{align}
Here, $e$ is the electric charge, $j_{\mu,R/L}$ is the charge (for $\mu=0$) and current (for $\mu=1$) operator for left/right chiral fermion, and $N$ can be identified as the total degeneracy of the effective one-dimensional system. The total charge and current density is respectively given by $j_0=\Psi^\dagger \Psi$ and $j_1=\Psi^\dagger \sigma_3 \Psi$, where $\Psi^\top= \left(\Psi_L, \Psi_R \right)$ is a two-component spinor, describing the emergent one-dimensional world. Due to the reduced dimensionality of the system, one can define $\sigma_3 \equiv \gamma_5$, and accordingly we can define axial or chiral charge and current density as $j^{ax}_0=\Psi^\dagger \gamma_5 \Psi=j_1$ and $j^{ax}_1=\Psi^\dagger \Psi=j_0$. Hence, one can express Eq.~(\ref{anomaly-1D}) as $\partial_\mu j^{ax}_{\mu}= N (e E)/(\hbar \pi)$. In the language of quantum field theory, such a violation of separate conservation laws for left and right fermions is known as \emph{chiral anomaly}, specific to odd spatial dimensions. In regular WSM, the degeneracy of the ZLL plays the role of $N$, whereas in WSMs with $n \neq 1$, $N= n \; \times$ LL degeneracy. Therefore, upon identifying $N$ as the \emph{total} degeneracy of the generalized WSM ZLL, we arrive at the quantum field theoretic \emph{Adler-Jackiw-Bell} anomaly equation~\cite{peskin, adler, bell}
\begin{equation}
\partial_\mu j^{ax}_\mu = n \; \frac{e^2}{2 \pi^2 \hbar} {\boldsymbol E} \cdot {\boldsymbol B},
\end{equation} 
now generalized for the generalized Weyl system, constituted by Weyl nodes with an arbitrary (integer) monopole charge $n$. Notice that here electric (${\boldsymbol E}$) and magnetic (${\boldsymbol B}$) fields are strictly \emph{static Abelian background fields}. 
We also stress that in the above expression ${\boldsymbol E} \cdot {\boldsymbol B}$ can \emph{never} be negative as the $B$-linear dependence arises from the LL DOS, and is insensitive to the direction of applied electric and magnetic field.

Such a tantalizing quantum field theoretic chiral anomaly analogy has led to considerable theoretical~\cite{Osada, fukushima, xu, Cho, Grushin, aji, Zyuzin3, GoswamiTewari, burkovsu, son, vazifeh, vishwanathsid, shovkovy, burkov-prl, Pesin, Yawen, Burkov, NiNi, Shun-Qing, Xie, QingYang, Andreev, Zyuzin} and experimental~\cite{Tajima2, kim1, zhao, XiongOng, ong, XHuang, CZhang, kharzeev, balicas, Shekhar-2, Wang-Xia, Wang-Wang, Zheng-Tian,Wiedmann} activities aimed at demonstrating the manifestation of chiral anomaly in solid state Weyl materials through longitudinal magnetotransport (LMT) studies. 
It is, however, unclear that these chiral anomaly considerations apply directly in solid state materials since the nonconservation of the chiral current in the quantum field theoretic Adler-Jackiw-Bell anomaly crucially depends on the existence of an \emph{unbounded} linear dispersion of Weyl fermions~\cite{peskin, fujikawa, adler, bell}, whereas in all solid state systems the energy dispersion is unavoidably bounded by the natural lattice cut off. 
However in a pioneering work Nielsen and Ninomiya showed that certain gapless semiconductors accommodating linear touching of valence and conduction bands can give rise to LMT arising essentially from the chiral anomaly physics of current nonconservation~\cite{nielsen}. 
In this regard, we must recall that almost $30$ years before the proposal of Nielsen and Ninomiya, Adam and Argyres demonstrated the generic existence of LMT for conventional three-dimensional Fermi liquids, i.e., ordinary metals and doped semiconductors, without invoking the notion of chiral or axial anomaly~\cite{Argyres1956:Transport}. 
Their derivation of LMT solely relies on few generic features of a three-dimensional electronic system in a strong magnetic field: (i) the existence of one-dimensional dispersive LLs since momentum along the applied magnetic field is a conserved quantity, and (ii) at least one partially filled LL that cuts Fermi energy at two isolated points (or in general even number of points in the Brillouin zone). 
Therefore, the existence of LMT in the absence of the classical Lorentz force (when ${\boldsymbol E}$ and ${\boldsymbol B}$ are parallel) necessarily points toward subtle quantum mechanical effects, but not necessarily has a direct connection to  chiral anomaly. 
Specifically, the chiral anomaly implies an LMT through the Nielsen-Ninomiya mechanism, but the reverse is not true, i.e., the existence of an LMT does not necessarily imply a chiral anomaly---LMT is necessary, but by no means sufficient, for the existence of the chiral anomaly.

Recently Goswami \emph{et al.} in Ref.~[\onlinecite{Goswami2015}] have demonstrated the universal existence of LMT for ``\emph{a generic three-dimensional metal}."  Obviously, this LMC of Ref.~[\onlinecite{Goswami2015}] has nothing whatsoever to do with any chiral anomaly in the quantum field theoretic sense, demonstrating that the existence of LMT by itself cannot be construed to necessarily imply the existence of a chiral anomaly in the parent material. 
We here follow the philosophy of Ref.~[\onlinecite{Goswami2015}] and show that any WSM with arbitrary monopole charge displays LMT when placed in a strong magnetic field without invoking the physics of chiral anomaly. 
For the sake of technical simplicity we assume that only the manifold of ZLLs is partially filled, i.e., the system is in the strong-field limit. 
However, our analysis can be generalized to demonstrate generic LMT in Weyl materials even when multiple LLs are partially filled. 
The LMT we predict theoretically is a generic property of impurity scattering in the strong-field limit as was originally pointed out in Ref.~[\onlinecite{Argyres1956:Transport}] a long time ago in the context of doped bulk semiconductors.
Our current work is a generalization of the work of Goswami \emph{et al.}~\cite{Goswami2015}, who considered an ordinary 3D electron gas model for a simple metal subjected to a strong magnetic field, whereas we consider a WSM with arbitrary monopole charge $n$ ($=1,2,3$) subjected to a magnetic field. 
Our results support and further reinforce the conclusion of Ref.~[\onlinecite{Goswami2015}], showing that LMC may very well be a generic property of 3D systems subjected to a strong magnetic field and impurity scattering.

\begin{figure}[!]
\includegraphics[width=5cm, height=5cm]{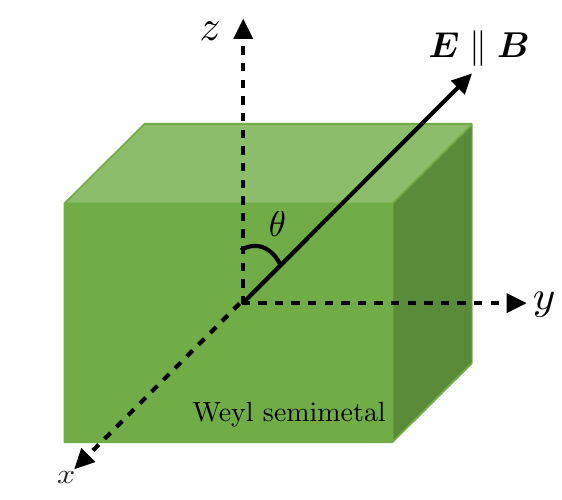}
\caption{(Color online) Schematic representation of the applied electric ($\boldsymbol E$) and magnetic ($\boldsymbol B$) fields (always parallel). Quasiparticle dispersion along $z$ direction is $v_z k_z$, while in the $xy$ plane is proportional to $|k_\perp|^{n/2}$, where ${\boldsymbol k}_\perp=(k_x, k_y)$. Here, $\theta$ parametrizes the tilting of magnetic field away from the $z$ direction. Weyl nodes are separated along the $z$ direction. }\label{geometry}
\end{figure}

In the presence of the periodic potential in a solid, arising from the ionic lattice, quasiparticles perform Bloch oscillations and consequently the system cannot sustain any finite steady-state current~\cite{ashcroft}. 
Nevertheless, in the presence of a momentum relaxation mechanism, which is naturally offered by impurity scattering, quasiparticles lose the freedom to perform a complete Bloch oscillation, and consequently a metallic system can support finite steady-state current, when the relaxation or transport lifetime ($\tau$) is much shorter than the Bloch oscillation period. (In this work, we focus on the low-temperature situation where impurity scattering is the primary scattering mechanism for transport as phonons are mostly frozen out.) 
However, in the presence of a strong magnetic field, i.e.,  when $\omega_c \tau \gg 1$ where $\omega_c$ is the cyclotron frequency, the orbit of the carriers is so curved that the period of completing a cyclotron motion becomes comparable (or actually much longer than) to the time between two successive impurity collisions. Then, the transport/relaxation time acquires a strong magnetic field dependence and $\tau =\tau(B)$. 
In this strong-field limit, one cannot assume the relaxation time to be field-independent which is the standard approximation in metallic transport since in metals one is often in the weak-field semiclassical limit where $\omega_c\tau\ll 1$.
Such a field dependence of $\tau$ is extracted here by using the \emph{quantum Boltzmann equation}, but omitted in the semiclassical theory (where the transport relaxation time is taken to be field-independent), which therefore is applicable for extremely weak magnetic fields ($\omega_c \tau \ll 1$). 
But, for sufficiently weak magnetic fields one needs to account for yet another pure quantum mechanical effect, the \emph{weak localization}~\cite{hikami, bergmann1984weak}, which arises from quantum interference of electron paths---our theory neglects the weak localization physics since the strong magnetic field completely breaks the time reversal invariance, suppressing all weak localization effects. 
Presently it is unknown how to take into account all of these quantum effects at an arbitrary magnetic field in a unified theoretical framework, and restricting ourselves to the strong-field quantum limit $\omega_c \tau \gg 1$ we establish the following theoretically:

\begin{enumerate}

\item{In the presence of both Gaussian and Coulomb impurities, we extract the magnetic field dependence of transport lifetime using quantum Boltzmann equation, and show that all WSMs manifest \emph{positive} longitudinal magnetoconductivity (LMC) or negative longitudinal magnetoresistivity (LMR), when the magnetic and electric fields are applied parallel to each other (see Fig.~\ref{geometry}). }

\item{Due to the carrier-induced screening of Coulomb potential (arising from finite DOS of LLs), we mainly focus on Gaussian impurities and show that when the fields are applied along a high-symmetry axis separating the Weyl nodes, the LMC increases linearly with $B$ in all WSMs (see Fig.~\ref{Fig:Conductivity-SingleWeyl}). The LMC in the presence of only Coulomb impurities grows as $B^2$, but only in extreme strong field limit. (We believe that the same result applies to screened Coulomb impurity potential also since screening renders the Coulomb potential into a short-range scattering potential similar to the Gaussian case.).}

\item{Although such a linear increase of LMC is insensitive to the field orientation in single WSMs (with monopole charge $n=1$), LMC develops a \emph{nonlinear} dependence on the magnetic field in double and triple WSMs when the magnetic field is tilted away from the high-symmetry axis (see Figs.~\ref{Fig:Conductivity-1} and ~\ref{Fig:Conductivity-2}). 
Thus, the quantitative field dependence of LMC depends on the WSM monopole charge, whereas the qualitative existence of the negative LMR itself (positive LMC) is generic in all WSMs in the strong-field regime.} 

\end{enumerate}

We also investigate the role of electronic interaction on LMT as the temperature (magnetic field) is gradually decreased (increased). The DOS in a pristine WSM scales as $E^{2/n}$ and the Weyl nodes are extremely robust against weak short-range electron-electron interaction. However, due to the Landau quantization induced by the external magnetic field, the DOS for the emergent one-dimensional system (along the field direction due to the quantization of the transverse motion) is \emph{constant}, which in turn can trigger various density-wave instabilities even when the electronic interaction is sufficiently weak. The main interaction effects of interactions can be summarized as the following: 

\begin{enumerate}

\item{Depending on the microscopic details, ZLLs can undergo a weak coupling (BCS-like) instability in the charge-density-wave (CDW) or spin-density-wave (SDW) channel as common in one-dimensional electron systems. Irrespective of the actual nature of the ordering the density-wave orders break the translational symmetry and gap out the ZLLs.}

\item{When the field is applied along the high-symmetry axis all $n$ number of ZLLs are expected to undergo density-wave ordering \emph{simultaneously}. Thus, below the transition temperature ($T_c$) LMC (LMR) becomes \emph{negative (positive)}, reversing the trend compared with the normal phase with no density-wave ordering. However, upon tilting the field away from the high-symmetry axis, the exact $n$-fold degeneracy of ZLLs is lifted, and the system can undergo a \emph{cascade} of density-wave transitions with distinct transition temperatures. Hence, LMC/LMR in a multi-WSM is expected to display $n$-fold discontinuity before it becomes negative/positive.}

\end{enumerate}   

Although such ordering in a clean system (devoid of any impurity) should take place for arbitrarily weak interaction, disorder naturally reduces the propensity towards such translational symmetry breaking ordering~\cite{Imry}. (We mention that there is no Anderson theorem here protecting the symmetry-broken phase against impurity scattering.) 
Thus clear signatures of CDW or SDW ordering can only be observed in clean systems at sufficiently low temperatures and for strong enough magnetic fields (and at low enough disorder). 

\begin{figure*}[!]
\subfigure[]{
\includegraphics[scale=0.3]{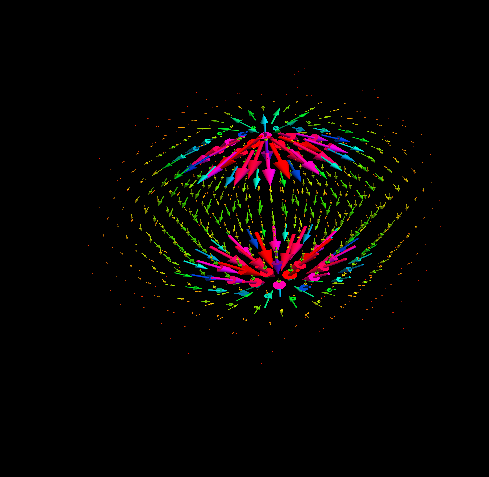}
\label{topology-SW}
}
\subfigure[]{
\includegraphics[scale=0.265]{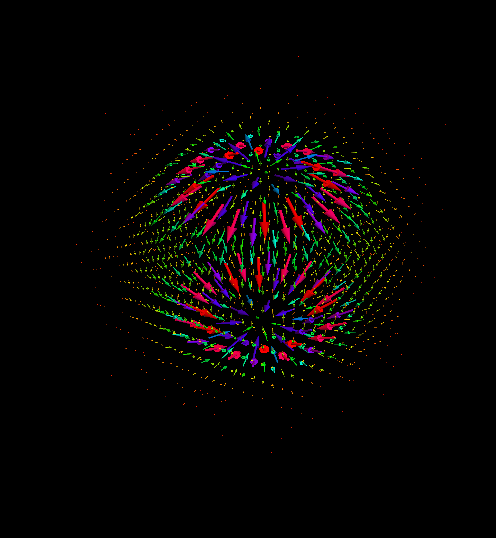}
\label{topology-DW}
}
\subfigure[]{
\includegraphics[scale=0.28]{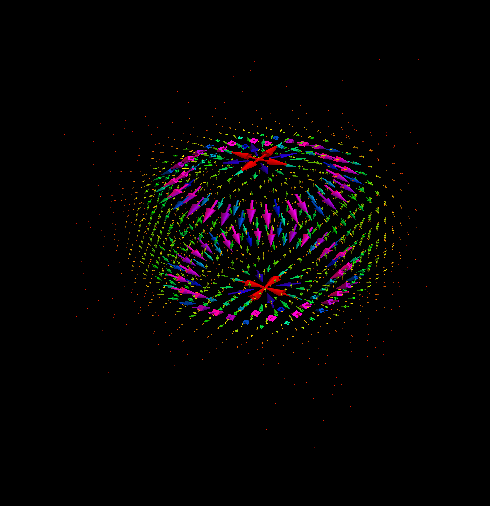}
\label{topology-TW}
}
\caption[]{(Color online) Vector field plots for the Berry curvature for Weyl fermion, (a) when the Weyl nodes are monopole (source) and anti-monopole (sink) of Berry curvature, with unit charge. In double and triple WSMs the Weyl nodes are source and sink of Berry curvature with charge two and three, respectively, and the vector field plots are shown in (b) and (c). }\label{topology}
\end{figure*}

The rest of the paper is organized in the following way. In Sec.~\ref{weylintro}, we discuss the possibility of realizing Weyl semimetals with different monopole charge and address their topological properties (such as bulk-boundary correspondence through degeneracy of Fermi arc). Section~\ref{landaulevel} is devoted to the demonstration of LL spectra in single, double and triple WSMs for arbitrary orientation of the magnetic field. DOS in these systems in the presence of a strong magnetic field is presented in Sec.~\ref{DOS}. Longitudinal magnetotransport for all members of the Weyl family constitutes the central theme of Sec.~\ref{transport}.
Effects of various density-wave ordering on LMT are discussed in Sec.~\ref{densitywave}. We summarize the main findings in Sec.~\ref{summary}. Some technical details are relegated to the Appendixes.

\section{Weyl semimetals: a general construction and topology}~\label{weylintro}

A WSM is realized when a three-dimensional solid state system lacks inversion and/or time-reversal symmetry, and the Kramers non-degenerate valence and conduction bands touch each other at isolated points in the Brillouin zone, known as Weyl nodes. The simplest realization (composed of only two Weyl points) arises from the following tight-binding model in a cubic lattice
\begin{equation}~\label{lattice}
H= \sum_{\boldsymbol k} \Psi^\dagger_{\boldsymbol k} \; \big[ \sigma_1 N_1({\boldsymbol k}_\perp) + \sigma_2 N_2({\boldsymbol k}_\perp) + \sigma_3 N_3(\boldsymbol k) \big] \Psi_{\boldsymbol k},
\end{equation}
where ${\boldsymbol k}_\perp=(k_x,k_y)$, $\Psi^\top ({\boldsymbol k})= (c_{\uparrow, \boldsymbol k}, c_{\downarrow, \boldsymbol k})$ is a two component spinor and $ \boldsymbol \sigma_i$ are three Pauli matrices. With the following choice: 
\begin{align}
N_3( \boldsymbol k)= -t [2-\cos (k_x a) -\cos (k_ya) ] -t \cos (k_z a)+m_z, \notag
\end{align}
where $a$ is the lattice spacing, two Weyl nodes are located along one of the $C_{4v}$ axes, namely at $k_z= \pm \cos^{-1}[m_z/(t_z a)]$. Now, if we choose $N_1 (\boldsymbol k)=t \sin (k_x a)$ and $N_2 (\boldsymbol k)=t \sin (k_ya)$, two Weyl nodes with monopole charge $\pm 1$ are realized at $\pm \bm{K}_0$, where $\bm{K}_0=(0,0,\cos^{-1}[m_z/(t_z a)])$. The linearized Hamiltonian for single WSM in the vicinity of the $\Gamma=(0,0,0)$ point is 
\begin{equation}\label{SWSM_Top}
H_{1}= \hbar(v_x \sigma_1 k_x + v_y \sigma_2 k_y) - \sigma_3  \frac{\hbar^2 \left(k^2_z+k^2_\perp \right)}{2m} +\Delta, 
\end{equation} 
where $v_{j}$s are the Fermi velocity of Weyl quasiparticles, $v_x=v_y=t a$, $m^{-1}=t a^2$, $\Delta=m_z-t$, and the momentum $({\boldsymbol k})$ is measured from the $\Gamma$ points. In the close proximity to these two Weyl points the low-energy excitations are respectively described by left and right chiral fermion, constituting source and sink of the Abelian Berry curvature, respectively. In an appropriate crystallographic environment it is also conceivable to realize Weyl points with higher monopole charge~\cite{Fang2012:Weyl,nagaosa-yang}. In particular Weyl nodes with monopole charge \emph{two} and \emph{three} can respectively be stabilized when the underlying lattice possesses tetragonal $C_{4v}$ and $C_{6v}$ symmetries, respectively~\cite{Fang2012:Weyl, nagaosa-yang}. A double WSM, for example, can be realized from Eq.~(\ref{lattice}) by choosing $N_1(\boldsymbol k) = t [\cos(k_x a) -\cos(k_y a)]$, $N_2 (\boldsymbol k) = t \sin(k_x a)\sin(k_y a)$, while leaving $N_3 (\boldsymbol k)$ unchanged~\cite{roy-bera}. But, in three spatial dimensions it is impossible to realize Weyl nodes with monopole charge larger than three from an underlying lattice model.

The above construction can also be viewed in the following way that will allow us to identify the topological invariant of these systems. Note that WSMs can be constructed by appropriately stacking two-dimensional layers of \emph{quantum anomalous Hall insulators} (QAHI) in the Brillouin zone along the $k_z$-direction. A two-dimensional QAHI supports quantized Hall conductivity $\sigma_{xy}=n e^2/h$, where $n$ represents the number of one-dimensional chiral edge states. For each $k_z$, satisfying $-\frac{\sqrt{2 m \Delta}}{\hbar}<k_z<\frac{\sqrt{2 m \Delta}}{\hbar}$, the pseudospin texture is \emph{skyrmion}, with skyrmion number $n$, and two Weyl nodes appear as \emph{singularities} in the Brillouin zone, across which the skyrmion number \emph{jumps} by $\pm n$. When stacked along the $k_z$ direction, the chiral edge modes from two-dimensional layers of QAHI produce the Fermi arc, \emph{possessing an $n$-fold orbital degeneracy}. The topological invariant of a WSM can be assessed from the gauge-invariant Berry curvature defined as~\cite{goswami-topology} 
\begin{align}
\Omega_{n,{\boldsymbol k},a}= \frac{(-1)^n}{4} \epsilon_{abc} {\boldsymbol n}_{\boldsymbol k} \cdot \left[ \frac{\partial {\boldsymbol n}_{\boldsymbol k}}{\partial k_b} \times \frac{\partial {\boldsymbol n}_{\boldsymbol k}}{\partial k_c} \right],
\end{align} 
where ${\boldsymbol n}_{\boldsymbol k}={\boldsymbol N}_{\boldsymbol k}/|{\boldsymbol N}_{\boldsymbol k}|$, and $n=\pm$ corresponds to valence and conduction band, respectively. Upon evaluating the components of the Abelian Berry curvature we plot it in Figs.~\ref{topology-SW}, ~\ref{topology-DW} and ~\ref{topology-TW} respectively for single, double, and triple WSM. The components of Berry curvature $\Omega_{n,{\boldsymbol k},x}$ and $\Omega_{n,{\boldsymbol k},y}$ are odd function of $k_x$ and $k_y$,  respectively. 
Hence, the number of field lines coming in and out of the $xz$ or $yz$ planes are equal, and the net Berry flux through these planes is zero. On the other hand, $\Omega_{n,{\boldsymbol k},z}$ is an even function of $\boldsymbol k$ and the net Berry flux through the $xy$ plane is $2 \pi {\mathcal S}(k_z)$, where ${\mathcal S}(k_z)$ is the Chern number of an effective two-dimensional system for a fixed $k_z$. For $-\frac{\sqrt{2 m \Delta}}{\hbar}<k_z<\frac{\sqrt{2 m \Delta}}{\hbar}$, ${\mathcal S}(k_z)=n$, while ${\mathcal S}(k_z)=0$ when $|k_z|>\frac{\sqrt{2 m \Delta}}{\hbar}$. Therefore, two Weyl points in a general WSM act as source and sink of Abelian Berry curvature, across which the Chern number of the underlying two-dimensional system jumps by an integer amount $n$.

In the low-energy limit the Hamiltonian for an WSM, constituted by Weyl nodes with monopole charge $\pm n$ is compactly written as 
\begin{equation}\label{hamillowenergy}
H_{\pm n}= \; \left( \begin{array}{cc}
\pm v_z \hbar k_z + \frac{\hbar^2 k^2_\perp}{2m} & \alpha_n \; (\hbar k_\perp)^n \; e^{-i n \theta_k} \\
\alpha_n \; (\hbar k_\perp)^n \; e^{i n \theta_k} & \mp v_z \hbar k_z-\frac{\hbar^2 k^2_\perp}{2m} 
\end{array}
\right),
\end{equation}    
after linearizing the generalized version of the Hamiltonian from Eq.~(\ref{SWSM_Top}) for general WSM around $\pm k^0_z$, where $k^0_z= \sqrt{2 m \Delta}$. In the above equation, therefore, $k_z$ is measured from $\pm k^0_z$. Thus, $\alpha_1$ and $\alpha_2$ respectively bear the unit of Fermi velocity and inverse of mass.

The majority of the known Weyl materials only break the inversion symmetry and the Weyl nodes are placed at different energies. Notice that the WSM arising from the tight-binding model, defined in Eq.~(\ref{lattice}), breaks both time-reversal and inversion symmetries, but the Weyl nodes are still located at the same energy. With simple modifications in this tight-binding model, for example by adding a term $[\Delta_0+ \Delta_{ch} \sin (k_z c)] \sigma_0$ to Eq.~(\ref{lattice}), the Weyl nodes can be placed at different energies, namely at $\Delta_0 \pm \Delta_{ch}$. Here, $\Delta_0$ and $\Delta_{ch}$ are respectively regular and chiral chemical potentials. However, our following discussion on various aspects of Weyl fermions in a strong magnetic field is qualitatively insensitive to such details of the system. Next we demonstrate how the underlying topological invariant of a WSM manifests through the LL spectrum when the system is subjected to a strong magnetic field.

\section{Landau levels}~\label{landaulevel}
\begin{figure}[!]
\includegraphics[scale=1]{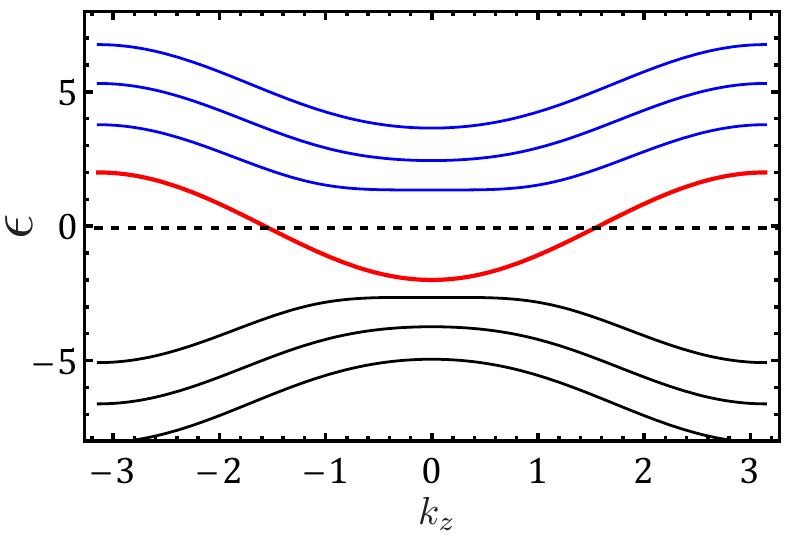}
\caption{\label{Fig:Spectrum-LLLattice} (Color online) Landau level (LL) spectrum in single Weyl semimetal, obtained from Eq.~(\ref{hamillowenergy}) when $n=1$. For details see Appendix~\ref{landaulevel_append}. Here the zeroth LL (red curve) is nondegenerate, and the energy $\epsilon$ (in units of $2 \alpha^2_1 \hbar^2/\ell_B^2$) is measured about a reference or zero point energy $\hbar^2/(2m \ell_B^2)$. At half-filling the zeroth LL cuts the Fermi energy (dotted line) at $k_z=\pm \pi/(2a)$ (we set $a=1$ for convenience)~\cite{Goswami2015}. Similar LL spectrum is found for double and triple Weyl semimetals, for which, however, the zeroth LL has two and three fold degenerate, respectively. }
\end{figure}

\begin{figure*}[!]
\subfigure[]{
\includegraphics[width=8.5cm, height=6.5cm]{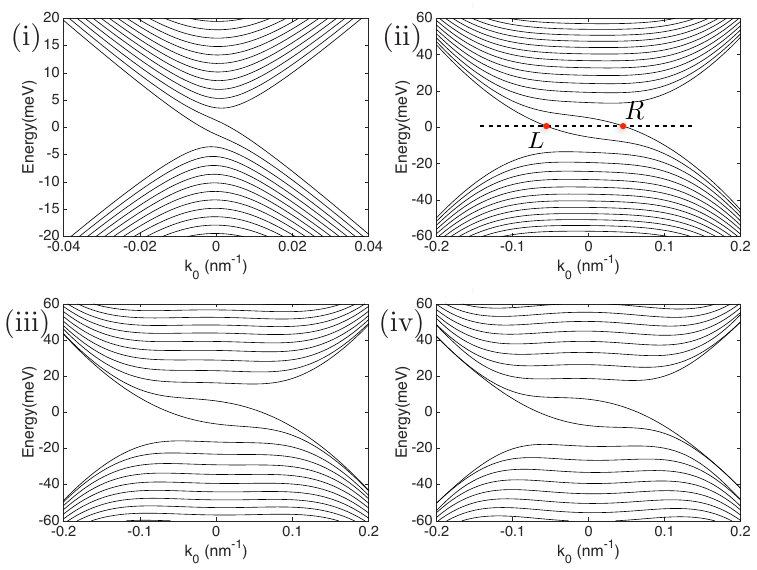}
\label{LL_DW_general}
}
\subfigure[]{
\includegraphics[width=8.5cm, height=6.5cm]{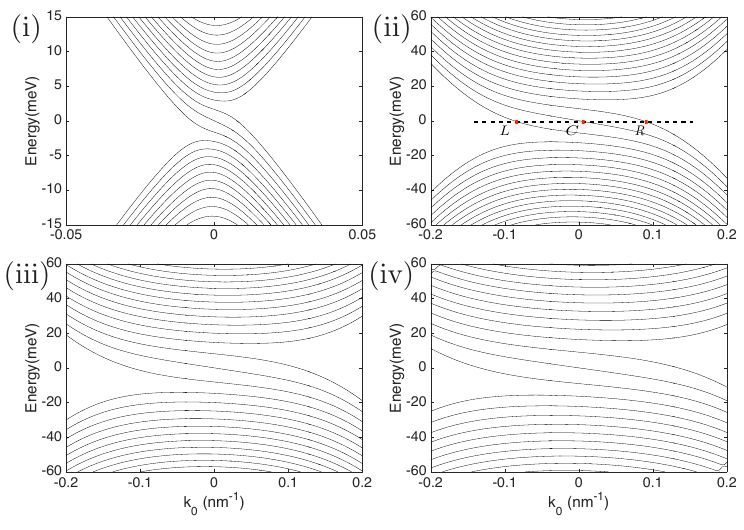}
\label{LL_TW_general}
}
\caption{\label{Fig:Spectrum-1} (Color online) (a) Landau level spectrum for double-Weyl semimetals when the magnetic field is $B = \SI{1}{T}$ and the direction of applied field is characterized by (i) $\theta = 3\pi/100$, (ii) $\theta = \pi/4$, (iii) $\theta = \pi/3$, and (iv) $\theta=\pi/2$ for $v_z = 658.2$\,meV$\cdot$nm and $\alpha_2 = 500$\,meV$\cdot$nm$^{2}$. (b) Landau level spectrum for triple-Weyl semimetals for same orientations of the magnetic field for $v_z = 658.2$\,meV$\cdot$nm and $\alpha_3 = 1782.29$\,meV$\cdot$nm$^{3}$. }
\end{figure*} 

Effects of the magnetic field applied in an arbitrary direction, can be captured through minimal coupling ${\boldsymbol k} \to {\boldsymbol k}-e {\boldsymbol A} \equiv {\boldsymbol \pi}$, where ${\boldsymbol A}$ is the vector potential, $e$ is electronic charge and the magnetic field is given by ${\boldsymbol B}=\nabla \times {\boldsymbol A}$. Before delving with the situation when the magnetic field is tilted away from the $z$ direction (parametrized by $\theta$, see Fig.~\ref{geometry}), let us first focus on a simpler situation when the field is applied along the $z$-direction, i.e. ${\boldsymbol B}=(0,0,1) B$, for which momentum $k_z$ is a conserved quantity. We can analytically obtain the LL spectrum~\cite{roy-sau, fiete}. For simplicity, we work with the Landau gauge $\bm{A} = (0, Bx, 0)$, and define the raising and lowering operators as $a =\ell(\pi_x - i\pi_y)/(\sqrt{2}\hbar)$ and $a^{\dagger} = \ell(\pi_x + i\pi_y)/(\sqrt{2}\hbar)$, where $\ell = \sqrt{\hbar/eB}$ is the magnetic length, and $[\pi_x, \pi_y] =-i\hbar eB$. The low energy Hamiltonian for a general Weyl semimetal then reads as 
\begin{align}
	H_{n,\theta=0} = \TwoDMatrix{\hbar v_z k_z}{\alpha_n\dfrac{(\sqrt{2}\hbar a)^n}{\ell^n}}{\alpha_n\dfrac{(\sqrt{2}\hbar a)^n}{\ell^n}}{- \hbar v_z k_z}. 
\end{align}
We here omit the term $\hbar^2 k^2_\perp/(2m)$ with respect to $k_x$ and $k_y$ for small momentum (near the Weyl nodes), mainly since inclusion of such term causes an overall shift in the LL energy at least when ${\boldsymbol B}=B \hat{z}$. The LL spectrum can then be readily obtained, yielding
\allowdisplaybreaks[4]
\begin{align}
	& E_{s,n, m}(k_z) \notag\\
	&= s\sqrt{v_z^2\hbar^2 k_z^2 + \dfrac{2\hbar^2\alpha_1^2}{\ell^2} m}, \, (m \geq 1) \nonumber \\
	&= s\sqrt{v_z^2 \hbar^2 k_z^2 + \dfrac{4\hbar^4\alpha_2^2}{\ell^4} m(m-1)}, \, (m \geq 2)  \\
	&= s\sqrt{v_z^2\hbar^2k_z^2 + \dfrac{8 \hbar^6\alpha_3^2}{\ell^6} m(m-1)(m-2)}, \, (m \geq 3),\nonumber
\end{align}  
for $s=\pm$ for single, double and triple WSM, respectively. For simple WSM the ZLL is comprised of a single branch with energy $E_{n=0} (k_z)=-\hbar v k_z$, while for double WSM $E_{n=0}(k_z) = E_{n=1}(k_z) = -\hbar vk_z$ and for a triple WSM $E_{n=0}(k_z) = E_{n=1}(k_z)=E_{n=2}(k_z) = -\hbar vk_z$. Thus, the ZLL for double and triple WSM enjoys additional \emph{two} and \emph{three} fold orbital degeneracy, respectively. We obtain exactly $n$ number of ZLL in the presence of bounded dispersion from a lattice model, as shown in Fig.~\ref{Fig:Spectrum-LLLattice} when ${\boldsymbol B}=B \hat{z}$. Details of the calculation are provided in Appendix~\ref{landaulevel_append}. \emph{Thus, the integer topological invariant of a WSM or monopole charge of the Weyl nodes sets the orbital degeneracy of the ZLL, at least when the field is applied along the separation of the Weyl nodes}. When the underlying lattice potential is taken into account, giving rise to bounded dispersion, the ZLL cuts the zero energy at $k_z=\pm \pi/(2 a)$, and in the following we focus near one such Weyl point.

Next we will show that the orbital degeneracy of ZLL remains unaffected with the tilting of the magnetic field away from the $z$-direction ($\theta \neq 0$), at least within the continuum description. For concreteness, we take the following vector potential $\bm{A} = (0, Bx\cos\theta, -Bx\sin\theta)$, giving rise to $\bm{B} = (0, B\sin\theta, B\cos\theta)$. The commutation relations between different momentum operators are given by $[\pi_x, \pi_y] = -i\hbar eB \cos\theta$ and $[\pi_x, \pi_z] = i\hbar eB \sin\theta$. We can define a pair of raising and lowering operators as
\begin{align}
	a &= \dfrac{\ell}{\sqrt{2}\hbar} [\pi_x - i(\pi_y \cos\theta - \pi_z \sin\theta)], \notag\\
	a^\dagger &= \dfrac{\ell}{\sqrt{2}\hbar} [\pi_x + i(\pi_y \cos\theta - \pi_z \sin\theta)], 
\end{align}
which satisfies the standard commutation relation $[a, a^\dagger] = 1$. The momentum along the magnetic field, defined as $k_0 \equiv k_y \sin\theta + k_z \cos\theta$, is a conserved quantity, and we find
\begin{align}
	\pi_x &= \dfrac{\hbar}{\sqrt{2}\ell}(a + a^{\dagger}),\;  
	\pi_y = \hbar k_0 \sin\theta + \dfrac{i\hbar}{\sqrt{2}\ell}(a-a^{\dagger})\cos\theta,\notag \\
	\pi_z &= \hbar k_0 \cos\theta - \dfrac{i\hbar}{\sqrt{2}\ell}(a-a^{\dagger})\sin\theta. \label{Eq:MomentumSubstitution}
\end{align}
The LL spectrum for an arbitrary orientation of the magnetic field $(\theta \neq 0)$ for arbitrary monopole charge $n$ cannot be obtained analytically. The numerically obtained LL spectra for double and triple WSM are respectively shown in Fig.~\ref{LL_DW_general} and Fig.~\ref{LL_TW_general}, for various choices of $\theta$. For simplicity, we have only shown the LL structure near one Weyl node, hosting a left chiral fermion. 
A similar structure is also realized for a right chiral fermion, and all LLs (including the zeroth one) are bounded due to the underlying lattice. Therefore, when $\theta \neq 0$ the exact degeneracy among the chiral ZLLs is lifted. However, there is always $n$ number of chiral ZLL crossing the zero energy. Such an outcome can be substantiated from the fact that in the absence of any magnetic field the monopole charge of the Weyl nodes is $\pm n$, which in turn determines the orbital degeneracy of the ZLL.

Although the LL spectrum obtained from the continuum description of WSM captures most of the essential features, a comment in this context is in order. When the field is tilted from the $z$ direction the two copies of ZLLs in double WSM cuts the zero energy at momentum $\pm [\pi/2 \pm \delta(\theta)] a^{-1}$, where $\delta(\theta)$ is dependent on the tilting angle, as shown in Fig.~\ref{LL_DW_general}, when the system is at half-filling. Similarly, in triple WSM the ZLL cuts the zero energy at $\pm [\pi/2 + j \delta(\theta)]a^{-1}$ for $j=-1,0,1$. In Figs.~\ref{LL_DW_general} and \ref{LL_TW_general}, the conserved momentum $k_0$ is measured from one of the emergent Weyl nodes in magnetic fields, located at $\pi/(2a)$. Next we discuss the resulting WSM DOS in the presence of a magnetic field.

\section{Density of states~\label{DOS}}

It is instructive to study separately the DOS in single, double, and triple WSM, when placed in a magnetic field, as it can be directly tested in angle-resolved quantum oscillation measurements~\cite{Pippard, behnia3}. In the presence of strong magnetic fields the kinetic energy in the plane perpendicular to ${\boldsymbol B}$ is completely quenched, while the LLs remain dispersive in the one dimension along the applied magnetic field. Thus the magnetic field causes an effective dimensional reduction of the system, and the WSM in the external magnetic field can be viewed as a \emph{collection of one-dimensional systems with multiple subbands}. The DOS of such a system can be found from the following definition: 
\begin{align}
	D(\mu) = \dfrac{L}{2\pi}\sum_{m}\int_{-\infty}^{+\infty} dk\,\delta[E_{m}(k)-\mu], 
\end{align}
where $m$ labels different subbands, and $k$ is the conserved momentum of the one-dimensional conducting channel along the magnetic field direction, and $L$ is the linear dimension along the field direction.

The DOS for a single WSM can be obtained in a closed-form. We first note that the dispersion for a single WSM can be cast in the following form: 
\begin{align}
	E_{m}(k) = \sqrt{\hbar^2 v_z^2 k^2 + w^2}, \label{Eq:WeylDispersion}
\end{align}
where $w = \sqrt{m}\epsilon_1$ and $\epsilon_1 = \sqrt{2}\hbar\alpha_1/\ell$. In such a system each $m \geq 1$ LLs crosses the Fermi energy $\mu$ twice, at $k_c = \pm \sqrt{\mu^2-w^2}/(\hbar v_z)$, whereas the $m=0$ LL only cuts the Fermi energy once, at $k = -\mu/(\hbar v_z)$. Therefore, the DOS for this single WSM is given by 
\begin{align}
	D(\mu) = D_0 \left[1+\sum_{m\geq1}\dfrac{2\mu\,\Theta(\mu-\sqrt{m}\epsilon_1)}{\sqrt{\mu^2-m\epsilon_1^2}}\right], 
\end{align}
where $D_0 = L/(2\pi\hbar v_z)$ and $\Theta(x)$ is the Heaviside Theta function. 

\begin{figure}[!]
\includegraphics[scale=1.1]{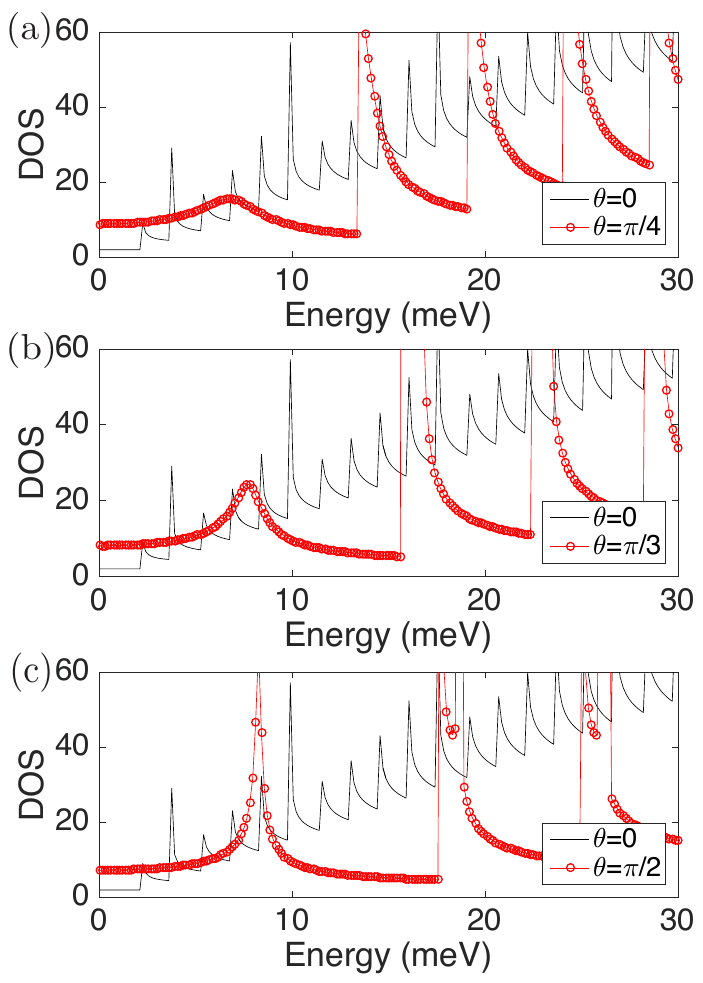}
\caption{\label{Fig:DOS_DW} (Color online) DOS for Landau levels in a double-Weyl semimetal, in units of $L/(2\pi\hbar v_z)$. The parameters are the same as those in Fig.~\ref{LL_DW_general}. The black curves are the DOS when the applied magnetic field is along the $z$ direction ($\theta=0$), see Eq.~\eqref{Eq:DOSExact}. }
\end{figure}  

\begin{figure}[!]
\includegraphics[scale=1.1]{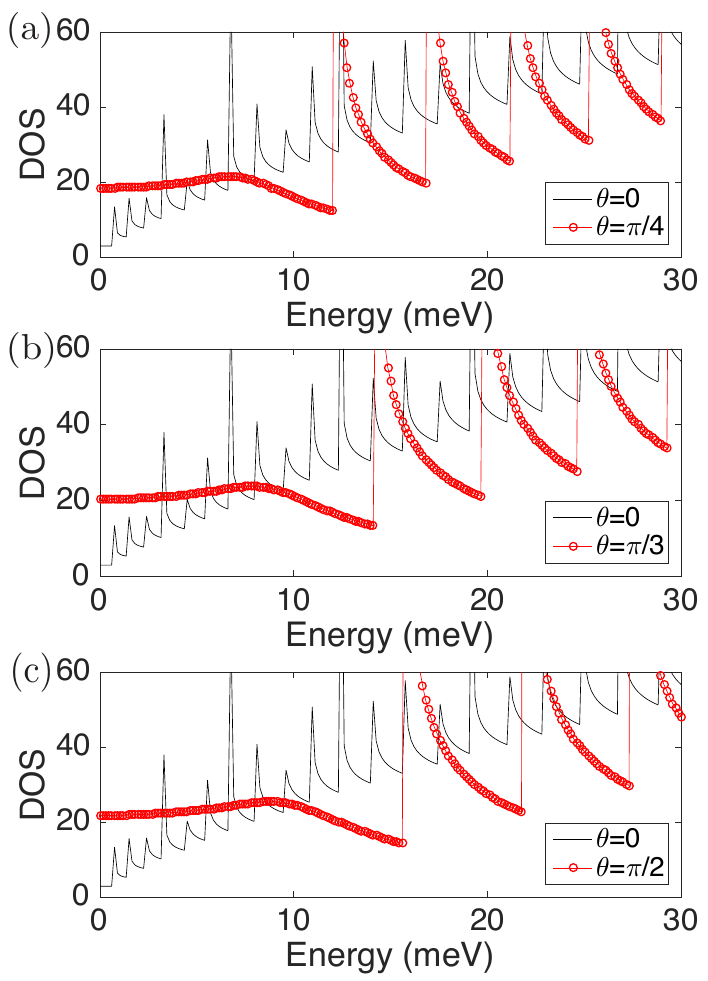}
\caption{\label{Fig:DOS_TW} (Color online) DOS for Landau levels in a triple-Weyl semimetal, in units of $L/(2\pi\hbar v_z)$. The parameters are the same as those in Fig.~\ref{LL_TW_general}. The black curves are the DOS when the applied magnetic field is along the $z$ direction ($\theta=0$), see Eq.~\eqref{Eq:DOSExact_TW}. }
\end{figure}  

Similarly we can find the DOS for double WSM analytically when the field is along the $k_z$ direction. The dispersion relation is then readily obtained from Eq.~\eqref{Eq:WeylDispersion} after replacing $w$ by $w=\sqrt{m(m-1)}\epsilon_2$, where $\epsilon_2 = 2\hbar^2 \alpha_2/\ell^2$, and the DOS for such a system is given by 
\begin{align}
	D(\mu) = D_0\left[2+\sum_{m\geq2}\dfrac{2\mu\,\Theta(\mu-\sqrt{m(m-1)}\epsilon_2)}{\sqrt{\mu^2-m(m-1)\epsilon_2^2}}\right], \label{Eq:DOSExact}
\end{align}
where we have accounted for the two-fold degeneracy of the ZLL. The above expression for the double WSM is plotted as black lines in Fig.~\ref{Fig:DOS_DW}. Similarly, it is easy to show that when the field is applied along the $z$ direction the DOS in triple WSM is given by 
\begin{align}
	D(\mu) = D_0\left[3+\sum_{m\geq 3}\dfrac{2\mu\,\Theta\left[\mu-\sqrt{m(m-1)(m-2)}\epsilon_3\right]}{\sqrt{\mu^2-m(m-1) (m-2)\epsilon_3^2}}\right], \label{Eq:DOSExact_TW}
\end{align}
where $\epsilon_3 = (\sqrt{2}\hbar)^3 \alpha_3/\ell^3$, as shown by the black lines in Fig.~\ref{Fig:DOS_TW}.

For arbitrary orientations of the magnetic field, we compute the DOS numerically. In Fig.~\ref{Fig:DOS_DW} we display the calculated DOS for three different orientations of the field in a double WSM. We find that the LL spacing continues to increase when the field is gradually tilted from $0$ to $\pi/2$. We also notice that as $\theta \to \pi/2$ an additional peak in DOS gradually develops within the manifold of ZLLs, which, however, can be an artifact of the continuum model. A similar result also holds for triple WSM, as shown in Fig.~\ref{Fig:DOS_TW}.

The DOS dictates the pattern of quantum oscillation measurements (through Shubnikov-de Haas effect or De Haas-van Alphen effect) of all transport and thermodynamic quantities. Therefore, the proposed anisotropic structure in DOS for multi-WSM can be detected from angle resolved quantum oscillations~\cite{Pippard, behnia3}. Such experiments can also test the robustness of the emergent peak in DOS within the ZLL of double WSM by tilting the magnetic field from the $z$ direction toward the $xy$ plane.

\section{Longitudinal magnetotransport}~\label{transport}

As shown in Sec.~\ref{landaulevel}, in the presence of strong magnetic fields the quasiparticle spectrum in any WSMs breaks into a set of LLs that remains dispersive along the direction of the applied field. 
Therefore, the application of an external magnetic field effectively breaks the system into a set of one-dimensional conducting wires, which is responsible for LMT. 
For the sake of simplicity and specificity, we assume that the external magnetic and electric (${\boldsymbol E}$) fields are always parallel to each other (see Fig.~\ref{geometry}). 
When ${\boldsymbol B}$ and ${\boldsymbol E}$ are not parallel, the transport occurs through both longitudinal and transverse components. However, the analysis of the transport properties for an arbitrary relative orientation of ${\boldsymbol B}$ and ${\boldsymbol E}$ fields falls outside the scope of the current work. 
To sustain a steady-state current, we take into account impurity-induced backscattering between the magnetic field induced emergent one-dimensional Weyl nodes, while the forward scattering contributes to the \emph{Dingle factor} for quantum oscillations. When such scattering is accounted for, the LMC (intra-band) can be written as 
\begin{align}
	\sigma(B) = \sum_{\alpha} g_\alpha \dfrac{e^2v_{F,\alpha}(B)\tau_{\alpha}(B)}{2\pi^2\hbar \ell^2}, 
	\label{Eq:ConductivityDefinition}
\end{align}
where the sum is over all partially filled Landau levels labeled by $\alpha$, $g_\alpha$ is the additional degeneracy due to internal degrees of freedom (such as spin), $v_{F,\alpha}(B)$ is the field dependent Fermi velocity obtained by linearizing the LL spectrum around the emergent one-dimensional Weyl nodes, and $\tau_{\alpha}(B)$ is the transport lifetime (obtained from backscattering, since forward scattering does not contribute to the resistivity). 
We here use the linearized Boltzmann equation to compute the lifetime and numerically extract the effective field dependent Fermi velocity by linearizing the spectrum of the ZLLs around the emergent one-dimensional Weyl nodes. 
We assume that only the subspace of ZLLs is partially filled and contributes to LMT. Two sources of elastic scattering are taken into account: (i) Gaussian disorder with 
\begin{align}
	U^{(G)}(\bq_{\perp},q_z) = U_0 e^{-R_0^2(q_{\perp}^2+q_z^2)/2}, \label{Eq:Gaussian}
\end{align}
where $R_0$ is the range of the impurity potential, and (ii) long-range ionic impurity scattering characterized by the screened Coulomb potential 
\begin{align}
U^{(C)}(\bq_{\perp},q_z) = \dfrac{U_c}{q_{\perp}^2+q_z^2+\qtf^2}, \label{Eq:CoulombPotential}
\end{align}
where $U_c\simeq 4\pi e^2/\kappa$ is the strength of the Coulomb impurity, and $\qtf$ is the Thomas-Fermi wave vector $\qtf$, 
\begin{align}
	\qtf(B) = \dfrac{e}{\sqrt{2\pi^2\hbar v_F(B)\ell_B^2}}, 
\end{align}
which is also a function of the magnetic field. Therefore with an increasing magnetic field, the screened Coulomb impurity potential gets more and more short-ranged as $\qtf$ increases. The increase in screening with increasing magnetic field follows from the enhancement of the DOS by the magnetic field. Note that in the extreme strong field limit the screened Coulomb potential becomes very small as $\qtf$ diverges. 

\subsection{Linearized Boltzmann equation}

The transport lifetime $\tau_{\alpha}(B)$ can be obtained from the linearized Boltzmann equation, which reads as~\cite{Goswami2015} 
\begin{align}
	1=\sum_{\beta,k_y',k_z'} W(k_y, k_z, \alpha; k_y', k_z', \beta)\biggl[ \tau_{\alpha}(k_z) 
	- \dfrac{v_{\beta}(k_z')}{v_{\alpha}(k_z)} \tau_{\beta}(k_z')\biggr], \notag
\end{align}
where the impurity matrix element squared is given by 
\allowdisplaybreaks[4]
\begin{align}
	W( k_y, &k_z, \alpha; k_y', k_z', \beta) 
	 = \dfrac{2\pi n_i}{\hbar}\int \dfrac{d^3q}{(2\pi)^3}|U(\bq)|^2 \label{Eq:W-Potential}\\ & 
\times |\bra{k_y, k_z, \alpha} e^{i\bq\cdot\br}\ket{k_y', k_z', \beta}|^2 \delta[\epsilon_{\alpha}(k_z)-\epsilon_{\beta}(k_z')]. \notag
\end{align}
After some involved algebra, the linearized Boltzmann equation can be cast in the following compact form 
\begin{widetext}
\begin{align}
	\dfrac{\hbar^2}{n_i\ell_{B}^2} &= \tau_{\alpha}(\kfa) 
\biggl[ \dfrac{2|U_{\alpha,\alpha}(2\kfa)|^2}{|v_{\alpha}(\kfa)|}
+ \sum_{\beta\neq\alpha}  
\dfrac{|U_{\alpha,\beta}(\kfa+\kfb)|^2+|U_{\alpha,\beta,}(\kfa-\kfb)|^2}{|v_{\beta}(\kfb)|}\biggr] \notag\\
&+ \sum_{\beta\neq\alpha}  \dfrac{\tau_{\beta}(\kfb)}{|v_{\alpha}(\kfa)|} \biggl[ |U_{\alpha,\beta}(\kfa+\kfb)|^2 - |U_{\alpha,\beta}(\kfa-\kfb)|^2 \biggr], \label{Eq:SimplifiedBoltzmann}
\end{align}
\end{widetext}
where $\kfa>0$ is the magnitude of the one-dimensional Fermi momentum in Landau level $\alpha$, $v_{\alpha}$ is the corresponding Fermi velocity and $n_i$ is impurity density. The effective one-dimensional interaction potential $U_{\alpha, \beta}(q_z)$ has the following form
\begin{align}
\!\!\!\!
\abs{U_{\alpha, \beta}(q_z)}^2 = \int \dfrac{d^2\bq_{\perp}}{(2\pi)^2\ell_B^2} \abs{U(\bq_{\perp}, q_z)}^2\,\abs{S_{\alpha,\beta}(\bq_{\perp})}^2, \label{Eq:EffectivePotential}
\end{align}
where $U(\bq_{\perp}, q_z)$ is the impurity potential, and $S_{\alpha, \beta}(\bq_{\perp})$ satisfies the relation
\begin{align}
\bra{k_y, k_z, \alpha} &e^{i\bq\cdot\br}\ket{k_y', k_z', \beta} \\ &\equiv S_{\alpha, \beta}(\bq_{\perp})\delta(k_y-k_y'+q_y)\delta(k_z-k_z'+q_z). \notag
\end{align}
One can then solve the coupled equations in Eq.~\eqref{Eq:SimplifiedBoltzmann} and obtain the transport lifetime $\tau_{\alpha}$ for each LL contributing to LMC. 

We can obtain the analytical expression for effective one-dimensional potentials from Eq.~\eqref{Eq:EffectivePotential}, when the field is applied along the $z$ direction (along which the Weyl nodes are separated in the absence of magnetic field). Let us first consider only the Gaussian impurity potential, for which the effective one-dimensional potential in a single WSM is given by  
\begin{align}
	|U_{0,0}^{(G)}(q_z)|^2 = \dfrac{U_0^2e^{-R_0^2q_z^2}}{2\pi\ell_B^4}\dfrac{1}{1+2\lambda }, 
\end{align}
where $\lambda = R_0^2/\ell_B^2$. However, with additional branches within the subspace of ZLLs, such as in double WSMs, supporting two ZLLs, there are the following two additional components for the effective potential 
\begin{align}
	|U_{1,1}^{(G)}(q_z)|^2 &= \dfrac{U_0^2e^{-R_0^2q_z^2}}{2\pi\ell_B^4} \dfrac{4\lambda^2+1}{(2\lambda +1)^3}, \\
	|U_{0,1}^{(G)}(q_z)|^2 	&= \dfrac{U_0^2e^{-R_0^2q_z^2}}{2\pi\ell_B^4} \dfrac{1}{(2\lambda +1)^2},
\end{align}
which respectively describes scattering potential within the $N=1$ ZLL and that between $N=0$ and $N=1$ ZLLs. The above formalism can immediately be generalized for triple WSMs, supporting three copies of ZLL.

Similarly the effective one-dimensional potential arising from Coulomb impurities is given by 
\begin{align}
|U_{\alpha,\beta}^{(C)}(q_z)|^2 = \dfrac{U_c^2}{2\pi}\int_0^{\infty}rdr \dfrac{|S_{\alpha\beta}(r)|^2}{\left(r^2+\eta^2 \right)^2}, 
\end{align}
where $\eta^2 = Q^2\ell_B^2 \equiv (q_z^2 + \qtf^2)\ell_B^2$. In single WSM  
\begin{align}
	|U_{0,0}^{(C)}(z)|^2 = \dfrac{U_c^2}{8\pi}e^{y}\Gamma(-1,y), 
\end{align} 
where $y=\eta^2/2$ and $\Gamma(s,z)$ is the \emph{upper incomplete Gamma function}. For double WSMs, the two additional components of the one-dimensional potential are 
\begin{align}
	|U_{1,1}^{(C)}(q_z)|^2 &= \dfrac{U_c^2}{8\pi}\left[e^{y}(y+1)(y+3)\Gamma(-1,y)-(1+2y^{-1})\right], \notag\\
	|U_{0,1}^{(C)}(q_z)|^2 &= \dfrac{U_c^2}{8\pi}\left[-e^{y}(y+1)\Gamma(-1,y)+y^{-1}\right], 
\end{align}
where the subscript notations are the same as that for Gaussian impurities. Note that in the extreme strong field limit, i.e., $y\to \infty$, all three effective potentials have an identical asymptotic form $|U_{0,0}^{(C)}(q_z)|^2=|U_{0,1}^{(C)}(q_z)|^2=|U_{1,1}^{(C)}(q_z)|^2 \simeq U_c^2/(8\pi y^2)$. For an arbitrary orientation of the magnetic field, we extract the effective one-dimensional potentials numerically. Next we discuss the LMC in single, double and triple WSM separately. 

\subsection{Magnetotransport in a single Weyl semimetal}


\begin{figure}[!]
\includegraphics[scale=1]{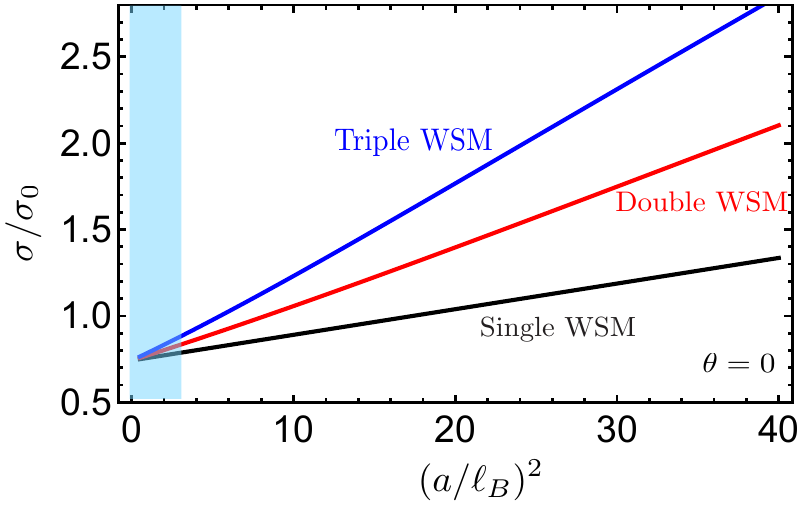}
\caption{\label{Fig:Conductivity-SingleWeyl} A comparative plot of longitudinal magnetoconductivity (LMC) in single (black), double (red) and triple (blue) Weyl semimetals in the presence of only Gaussian impurities. The field is applied along the axis separating the Weyl nodes in pristine system (the $z$ direction). In the shaded region Landau level are not sharp and one needs to account for weak localization effects. In this region the semiclassical theory for magnetotransport may also be applicable. Here, $\sigma_0=e^2\hbar v_z^2/(2\pi n_iU_0^2)$, and for discussion see Eq.~(\ref{Eq:LMC_SingleWeyl}).}
\end{figure}


To set the stage, we begin with the discussion of LMT in single WSM. In the presence of only Gaussian impurities, the scattering lifetime in single WSM is given by 
\begin{align}
\tau_G^{\text{(S)}} = \dfrac{2\pi\hbar^2\ell_B^2|v_F|}{2n_iU_0^2 I_{RR}}= \dfrac{\pi\hbar^2 |v_F|\ell_B^2}{n_iU_0^2}(1+2\lambda )e^{4\pi^2R_0^2/a^2}, \notag
\end{align} 
and the corresponding LMC is 
\begin{align}
	\sigma_G^{\text{(S)}} = \dfrac{e^2|v_F|}{2\pi^2\hbar\ell_B^2}\tau_G = \sigma_0 \; \left( 1+2\lambda \right) \; e^{4\pi^2t^2}, \label{Eq:LMC_SingleWeyl}
\end{align} 
where $\sigma_0=e^2\hbar v_z^2/(2\pi n_iU_0^2)$ and $t = R_0/a$. We have used the fact that the Fermi wave vector for backscattering in a single WSM is $q_z = \pi/a$. Therefore, LMC in single WSM, subject to only Gaussian impurities, increases linearly with the magnetic field, as shown in Fig.~\ref{Fig:Conductivity-SingleWeyl} (black curve). 

The scattering lifetime due to Coulomb impurities in a single WSM is given by 
\begin{align}
	\tau_C^{\text{(S)}} = \dfrac{\hbar^2|v_F|}{2n_i\ell_B^2|U_{0,0}^{(C)}(2k_F)|^2} 
	= \dfrac{\hbar^2|v_F|}{2n_iU_c^2\ell_B^2}\dfrac{8\pi e^{-y_c}}{\Gamma(-1,y_c)}, 
\end{align}
where	$y_c= \left[ (2k_F)^2\ell_B^2 + \qtf^2\ell_B^2 \right]/2$. Therefore, the total scattering lifetime ($\tau$) in the presence of both Gaussian and Coulomb impurities is given by 
\begin{align}
	\tau^{-1} = \tau_G^{-1} + \tau_C^{-1}, \label{Eq:Matthiessen} 
\end{align}
following the \emph{Matthiessen's rule} which is well-valid at low temperatures. Thus the total LMC in a single WSM is 
\begin{align}
	\sigma^{(S)} = \dfrac{e^2\hbar v_F^2}{2\pi n_ia^4U_c^2}Z^2\left[\dfrac{e^{y_c}}{4}\Gamma(-1,y_c) +\dfrac{ Z^2\mathcal{A}e^{-4\pi^2 t^2}}{(1+2t^2 Z)} \right]^{-1}, 
\end{align}
where $Z = a^2/\ell_B^2 \propto B$. Notice that in the presence of only Coulomb impurities the LMC grows as $B^2$ in the extremely strong field limit. However, in the strong magnetic field limit Gaussian impurities dominate the LMT (since typically $\tau_C \gg \tau_G$) and from here onward we only take into account the scattering lifetime arising from Gaussian impurities since the screened Coulomb disorder induced magnetoresistance vanishes in the strong-field limit as $1/B^2$.

\subsection{Magnetotransport in a double Weyl semimetal}

\begin{figure}[!]
\includegraphics[scale=1]{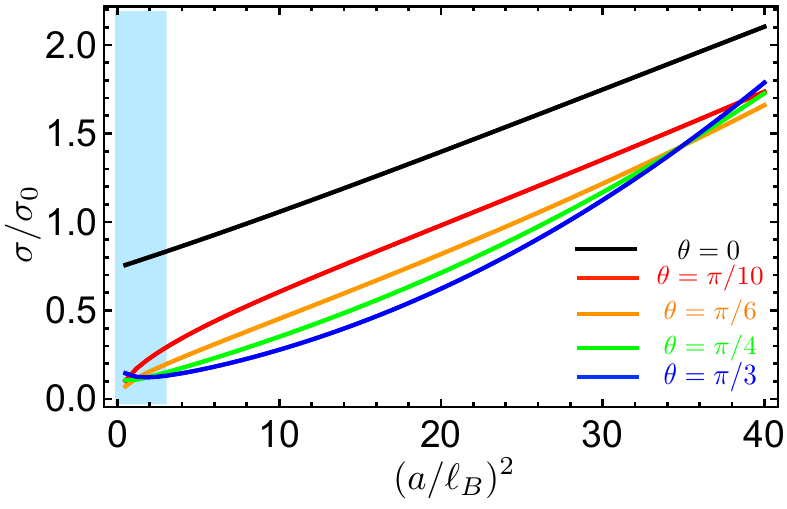}
\caption{\label{Fig:Conductivity-1} A plot of longitudinal magnetoconductivity (LMC) in double Weyl semimetal for various orientations of the applied magnetic  field (captured by the parameter $\theta$). We here only account for Gaussian impurities. In the shaded region weak localization effect dominates the LMC, where our analysis is qualitatively inapplicable. Here, $\sigma_0=e^2\hbar v_z^2/(2\pi n_iU_0^2)$, and for discussion see Eq.~(\ref{Eq:LMC_SingleWeyl}).}\label{LMC_DWS_1}
\end{figure} 

For a double WSM \emph{at half filling}, the coupled equations in Eq.~\eqref{Eq:SimplifiedBoltzmann} can be simplified since the Fermi velocities of the two ZLLs are exactly the same by symmetry. As a result the two coupled equations for the transport lifetime can be rewritten as 
\begin{align}
&\dfrac{\hbar^2|v_F|}{n_i\ell_B^2} = \tau_{R}\left[2\mathcal{I}_{RR}+\mathcal{I}_{RL}^{(-)}+\mathcal{I}_{RL}^{(+)}\right] + \tau_{L} \left[\mathcal{I}_{RL}^{(+)}-\mathcal{I}_{RL}^{(-)}\right], \notag\\
&= \tau_{L}\left[2\mathcal{I}_{LL}+\mathcal{I}_{LR}^{(-)}+\mathcal{I}_{LR}^{(+)}\right] + \tau_{R} \left[\mathcal{I}_{LR}^{(+)}-\mathcal{I}_{LR}^{(-)}\right],
\end{align}
where $v_F$ represents the same Fermi velocity for the two ZLLs, which we label $R$ and $L$, as shown in Fig.~\ref{LL_DW_general}. The backscattering potentials can be written as
\begin{align}
\mathcal{I}_{RR} = \abs{U_{R,R}(2k_{F,R})}^2, & \: \mathcal{I}_{RL}^{(\pm)} = \abs{U_{R,L}(k_{F,R}\pm k_{F,L})}^2,  \notag\\
\mathcal{I}_{LL} &= \abs{U_{L,L}(2k_{F,L})}^2, \label{Eq:EffectivePotential_DoubleWeyl}
\end{align}
where $U_{i,j} (x)$ is defined in Eq.~(\ref{Eq:EffectivePotential}) and by symmetry $\mathcal{I}_{LR}^{(\pm)} = \mathcal{I}_{RL}^{(\pm)}$. Therefore, one of the scattering lifetimes is 
\begin{align}
&\tau_{R} = \\
&\dfrac{\left(\mathcal{I}_{LL}+\mathcal{I}_{RL}^{(-)}\right)\left({\hbar^2|v_F|}/{n_i\ell_B^2}\right)}{(\mathcal{I}_{RR}+\mathcal{I}_{LL})\left(\mathcal{I}_{RL}^{(+)}+\mathcal{I}_{RL}^{(-)}\right)+2\left(\mathcal{I}_{RR}\mathcal{I}_{LL} + \mathcal{I}_{RL}^{(+)}\mathcal{I}_{RL}^{(-)}\right)}, \notag
\end{align}
 and $\tau_L$ is obtained after taking $L \leftrightarrow R$ in the above expression. Since the Fermi velocities of the two ZLLs are identical at half-filling, the total transport lifetime is $\tau_{L}+\tau_{R}$ and the LMC assumes the form
\begin{align}
	\sigma_G(B) = \dfrac{e^2\hbar }{\pi n_iU_0^2} v_F^2(B)K_G(B,\theta), 
\end{align}
where $K_G(B,\theta)$ is a dimensionless quantity, which for $\theta=0$ reads as  
\begin{align}
&K_G(B,0) \notag \\
&=\dfrac{e^{4\pi^2t^2}}{2}\dfrac{(2\lambda +1)^2[4\lambda^2+2\lambda +1+e^{4\pi^2t^2}(2\lambda +1)]}{2\lambda +4\lambda^2(\lambda +1)+1+e^{4\pi^2t^2}(1+2\lambda +2\lambda^2)},\notag\\
&\simeq \dfrac{e^{4\pi^2 t^2}}{2}
\begin{cases}
   1+4\lambda & \lambda \ll 1, \\
   2+4\lambda & \lambda \gg 1. 
\end{cases}
\end{align}
We display $\sigma_G(B)$ for $\theta=0$ in Fig.~\ref{Fig:Conductivity-SingleWeyl} (red curve). Thus for zero-range ($R_0=0$) impurities the LMC is constant, whereas for Gaussian impurities it increases monotonically and linearly with magnetic field $B$, thus giving rise to \emph{positive} LMC. Also notice that in the limit $t \gg 1$ (corresponding to fat Gaussian impurities) or $\lambda \gg 1$ (corresponding to strong magnetic field) the LMC for a double WSM is exactly twice that of a single-Weyl semimetal [see Eq.~\eqref{Eq:LMC_SingleWeyl} and Fig.~\ref{Fig:Conductivity-SingleWeyl}], since the scattering between two Landau levels with different indices ``L" and ``R" will be suppressed in comparison to that within the same LL. Thus in the strong field limit ($\lambda \gg1$) or fat Gaussian impurity ($t \gg 1$) limit the two ZLLs become effectively decoupled and conduct independently.

For general orientations of the magnetic field we cannot obtain the LMC analytically and we have to resort to a numerical approach. The results are shown in Fig.~\ref{LMC_DWS_1}. It is interesting to notice that when $\theta \neq 0$, LMC develops \emph{nonlinear} dependence on the field, and only at sufficiently strong fields the $B$-linear LMC is recovered. Such non-trivial dependence on magnetic field and tilted angle should be experimentally observable in transport measurements. For LMC in double WSM in the presence of only Coulomb or ionic impurities see Appendix~\ref{LMT_Coulomb_append}.

\subsection{Magnetotransport in a triple Weyl semimetal}

Finally we discuss LMC in triple WSM by restricting ourselves to Gaussian impurities. Unlike double WSM, the Fermi velocities of the three ZLLs (at and near half filling) are no longer equal: in general we have $v_L=v_R\neq v_C$, where $L$, $C$, and $R$ are the labels for the three branches of ZLL, shown in Fig.~\ref{LL_TW_general}. The LMC then assumes the following form: 
\begin{align}
	\sigma(B) &= \dfrac{e^2\hbar v_L}{\pi n_iU_0^2}\left[v_L(\tau_L+\tau_R) + v_C\tau_C\right] \notag\\ &= \sigma_0\left\{\dfrac{v_L^2(B)}{v_z^2}\left[\tilde{\tau}_L+\tilde{\tau}_R + w^{-1}\tilde{\tau}_C\right]\right\}, \label{LMC_TWSM} 
\end{align}
where  $w = v_L/v_C$ and $\tilde{\tau}_i = n_iU_0^2 \tau_i/(2\pi\ell^2\hbar^2v_L)$ is a dimensionless quantity.

The LMC in a triple WSM can be obtained analytically when the magnetic field is applied along the $z$ direction ($\theta=0$), see Fig.~\ref{Fig:Conductivity-SingleWeyl} (blue curve). Notice all three Fermi velocities are then equal ($w=1$). For sufficiently strong magnetic fields ($\lambda \gg1$) or fat Gaussian impurity potential ($t \gg 1$), three ZLLs again behave as decoupled one-dimensional wires and the LMC in a triple WSM is three times that of a single WSM; compare black and blue curves in Fig.~\ref{Fig:Conductivity-SingleWeyl}. 

\begin{figure}[!]
\includegraphics[scale=1]{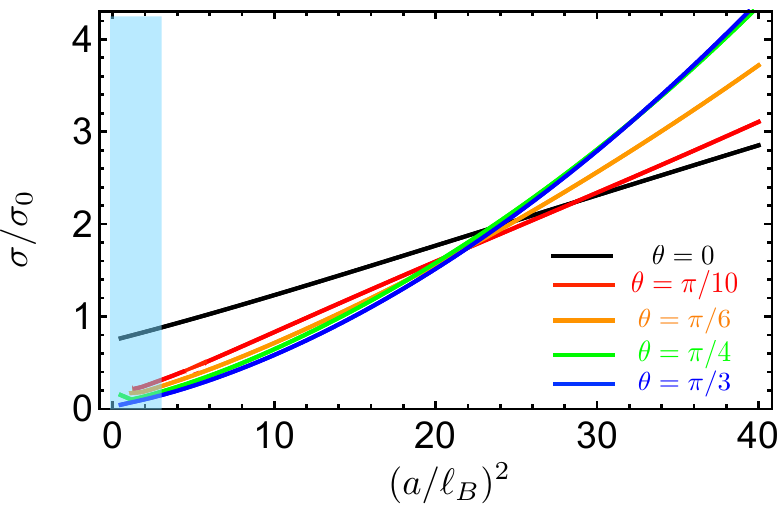}
\caption{\label{Fig:Conductivity-2} A plot of longitudinal magnetoconductivity (LMC) in triple Weyl semimetal for various orientations of the applied magnetic field (captured by the parameter $\theta$). We here only account for Gaussian impurities. In the shaded region weak localization effect dominates the LMC, where our analysis is qualitatively inapplicable. Here, $\sigma_0=e^2\hbar v_z^2/(2\pi n_iU_0^2)$, and for discussion see Eq.~(\ref{Eq:LMC_SingleWeyl}).}
\end{figure} 

For arbitrary orientations of the magnetic field, we compute various scattering elements, shown in Eq.~(\ref{LMC_TWSM}), numerically. In particular, notice that three Fermi vectors for backscattering have the magnitudes of $\pi/a$, and $\pi/a\pm k_c$, respectively, where $k_c$ is again a function of the direction and strength of the applied magnetic field. The resulting LMC in triple WSM is shown in Fig.~\ref{Fig:Conductivity-2}. Similar to the situation in double WSM, the LMC in triple WSM also exhibits an anisotropic angular dependence on the field orientations. In particular, as the tilting angle $\theta$ increases LMC displays a smooth and monotonic crossover from $B$-linear to \emph{nonlinear} in $B$ dependence, but LMC always remains positive. 

The magnetotransport results presented in this section are valid only when the WSM remains in its normal ``Fermi liquid'' phase, i.e., there is no symmetry-breaking transition induced by interactions. Any quantitative interaction effects can be included in the theory simply by reinterpreting each parameter (e.g., Fermi velocity, DOS) as the corresponding renormalized quantity within a Fermi liquid type many-body renormalization without any quantum phase transition. 
In particular, the theory of this section would not apply in the presence of any field-induced density-wave ordering transition which might happen because of electron-electron interactions. 
Next we will discuss the effect of various density-wave ordering within the manifold of ZLLs and its effects on LMT. 

\section{Electronic interaction and density-wave orderings}~\label{densitywave}

The DOS in a WSM, constituted by Weyl nodes with monopole charges $\pm n$, vanishes as $E^{2/n}$ as one approaches the Weyl points. Here the energy $E$ is measured from the Weyl points. Therefore, any weak electron-electron interaction (restricting ourselves to Hubbard-like local interactions, as appropriate for tight-binding lattice models) is an \emph{irrelevant} perturbation around the noninteracting Gaussian fixed point, and Weyl semimetals describe an infrared stable fixed point in the language of renormalization group. However, a strong magnetic field quenches  the quasiparticle dispersion into a set of discrete LLs, all of them dispersing along the applied magnetic field. The DOS for such effective one-dimensional systems is \emph{constant}, significantly enhancing the effects of electron-electron interaction. Therefore, sufficiently strong magnetic fields can hybridize the emergent Weyl nodes and gap them out through the formation of a density-wave order that breaks the \emph{translational symmetry}. This density wave ordering arises from the system minimizing the interaction energy at the cost of kinetic energy which is quenched by the magnetic field. 
Such a mechanism of developing density-wave order remains operative in any interacting three-dimensional electronic system in the presence of a strong external magnetic field~\cite{fukuyama, Sankar1, bryant, roy-sau}, and is akin to the density wave ordering in effective one-dimensional electron systems. In the continuum limit the CDW or SDW order breaks an emergent \emph{continuous} $U(1)$ translational symmetry~\cite{roy-sau, Li-Roy-DasSarma, Xiao-SnTe,Xiao-DW}, and disorder can couple to density-wave order as \emph{random field}. The Imry-Ma argument then prohibits such an order from acquiring a true long-range order~\cite{Imry}. However, in the presence of an underlying lattice, the density-wave orders only break \emph{discrete} translational symmetry. Hence, our proposed density-wave order (CDW or SDW) can exhibit a true long range ordering and may be observable at moderately strong magnetic fields. For the sake of simplicity, we here assume that only the manifold of ZLLs are partially filled, and concomitantly, density-wave ordering develops within this manifold. 
The following discussion can easily be generalized when multiple LLs are partially filled, but the quantitative details will be different with the density wave ordering trend being suppressed as more LLs participate in the energetics. 
The actual nature of the density-wave ordering (CDW or SDW) within ZLLs depends on the microscopic details of the Weyl system, as we discuss below.

To capture the low energy physics in such WSMs we appropriately define a four component spinor as $\Psi^\top(\vec{k})=[\Psi_{L}(\vec{k}), \Psi_{R}(\vec{k})]$, where $\Psi_{X}(\vec{k})$ are two component spinors for left ($L$) and right ($R$) chiral fermions, organized as $ \Psi_{X}^\top(\vec{k})=[ \Psi_{X, \uparrow} (\pm \vec{Q}+\vec{k}), \Psi_{X,\downarrow}(\pm \vec{Q}+\vec{k})]$ for $X=L,R$. In this notation, Weyl nodes are located at $\pm \vec{Q}$ and $\uparrow, \downarrow$ are the Kramers partners or two spin projections. Let us first focus on the low energy Hamiltonian for WSMs, shown in Eq.~(\ref{hamillowenergy}),  which can be written compactly as
\begin{align}
H_{0,n} = \tau_0 \otimes \left[  \sigma_1 \cos \theta_k +  \sigma_2 \sin \theta_k \right] \alpha_n k^{n}_\perp
+ \tau_3 \otimes \sigma_3 v_z k_z, \label{SWM_1} 
\end{align}       
where $\theta_k=\tan^{-1}(k_y/k_x)$. For such Weyl systems $\vec{Q}=Q \hat{z}$, hence the Weyl nodes are along the $z$ direction. Notice that two distinct types of density-wave orders, namely the CDW and SDW, can gap out the Weyl nodes at the cost of the translational symmetry. Respectively the effective single-particle Hamiltonian in the presence of these two orderings are
\begin{eqnarray}
H_{C/S} = \Delta_{C/S} \: \left( \tau_1 \otimes \cos \phi + \tau_2 \otimes \sin \phi \right) \otimes \sigma_{0/3}.
\end{eqnarray}      
In the continuum description $\phi$ is a continuous variable, and these two ordered phases break continuous $U(1)$ chiral symmetry generated by $U_c=\tau_3 \otimes \sigma_0$, representing the generator of translational symmetry~\cite{roy-sau}, and $[H_{0,n}, U_c]=0$,  whereas $\left\{U_c, H_{x} \right\}=0$ for $x=C$ and $S$. But, the underlying lattice potential lifts such continuous symmetry and prefers a certain locking angle for the order parameter. Therefore, the ordered phase is not accompanied by true massless Goldstone mode, similar to the situation with underlying valence bond solid or superconductor in two-dimensional graphene~\cite{roy-herbut}.

\begin{figure*}[!]
\includegraphics[scale=1.2]{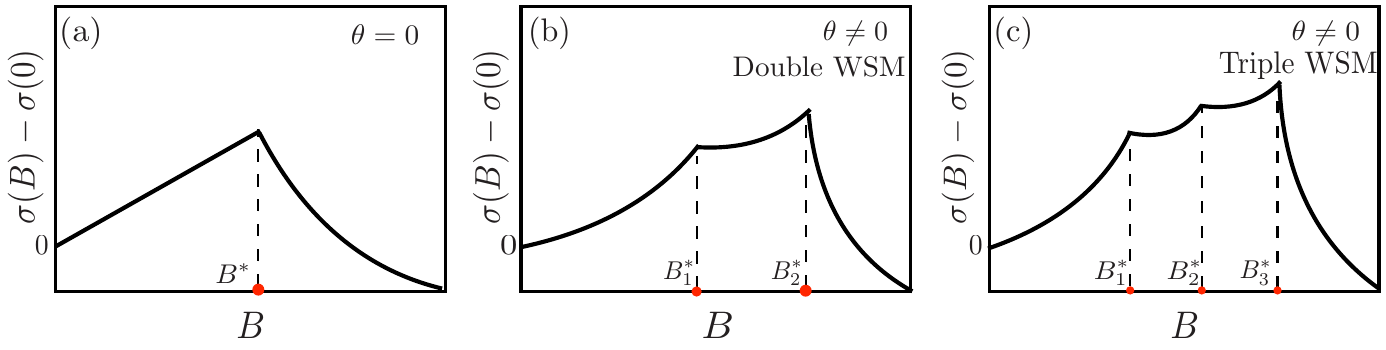}
\caption{A schematic variation of longitudinal magnetoconductivity (LMC) upon accounting for density-wave instabilities (either CDW or SDW) within the subspace of the zeroth Landau levels (ZLLs). (a) Qualitative variation of LMC is shown for any WSMs (constituted by Weyl nodes with monopole charge $n$) when the field is applied along the $z$ direction, and therefore $n$ number of ZLLs are exactly degenerate in a WSM. LMC becomes positive when $B>B^\ast$. Qualitative variation of LMC in (b) double and (c) triple Weyl semimetals when the field is tilted away from the $z$ direction (assuming the Weyl nodes are separated along the $z$ direction in the absence of magnetic field). Separate discontinuities in LMC arise since the $n$ ZLLs are no longer degenerate when $\theta \neq 0$. Each discontinuity is associated with the density-wave ordering in separate branches of the ZLLs, before LMC become negative when all ZLLs are gapped out. }\label{LMT-density-wave}
\end{figure*}

Even though both CDW and SDW orders gap out the ZLLs, a subtle competition between these two orders may ultimately determine the actual pattern of the symmetry breaking. Notice that $\left\{ H_{0,n}, H_{S} \right\}=0$ and the SDW order stands as a chiral symmetry breaking \emph{mass} for Weyl fermions. Thus, when SDW order sets in, besides gapping out the ZLLs, it also pushes down all filled LLs, following the spirit of \emph{magnetic catalysis}~\cite{miransky, kennett}. Hence, the SDW order appears to be energetically favored over CDW, when Weyl nodes are located along one of the $C_{4v}$ axes of a cubic system, for example $k_z$. 

It is also conceivable to realize the Weyl semimetals for which the low energy Hamiltonian is given by        
\begin{align}
\!\!\!\!
H_{0,n}= \tau_3 \otimes \left[ \alpha_n \; k^{n}_\perp \left[ \sigma_1 \cos \theta_k +  \sigma_2 \sin \theta_k \right] + \sigma_3 v_z k_z \right],   \label{SWM_2}
\end{align}
when the Weyl nodes are located along (1,1,1) direction in a cubic system, for example, and $\vec{Q}=Q (1,1,1)/\sqrt{3}$. Once again both CDW and SDW orders can generate a spectral gap within the manifold of ZLLs. However, the role of the CDW and SDW orders is now reversed; CDW representing a chiral symmetry breaking mass. Hence, in a WSM for which the Weyl nodes are placed along one of the $C_{3v}$ axes of a cubic system and the low energy Hamiltonian assumes the form in Eq.~(\ref{SWM_2}), the CDW order is energetically favored over the SDW one. 
Thus the nature of the density-wave ordering crucially depends on the microscopic details of the system. However, the signature of such translation symmetry breaking ordering on LMT is qualitatively insensitive to its actual nature since in both cases gaps open up in the spectrum.

As shown in the previous section, due to partially filled subspace of ZLLs, all WSMs manifest positive LMC or negative LMR, when these systems are placed in a strong magnetic field. On the other hand, electronic interaction can be conducive for translational symmetry breaking CDW or SDW order that produces a spectral gap within the ZLL. With the onset of such insulating ordering there is no partially filled LL in the system and LMR (LMC) can now manifest an upturn (downturn) and become positive (negative) in a strong magnetic field. Although such an instability occurs for an infinitesimal strength of interaction (as appropriate for the effective one-dimensional nature of the WSMs in the strong magnetic field) in a clean system for $T<T_c$, with $T_c$ being the transition temperature that scales as $T_c \sim \exp[-1/(g D(B))]$, where $g$ is the strength of interaction responsible for density-wave ordering and $D(B)$ is the DOS of ZLLs (following the BCS scaling law), the presence of disorder may hinder its onset. 
This can be seen most easily by realizing that the DOS $D(B)$ is suppressed by impurity scattering (e.g. Dingle temperature effect) leading to a strong suppression of $T_c$~\cite{Sankar1}. Thus only for sufficiently strong magnetic fields and at low temperature these orderings can set in and the system can display an upturn in LMR. Also, the system is more likely to manifest the density wave instability for larger interaction strengths since increasing $g$ enhances $T_c$.  
When the magnetic field is applied along $z$ direction, all ZLLs are degenerate in double and triple WSM (in a single WSM there is always only one ZLL); hence they simultaneously undergo manybody instability toward the formation of a density-wave order as temperature is gradually lowered below $T_c$. Concomitantly, all WSMs display negative LMC or positive LMR above a unique strength of magnetic field $B>B_\ast$, when the field is applied along the $z$ direction. The periodicity of such density-wave ordering is $2\pi/(2a)$. However, the situation changes dramatically when the field is tilted away from the $z$ direction in double and triple WSMs.

As shown in Sec.~\ref{landaulevel}, when the magnetic field is tilted away from the $z$ direction, the exact degeneracy among the ZLLs in double and triple WSM is lifted. As a result, the density-wave ordering in each ZLL is characterized by different transition temperatures, even though the degeneracy of the ZLL does not change with the field orientation, which can be qualitatively understood in the following way. For the sake of simplicity we can assume that short-range interactions arise from its long range Coulomb tail, which in three spatial dimensions scales as $V(q) \sim q^{-2}$. Then it is natural to assume that electronic interaction is stronger for density-wave ordering with smaller wave vectors since $V(q)$ increases for smaller $q$. Therefore, in double WSMs, as the temperature (magnetic field) is gradually lowered (increased) the Fermi points marked by $R$ get gapped out first, followed by the ones marked by $L$ [see Fig.~\ref{LL_DW_general}]. The periodicity of density-wave ordering in the $R$ and $L$ channels is respectively $2 \left[ \pi/2 \pm \delta (\theta) \right]/a$ where $\delta (\theta)$ depends on the orientation of the magnetic field. Similarly, in triple WSM, the mass generation first takes place among the $R$ points, followed by the $C$ points and finally the $L$ points, see Fig.~\ref{LL_TW_general}. The periodicity of density-wave ordering in these three channels are $2 \left[ \pi/2 + j \; \delta (\theta) \right]/a$, for $j=-1,0,1$ respectively for $R$, $C$ and $L$ channels. Thus as the temperature is gradually lowered in double and triple WSMs (with field tilted away from the $z$ direction), they undergo a cascade of transitions into density-wave phases with different periodicity, and below each such transition temperature the LMC displays \emph{discontinuity} and its rate of growth decreases, before it becomes negative when all ZLLs are gapped out, as shown in Figs.~\ref{LMT-density-wave}~(b) and~\ref{LMT-density-wave}~(c). 
In other words, in single WSM the LMR becomes positive once the density-wave ordering sets in, while in double (triple) Weyl semimetal LMR should display a two (three) stage transition toward a positive value. The separation among the transition temperature or onset magnetic field ($B^\ast_j$) among various density-wave transitions (hence the distance among various discontinuities in LMC/LMR) gets bigger as the magnetic field is gradually tilted away from the $z$ direction, since the emergent one-dimensional Fermi points gets further separated with increasing tilting of the magnetic field away from the axis separating two Weyl nodes [see Figs.~\ref{LL_DW_general} and~\ref{LL_TW_general}]. Such steps in LMR can possibly be detected in experiments and may serve as an indication for density-wave ordering in the system. 

\section{Summary and Discussion}~\label{summary}

To summarize, we here address several important aspects of Weyl semimetals when these systems are placed in a strong magnetic field. In the absence of magnetic fields, three-dimensional Weyl semimetals are constituted by Weyl nodes that can be classified according to the monopole charge $n=1$, $2$ and $3$. The integer monopole charge ($n$) determines the topological invariant of a Weyl semimetal, and the surface Fermi arc in such systems possesses an additional $n$-fold \emph{orbital} degeneracy. When placed in a magnetic field, the underlying topological invariant manifests itself through the number of chiral zeroth Landau levels. Thus double and triple Weyl semimetals respectively support two and three copies of the zeroth Landau level while a regular or single Weyl semimetal does not have any additional  degeneracy when the magnetic field is applied in the $z$ direction, along which the Weyl nodes are separated in the momentum space. The same conclusion also follows from a calculation based on a lattice model as shown in Appendix~\ref{landaulevel_append}. Within the framework of the continuum description of these system, we always find $n$ zeroth Landau levels, irrespective of the direction of the applied magnetic field, although their exact degeneracy breaks down when the field is tilted away from the $z$ direction. Also the density of states in the presence of Landau levels displays a strong angular dependence which can be observed in angle resolved quantum oscillation experiments.

We also show that the formation of Landau levels in Weyl materials can lead to longitudinal magnetotransport. 
To obtain the WSM magnetoconductivity we use the relaxation mechanism arising from the backscattering by impurities and taking into account two sources of elastic scattering (a) Gaussian impurity and (b) Coulomb impurity. However, due to finite density of states, the scattering potential in the effective one-dimensional Landau levels is always short ranged, and we mainly focus on the Gaussian disorder since the Coulomb disorder is screened by the carriers. 
Assuming that (i) the system is in the quantum limit ($\omega_c \tau >1$), and (ii) only the subspace of zeroth Landau level is partially filled, we calculate the magnetic field dependence of transport lifetime in the presence of strong backscattering using the quantum Boltzmann equation, showing that the longitudinal magnetoconductivity increases linearly with the applied magnetic field when the field is applied along the separation of Weyl nodes in the pristine system. Within the low-energy approximation, we find that such linear-$B$ dependence of the longitudinal magnetoconductance in double and triple Weyl semimetals smoothly crosses over to a nonlinear $B$ dependence when the field is tilted away from the $z$ direction, which can be tested experimentally. The longitudinal magnetoconductivity in all Weyl semimetals scales as $B^2$ when the magnetic field is applied along the separation of two pristine Weyl nodes, in the presence of Coulomb/ionic impurities, but in the extremely strong field limit.

In order to verify our predictions of LMC in systems dominant by Gaussian (short-ranged) and Coulomb (long-ranged) impurities, as well as the crossover between them, one can systematically introduce short-ranged or long-ranged impurities in a given sample, and vary their concentrations in a controlled way. 
The Gaussian and Coulomb disorder could arise from neutral and charged impurities/defects respectively, which are often invariably present in electronic materials anyway. 
For example, long-range disordered potentials can be generated by dopant charged impurities, while short-range disordered potentials can be introduced by radiation damage. Such an experimental technique provided the key evidence in finding out the relative importance of long-range versus short-range disorder potentials in the transport properties of graphene~\cite{EHHwang1,EHHwang2,adam2007self,Adam2008charged}. A similar technique could in principle be used in WSMs to test our predictions with respect to longitudinal magnetotransport.

Our current work on the magnetotransport properties of 3D WSMs, along with the earlier work of Goswami \emph{et al.}~\cite{Goswami2015} on magnetotransport in ordinary 3D metals which we follow, reinforce the point that positive longitudinal magnetoconductance may be a generic behavior in 3D systems subjected to a strong magnetic field and quenched disorder (i.e., impurity scattering).  
Much caution should therefore be used in interpreting the mere observation of a positive LMC as evidence in support of a chiral anomaly.  While the chiral anomaly indeed implies positive LMC, the reverse is not true.  Indeed, recent experiments~\cite{balicas,Wiedmann} find the presence of LMC in various 3D systems subjected to external magnetic fields where chiral anomalies are not expected to be operational.

Finally, we also address the effects of electron-electron interaction on longitudinal magnetotransport in Weyl semimetals. Due to their finite density of states of effective one-dimensional system, Weyl fermions are susceptible to various types of translational symmetry breaking density-wave ordering (charge or spin depending on microscopic details) even for infinitesimal strength of interaction in the presence of strong external magnetic fields. 
Since a single Weyl semimetal hosts one zeroth Landau level, the onset of a density-wave order immediately gives rise to positive longitudinal magnetoresistance. Such upturn in LMR can be observed below the transition temperature for sufficiently strong magnetic fields. A similar situation arises in double and triple Weyl semimetals when the field is applied along the $z$ direction, since all zeroth Landau levels are exactly degenerate. However, the situation changes dramatically as the applied field is gradually tilted away from the $z$ direction. Since, the two/three zeroth Landau levels in double/triple Weyl semimetals are no longer degenerate, they respectively undergo two and three transitions into density-wave phases, characterized by different periodicity as well as transition temperature. 
Thus, as the temperature is gradually lowered in the presence of a strong magnetic field (applied at a finite angle with respect to the $z$ direction), the decrease of the longitudinal magnetoresistance with magnetic field gets slower across each such transition, before it finally becomes positive when all members of the zeroth Landau level are gapped out by density-wave orderings. Thus future experiments in various members of Weyl semimetals, when placed in a magnetic field, can display intriguing confluence of Landau quantization, longitudinal magnetotransport (possibly manifesting one-dimensional chiral or axial anomaly), and interaction driven charge/spin-density-wave ordering.              
We note that the density wave ordering, particularly any charge density-wave ordering, should also generate a lattice Peierls instability in the WSM through the electron-phonon coupling, which could in principle be directly observed in a neutron scattering experiment in an external magnetic field. 

\acknowledgements

This work was supported by JQI-NSF-PFC and LPS-MPO-CMTC. We thank Pallab Goswami, M. Zahid Hasan for discussion. B.~R. is thankful to Nordita, Center for Quantum Materials for hospitality where part of this work was finalized.

\appendix

\section{Landau levels in a general Weyl semimetal from a tight-binding model}~\label{landaulevel_append}

\begin{figure}[!]
\includegraphics[scale=1]{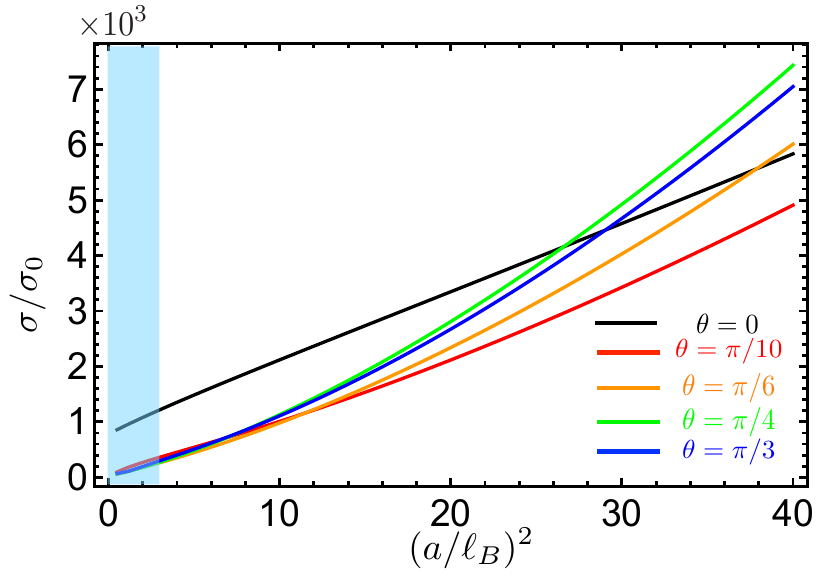}
\caption{ A plot of longitudinal magnetoconductivity (LMC) in double Weyl semimetal for various orientations of the applied magnetic  field (captured by the parameter $\theta$). We here only account for Coulomb impurities. In the shaded region weak localization effect dominates the LMC, where our analysis is qualitatively inapplicable.\label{Fig:LMC_Coulomb} }
\end{figure}

In this appendix we show the derivation of LL spectrum for single, double and triple WSM from a lattice model (bounded dispersion), when the magnetic field is applied along the separation of Weyl nodes, i.e. along the $z$ direction. Let us begin with the following model for a general WSM
\begin{align}
        H =& \left[ t_z\cos(k_z)-m_z + \dfrac{\hbar^2k_x^2+\hbar^2k_y^2}{2m}\right]\sigma_z \notag\\ &+
        \begin{pmatrix}
                0 & \alpha_n(\hbar k_x-i\hbar k_y)^n \\
                \alpha_n(\hbar k_x + i\hbar k_y)^n & 0
        \end{pmatrix},
\end{align}
where $n$ is the monopole charge. If the applied magnetic field is along the $z$ direction, the Landau level spectrum of this model can be expressed in a closed form. For example, for a single WSM $(n=1)$ we have for $N\geq 1$
\begin{align}
        E = \pm\sqrt{\left[\epsilon_z+\dfrac{\hbar^2}{2m\ell_B^2}(2N+1)\right]^2+\dfrac{2\alpha_1^2\hbar^2}{\ell_B^2}N},
\end{align}
where $\epsilon_z = t_z\cos(k_z)-m_z$, and $N$ is the Landau level index.
The energy dispersion for double $(n=2)$ and triple $(n=3)$ WSM can be similarly obtained as follows,
\allowdisplaybreaks[4]
\begin{align}
        & E_{N, m}(k_z) \\
        &= \pm \left[\left[\epsilon_z+\dfrac{\hbar^2}{2m\ell_B^2}(2N+1)\right]^2 + \dfrac{4\hbar^4\alpha_2^2}{\ell^4} N(N-1)\right]^{\frac{1}{2}}, \notag\\
        &= \pm \left[\left[\epsilon_z+\dfrac{\hbar^2}{2m\ell_B^2}(2N+1)\right]^2 + \dfrac{8 \hbar^6\alpha_3^2}{\ell^6} N(N-1)(N-2)\right]^{\frac{1}{2}},\notag
\end{align}
which is valid for $N\geq 2$ and $N\geq 3$, respectively. The resulting LL spectrum for single WSM has already been displayed in Fig.~\ref{Fig:Spectrum-LLLattice}. The energy of the ZLL (respectively one, two and three fold degenerate in single, double and triple WSM) is 
\begin{align}
E_{0}(k_z)=t_z\cos(k_z)-m_z +\dfrac{\hbar^2}{2m\ell_B^2}, 
\end{align}
and throughout we measured the energy from the reference value or zero point energy $\hbar^2/(2m\ell_B^2)$.

\section{Magnetoconductivity by Coulomb impurities}~\label{LMT_Coulomb_append}

In this appendix we want to briefly discuss the LMC in double WSMs only in the presence of Coulomb impurities. The results are displayed in Fig.~\ref{Fig:LMC_Coulomb} for various field orientations. We note that typically the transport lifetime due to Coulomb impurities are much longer than that for Gaussian impurities. Thus according to the Matthiessen's rule, shown in Eq.~\eqref{Eq:Matthiessen}, the transport lifetime is dominated by Gaussian impurities. However, in the complete absence of Gaussian impurities, one can expect to observe the LMC shown in Fig.~\ref{Fig:LMC_Coulomb}. We also want note that LMC driven by Coulomb impurities will show a $B^2$ dependence only in the limit of $B\to\infty$. We also find similar field dependence of LMC in the presence of only Coulomb impurities in single as well as triple WSMs. 

\bibliography{Bib-DoubleWeyl}

\begin{thebibliography}{108}%
\makeatletter
\providecommand \@ifxundefined [1]{%
 \@ifx{#1\undefined}
}%
\providecommand \@ifnum [1]{%
 \ifnum #1\expandafter \@firstoftwo
 \else \expandafter \@secondoftwo
 \fi
}%
\providecommand \@ifx [1]{%
 \ifx #1\expandafter \@firstoftwo
 \else \expandafter \@secondoftwo
 \fi
}%
\providecommand \natexlab [1]{#1}%
\providecommand \enquote  [1]{``#1''}%
\providecommand \bibnamefont  [1]{#1}%
\providecommand \bibfnamefont [1]{#1}%
\providecommand \citenamefont [1]{#1}%
\providecommand \href@noop [0]{\@secondoftwo}%
\providecommand \href [0]{\begingroup \@sanitize@url \@href}%
\providecommand \@href[1]{\@@startlink{#1}\@@href}%
\providecommand \@@href[1]{\endgroup#1\@@endlink}%
\providecommand \@sanitize@url [0]{\catcode `\\12\catcode `\$12\catcode
  `\&12\catcode `\#12\catcode `\^12\catcode `\_12\catcode `\%12\relax}%
\providecommand \@@startlink[1]{}%
\providecommand \@@endlink[0]{}%
\providecommand \url  [0]{\begingroup\@sanitize@url \@url }%
\providecommand \@url [1]{\endgroup\@href {#1}{\urlprefix }}%
\providecommand \urlprefix  [0]{URL }%
\providecommand \Eprint [0]{\href }%
\providecommand \doibase [0]{http://dx.doi.org/}%
\providecommand \selectlanguage [0]{\@gobble}%
\providecommand \bibinfo  [0]{\@secondoftwo}%
\providecommand \bibfield  [0]{\@secondoftwo}%
\providecommand \translation [1]{[#1]}%
\providecommand \BibitemOpen [0]{}%
\providecommand \bibitemStop [0]{}%
\providecommand \bibitemNoStop [0]{.\EOS\space}%
\providecommand \EOS [0]{\spacefactor3000\relax}%
\providecommand \BibitemShut  [1]{\csname bibitem#1\endcsname}%
\let\auto@bib@innerbib\@empty
\bibitem [{\citenamefont {Hasan}\ and\ \citenamefont
  {Kane}(2010)}]{kane-hasan-RMP}%
  \BibitemOpen
  \bibfield  {author} {\bibinfo {author} {\bibfnamefont {M.~Z.}\ \bibnamefont
  {Hasan}}\ and\ \bibinfo {author} {\bibfnamefont {C.~L.}\ \bibnamefont
  {Kane}},\ }\href {\doibase 10.1103/RevModPhys.82.3045} {\bibfield  {journal}
  {\bibinfo  {journal} {Rev. Mod. Phys.}\ }\textbf {\bibinfo {volume} {82}},\
  \bibinfo {pages} {3045} (\bibinfo {year} {2010})}\BibitemShut {NoStop}%
\bibitem [{\citenamefont {Qi}\ and\ \citenamefont
  {Zhang}(2011)}]{qi-zhang-RMP}%
  \BibitemOpen
  \bibfield  {author} {\bibinfo {author} {\bibfnamefont {X.-L.}\ \bibnamefont
  {Qi}}\ and\ \bibinfo {author} {\bibfnamefont {S.-C.}\ \bibnamefont {Zhang}},\
  }\href {\doibase 10.1103/RevModPhys.83.1057} {\bibfield  {journal} {\bibinfo
  {journal} {Rev. Mod. Phys.}\ }\textbf {\bibinfo {volume} {83}},\ \bibinfo
  {pages} {1057} (\bibinfo {year} {2011})}\BibitemShut {NoStop}%
\bibitem [{\citenamefont {Sau}\ \emph {et~al.}(2010)\citenamefont {Sau},
  \citenamefont {Lutchyn}, \citenamefont {Tewari},\ and\ \citenamefont {{Das
  Sarma}}}]{sau-prl}%
  \BibitemOpen
  \bibfield  {author} {\bibinfo {author} {\bibfnamefont {J.~D.}\ \bibnamefont
  {Sau}}, \bibinfo {author} {\bibfnamefont {R.~M.}\ \bibnamefont {Lutchyn}},
  \bibinfo {author} {\bibfnamefont {S.}~\bibnamefont {Tewari}}, \ and\ \bibinfo
  {author} {\bibfnamefont {S.}~\bibnamefont {{Das Sarma}}},\ }\href {\doibase
  10.1103/PhysRevLett.104.040502} {\bibfield  {journal} {\bibinfo  {journal}
  {Phys. Rev. Lett.}\ }\textbf {\bibinfo {volume} {104}},\ \bibinfo {pages}
  {040502} (\bibinfo {year} {2010})}\BibitemShut {NoStop}%
\bibitem [{\citenamefont {Hasan}\ \emph {et~al.}()\citenamefont {Hasan},
  \citenamefont {Xu},\ and\ \citenamefont {Neupane}}]{hasan-neupane}%
  \BibitemOpen
  \bibfield  {author} {\bibinfo {author} {\bibfnamefont {M.~Z.}\ \bibnamefont
  {Hasan}}, \bibinfo {author} {\bibfnamefont {S.-Y.}\ \bibnamefont {Xu}}, \
  and\ \bibinfo {author} {\bibfnamefont {M.}~\bibnamefont {Neupane}},\
  }\href@noop {} {\ }\Eprint {http://arxiv.org/abs/arXiv:1406.1040 (2014)}
  {arXiv:1406.1040 (2014)} \BibitemShut {NoStop}%
\bibitem [{\citenamefont {Ando}\ and\ \citenamefont {Fu}(2015)}]{fu-ando}%
  \BibitemOpen
  \bibfield  {author} {\bibinfo {author} {\bibfnamefont {Y.}~\bibnamefont
  {Ando}}\ and\ \bibinfo {author} {\bibfnamefont {L.}~\bibnamefont {Fu}},\
  }\href {\doibase 10.1146/annurev-conmatphys-031214-014501} {\bibfield
  {journal} {\bibinfo  {journal} {Annu. Rev. Condens. Matter Phys.}\ }\textbf
  {\bibinfo {volume} {6}},\ \bibinfo {pages} {361} (\bibinfo {year}
  {2015})}\BibitemShut {NoStop}%
\bibitem [{\citenamefont {Dzero}\ \emph {et~al.}(2016)\citenamefont {Dzero},
  \citenamefont {Xia}, \citenamefont {Galitski},\ and\ \citenamefont
  {Coleman}}]{galitski-coleman}%
  \BibitemOpen
  \bibfield  {author} {\bibinfo {author} {\bibfnamefont {M.}~\bibnamefont
  {Dzero}}, \bibinfo {author} {\bibfnamefont {J.}~\bibnamefont {Xia}}, \bibinfo
  {author} {\bibfnamefont {V.}~\bibnamefont {Galitski}}, \ and\ \bibinfo
  {author} {\bibfnamefont {P.}~\bibnamefont {Coleman}},\ }\href {\doibase
  10.1146/annurev-conmatphys-031214-014749} {\bibfield  {journal} {\bibinfo
  {journal} {Annu. Rev. Condens. Matter Phys.}\ }\textbf {\bibinfo {volume}
  {7}},\ \bibinfo {pages} {249} (\bibinfo {year} {2016})}\BibitemShut {NoStop}%
\bibitem [{\citenamefont {Volovik}(2009)}]{volovik}%
  \BibitemOpen
  \bibfield  {author} {\bibinfo {author} {\bibfnamefont {G.}~\bibnamefont
  {Volovik}},\ }\href@noop {} {\emph {\bibinfo {title} {The Universe in a
  Helium Droplet}}},\ International Series of Monographs on Physics\ (\bibinfo
  {publisher} {OUP Oxford},\ \bibinfo {year} {2009})\BibitemShut {NoStop}%
\bibitem [{\citenamefont {Burkov}(2015{\natexlab{a}})}]{burkov-review}%
  \BibitemOpen
  \bibfield  {author} {\bibinfo {author} {\bibfnamefont {A.~A.}\ \bibnamefont
  {Burkov}},\ }\href {\doibase 10.1088/0953-8984/27/11/113201} {\bibfield
  {journal} {\bibinfo  {journal} {J. Phys. Condens. Matter}\ }\textbf {\bibinfo
  {volume} {27}},\ \bibinfo {pages} {113201} (\bibinfo {year}
  {2015}{\natexlab{a}})}\BibitemShut {NoStop}%
\bibitem [{\citenamefont {Rao}()}]{rao-review}%
  \BibitemOpen
  \bibfield  {author} {\bibinfo {author} {\bibfnamefont {S.}~\bibnamefont
  {Rao}},\ }\href@noop {} {\ }\Eprint {http://arxiv.org/abs/arXiv:1603.02821
  (2016)} {arXiv:1603.02821 (2016)} \BibitemShut {NoStop}%
\bibitem [{\citenamefont {Weng}\ \emph {et~al.}(2015)\citenamefont {Weng},
  \citenamefont {Fang}, \citenamefont {Fang}, \citenamefont {Bernevig},\ and\
  \citenamefont {Dai}}]{TaAs1}%
  \BibitemOpen
  \bibfield  {author} {\bibinfo {author} {\bibfnamefont {H.}~\bibnamefont
  {Weng}}, \bibinfo {author} {\bibfnamefont {C.}~\bibnamefont {Fang}}, \bibinfo
  {author} {\bibfnamefont {Z.}~\bibnamefont {Fang}}, \bibinfo {author}
  {\bibfnamefont {B.~A.}\ \bibnamefont {Bernevig}}, \ and\ \bibinfo {author}
  {\bibfnamefont {X.}~\bibnamefont {Dai}},\ }\href {\doibase
  10.1103/PhysRevX.5.011029} {\bibfield  {journal} {\bibinfo  {journal} {Phys.
  Rev. X}\ }\textbf {\bibinfo {volume} {5}},\ \bibinfo {pages} {011029}
  (\bibinfo {year} {2015})}\BibitemShut {NoStop}%
\bibitem [{\citenamefont {Huang}\ \emph
  {et~al.}(2015{\natexlab{a}})\citenamefont {Huang}, \citenamefont {Xu},
  \citenamefont {Belopolski}, \citenamefont {Lee}, \citenamefont {Chang},
  \citenamefont {Wang}, \citenamefont {Alidoust}, \citenamefont {Bian},
  \citenamefont {Neupane}, \citenamefont {Zhang}, \citenamefont {Jia},
  \citenamefont {Bansil}, \citenamefont {Lin},\ and\ \citenamefont
  {Hasan}}]{TaAs2}%
  \BibitemOpen
  \bibfield  {author} {\bibinfo {author} {\bibfnamefont {S.-M.}\ \bibnamefont
  {Huang}}, \bibinfo {author} {\bibfnamefont {S.-Y.}\ \bibnamefont {Xu}},
  \bibinfo {author} {\bibfnamefont {I.}~\bibnamefont {Belopolski}}, \bibinfo
  {author} {\bibfnamefont {C.-C.}\ \bibnamefont {Lee}}, \bibinfo {author}
  {\bibfnamefont {G.}~\bibnamefont {Chang}}, \bibinfo {author} {\bibfnamefont
  {B.}~\bibnamefont {Wang}}, \bibinfo {author} {\bibfnamefont {N.}~\bibnamefont
  {Alidoust}}, \bibinfo {author} {\bibfnamefont {G.}~\bibnamefont {Bian}},
  \bibinfo {author} {\bibfnamefont {M.}~\bibnamefont {Neupane}}, \bibinfo
  {author} {\bibfnamefont {C.}~\bibnamefont {Zhang}}, \bibinfo {author}
  {\bibfnamefont {S.}~\bibnamefont {Jia}}, \bibinfo {author} {\bibfnamefont
  {A.}~\bibnamefont {Bansil}}, \bibinfo {author} {\bibfnamefont
  {H.}~\bibnamefont {Lin}}, \ and\ \bibinfo {author} {\bibfnamefont {M.~Z.}\
  \bibnamefont {Hasan}},\ }\href {\doibase 10.1038/ncomms8373} {\bibfield
  {journal} {\bibinfo  {journal} {Nat. Commun.}\ }\textbf {\bibinfo {volume}
  {6}},\ \bibinfo {pages} {7373} (\bibinfo {year}
  {2015}{\natexlab{a}})}\BibitemShut {NoStop}%
\bibitem [{\citenamefont {Zhang}\ \emph {et~al.}()\citenamefont {Zhang},
  \citenamefont {Yuan}, \citenamefont {Xu}, \citenamefont {Lin}, \citenamefont
  {Tong}, \citenamefont {Hasan}, \citenamefont {Wang}, \citenamefont {Zhang},\
  and\ \citenamefont {Jia}}]{TaAs4}%
  \BibitemOpen
  \bibfield  {author} {\bibinfo {author} {\bibfnamefont {C.}~\bibnamefont
  {Zhang}}, \bibinfo {author} {\bibfnamefont {Z.}~\bibnamefont {Yuan}},
  \bibinfo {author} {\bibfnamefont {S.-Y.}\ \bibnamefont {Xu}}, \bibinfo
  {author} {\bibfnamefont {Z.}~\bibnamefont {Lin}}, \bibinfo {author}
  {\bibfnamefont {B.}~\bibnamefont {Tong}}, \bibinfo {author} {\bibfnamefont
  {M.~Z.}\ \bibnamefont {Hasan}}, \bibinfo {author} {\bibfnamefont
  {J.}~\bibnamefont {Wang}}, \bibinfo {author} {\bibfnamefont {C.}~\bibnamefont
  {Zhang}}, \ and\ \bibinfo {author} {\bibfnamefont {S.}~\bibnamefont {Jia}},\
  }\href@noop {} {\ }\Eprint {http://arxiv.org/abs/arXiv:1502.00251 (2015)}
  {arXiv:1502.00251 (2015)} \BibitemShut {NoStop}%
\bibitem [{\citenamefont {Shekhar}\ \emph {et~al.}(2015)\citenamefont
  {Shekhar}, \citenamefont {Nayak}, \citenamefont {Sun}, \citenamefont
  {Schmidt}, \citenamefont {Nicklas}, \citenamefont {Leermakers}, \citenamefont
  {Zeitler}, \citenamefont {Skourski}, \citenamefont {Wosnitza}, \citenamefont
  {Liu}, \citenamefont {Chen}, \citenamefont {Schnelle}, \citenamefont
  {Borrmann}, \citenamefont {Grin}, \citenamefont {Felser},\ and\ \citenamefont
  {Yan}}]{Shekhar}%
  \BibitemOpen
  \bibfield  {author} {\bibinfo {author} {\bibfnamefont {C.}~\bibnamefont
  {Shekhar}}, \bibinfo {author} {\bibfnamefont {A.~K.}\ \bibnamefont {Nayak}},
  \bibinfo {author} {\bibfnamefont {Y.}~\bibnamefont {Sun}}, \bibinfo {author}
  {\bibfnamefont {M.}~\bibnamefont {Schmidt}}, \bibinfo {author} {\bibfnamefont
  {M.}~\bibnamefont {Nicklas}}, \bibinfo {author} {\bibfnamefont
  {I.}~\bibnamefont {Leermakers}}, \bibinfo {author} {\bibfnamefont
  {U.}~\bibnamefont {Zeitler}}, \bibinfo {author} {\bibfnamefont
  {Y.}~\bibnamefont {Skourski}}, \bibinfo {author} {\bibfnamefont
  {J.}~\bibnamefont {Wosnitza}}, \bibinfo {author} {\bibfnamefont
  {Z.}~\bibnamefont {Liu}}, \bibinfo {author} {\bibfnamefont {Y.}~\bibnamefont
  {Chen}}, \bibinfo {author} {\bibfnamefont {W.}~\bibnamefont {Schnelle}},
  \bibinfo {author} {\bibfnamefont {H.}~\bibnamefont {Borrmann}}, \bibinfo
  {author} {\bibfnamefont {Y.}~\bibnamefont {Grin}}, \bibinfo {author}
  {\bibfnamefont {C.}~\bibnamefont {Felser}}, \ and\ \bibinfo {author}
  {\bibfnamefont {B.}~\bibnamefont {Yan}},\ }\href {\doibase 10.1038/nphys3372}
  {\bibfield  {journal} {\bibinfo  {journal} {Nat. Phys.}\ }\textbf {\bibinfo
  {volume} {11}},\ \bibinfo {pages} {645} (\bibinfo {year} {2015})}\BibitemShut
  {NoStop}%
\bibitem [{\citenamefont {Borisenko}\ \emph {et~al.}()\citenamefont
  {Borisenko}, \citenamefont {Evtushinsky}, \citenamefont {Gibson},
  \citenamefont {Yaresko}, \citenamefont {Kim}, \citenamefont {Ali},
  \citenamefont {Buechner}, \citenamefont {Hoesch},\ and\ \citenamefont
  {Cava}}]{Borisenko}%
  \BibitemOpen
  \bibfield  {author} {\bibinfo {author} {\bibfnamefont {S.}~\bibnamefont
  {Borisenko}}, \bibinfo {author} {\bibfnamefont {D.}~\bibnamefont
  {Evtushinsky}}, \bibinfo {author} {\bibfnamefont {Q.}~\bibnamefont {Gibson}},
  \bibinfo {author} {\bibfnamefont {A.}~\bibnamefont {Yaresko}}, \bibinfo
  {author} {\bibfnamefont {T.}~\bibnamefont {Kim}}, \bibinfo {author}
  {\bibfnamefont {M.~N.}\ \bibnamefont {Ali}}, \bibinfo {author} {\bibfnamefont
  {B.}~\bibnamefont {Buechner}}, \bibinfo {author} {\bibfnamefont
  {M.}~\bibnamefont {Hoesch}}, \ and\ \bibinfo {author} {\bibfnamefont {R.~J.}\
  \bibnamefont {Cava}},\ }\href@noop {} {\ }\Eprint
  {http://arxiv.org/abs/arXiv:1507.04847 (2015)} {arXiv:1507.04847 (2015)}
  \BibitemShut {NoStop}%
\bibitem [{\citenamefont {Liu}\ \emph {et~al.}()\citenamefont {Liu},
  \citenamefont {Hu}, \citenamefont {Zhang}, \citenamefont {Graf},
  \citenamefont {Cao}, \citenamefont {Radmanesh}, \citenamefont {Adams},
  \citenamefont {Zhu}, \citenamefont {Cheng}, \citenamefont {Liu},
  \citenamefont {Phelan}, \citenamefont {Wei}, \citenamefont {Tennant},
  \citenamefont {DiTusa}, \citenamefont {Chiorescu}, \citenamefont {Spinu},\
  and\ \citenamefont {Mao}}]{Liu-Mao}%
  \BibitemOpen
  \bibfield  {author} {\bibinfo {author} {\bibfnamefont {J.~Y.}\ \bibnamefont
  {Liu}}, \bibinfo {author} {\bibfnamefont {J.}~\bibnamefont {Hu}}, \bibinfo
  {author} {\bibfnamefont {Q.}~\bibnamefont {Zhang}}, \bibinfo {author}
  {\bibfnamefont {D.}~\bibnamefont {Graf}}, \bibinfo {author} {\bibfnamefont
  {H.~B.}\ \bibnamefont {Cao}}, \bibinfo {author} {\bibfnamefont {S.~M.~A.}\
  \bibnamefont {Radmanesh}}, \bibinfo {author} {\bibfnamefont {D.~J.}\
  \bibnamefont {Adams}}, \bibinfo {author} {\bibfnamefont {Y.~L.}\ \bibnamefont
  {Zhu}}, \bibinfo {author} {\bibfnamefont {G.~F.}\ \bibnamefont {Cheng}},
  \bibinfo {author} {\bibfnamefont {X.}~\bibnamefont {Liu}}, \bibinfo {author}
  {\bibfnamefont {W.~A.}\ \bibnamefont {Phelan}}, \bibinfo {author}
  {\bibfnamefont {J.}~\bibnamefont {Wei}}, \bibinfo {author} {\bibfnamefont
  {D.~A.}\ \bibnamefont {Tennant}}, \bibinfo {author} {\bibfnamefont {J.~F.}\
  \bibnamefont {DiTusa}}, \bibinfo {author} {\bibfnamefont {I.}~\bibnamefont
  {Chiorescu}}, \bibinfo {author} {\bibfnamefont {L.}~\bibnamefont {Spinu}}, \
  and\ \bibinfo {author} {\bibfnamefont {Z.~Q.}\ \bibnamefont {Mao}},\
  }\href@noop {} {\ }\Eprint {http://arxiv.org/abs/arXiv:1507.07978 (2015)}
  {arXiv:1507.07978 (2015)} \BibitemShut {NoStop}%
\bibitem [{\citenamefont {Chang}\ \emph {et~al.}()\citenamefont {Chang},
  \citenamefont {Xu}, \citenamefont {Sanchez}, \citenamefont {Huang},
  \citenamefont {Lee}, \citenamefont {Chang}, \citenamefont {Zheng},
  \citenamefont {Bian}, \citenamefont {Belopolski}, \citenamefont {Alidoust},
  \citenamefont {Jeng}, \citenamefont {Bansil}, \citenamefont {Lin},\ and\
  \citenamefont {Hasan}}]{Chang-Hasan}%
  \BibitemOpen
  \bibfield  {author} {\bibinfo {author} {\bibfnamefont {G.}~\bibnamefont
  {Chang}}, \bibinfo {author} {\bibfnamefont {S.-Y.}\ \bibnamefont {Xu}},
  \bibinfo {author} {\bibfnamefont {D.~S.}\ \bibnamefont {Sanchez}}, \bibinfo
  {author} {\bibfnamefont {S.-M.}\ \bibnamefont {Huang}}, \bibinfo {author}
  {\bibfnamefont {C.-C.}\ \bibnamefont {Lee}}, \bibinfo {author} {\bibfnamefont
  {T.-R.}\ \bibnamefont {Chang}}, \bibinfo {author} {\bibfnamefont
  {H.}~\bibnamefont {Zheng}}, \bibinfo {author} {\bibfnamefont
  {G.}~\bibnamefont {Bian}}, \bibinfo {author} {\bibfnamefont {I.}~\bibnamefont
  {Belopolski}}, \bibinfo {author} {\bibfnamefont {N.}~\bibnamefont
  {Alidoust}}, \bibinfo {author} {\bibfnamefont {H.-T.}\ \bibnamefont {Jeng}},
  \bibinfo {author} {\bibfnamefont {A.}~\bibnamefont {Bansil}}, \bibinfo
  {author} {\bibfnamefont {H.}~\bibnamefont {Lin}}, \ and\ \bibinfo {author}
  {\bibfnamefont {M.~Z.}\ \bibnamefont {Hasan}},\ }\href@noop {} {\ }\Eprint
  {http://arxiv.org/abs/arXiv:1512.08781 (2015)} {arXiv:1512.08781 (2015)}
  \BibitemShut {NoStop}%
\bibitem [{\citenamefont {Xu}\ \emph {et~al.}(2015{\natexlab{a}})\citenamefont
  {Xu}, \citenamefont {Belopolski}, \citenamefont {Alidoust}, \citenamefont
  {Neupane}, \citenamefont {Bian}, \citenamefont {Zhang}, \citenamefont
  {Sankar}, \citenamefont {Chang}, \citenamefont {Yuan}, \citenamefont {Lee},
  \citenamefont {Huang}, \citenamefont {Zheng}, \citenamefont {Ma},
  \citenamefont {Sanchez}, \citenamefont {Wang}, \citenamefont {Bansil},
  \citenamefont {Chou}, \citenamefont {Shibayev}, \citenamefont {Lin},
  \citenamefont {Jia},\ and\ \citenamefont {Hasan}}]{TaAs3}%
  \BibitemOpen
  \bibfield  {author} {\bibinfo {author} {\bibfnamefont {S.-Y.}\ \bibnamefont
  {Xu}}, \bibinfo {author} {\bibfnamefont {I.}~\bibnamefont {Belopolski}},
  \bibinfo {author} {\bibfnamefont {N.}~\bibnamefont {Alidoust}}, \bibinfo
  {author} {\bibfnamefont {M.}~\bibnamefont {Neupane}}, \bibinfo {author}
  {\bibfnamefont {G.}~\bibnamefont {Bian}}, \bibinfo {author} {\bibfnamefont
  {C.}~\bibnamefont {Zhang}}, \bibinfo {author} {\bibfnamefont
  {R.}~\bibnamefont {Sankar}}, \bibinfo {author} {\bibfnamefont
  {G.}~\bibnamefont {Chang}}, \bibinfo {author} {\bibfnamefont
  {Z.}~\bibnamefont {Yuan}}, \bibinfo {author} {\bibfnamefont {C.-C.}\
  \bibnamefont {Lee}}, \bibinfo {author} {\bibfnamefont {S.-M.}\ \bibnamefont
  {Huang}}, \bibinfo {author} {\bibfnamefont {H.}~\bibnamefont {Zheng}},
  \bibinfo {author} {\bibfnamefont {J.}~\bibnamefont {Ma}}, \bibinfo {author}
  {\bibfnamefont {D.~S.}\ \bibnamefont {Sanchez}}, \bibinfo {author}
  {\bibfnamefont {B.}~\bibnamefont {Wang}}, \bibinfo {author} {\bibfnamefont
  {A.}~\bibnamefont {Bansil}}, \bibinfo {author} {\bibfnamefont
  {F.}~\bibnamefont {Chou}}, \bibinfo {author} {\bibfnamefont {P.~P.}\
  \bibnamefont {Shibayev}}, \bibinfo {author} {\bibfnamefont {H.}~\bibnamefont
  {Lin}}, \bibinfo {author} {\bibfnamefont {S.}~\bibnamefont {Jia}}, \ and\
  \bibinfo {author} {\bibfnamefont {M.~Z.}\ \bibnamefont {Hasan}},\ }\href
  {\doibase 10.1126/science.aaa9297} {\bibfield  {journal} {\bibinfo  {journal}
  {Science}\ }\textbf {\bibinfo {volume} {349}},\ \bibinfo {pages} {613}
  (\bibinfo {year} {2015}{\natexlab{a}})}\BibitemShut {NoStop}%
\bibitem [{\citenamefont {Lv}\ \emph {et~al.}(2015)\citenamefont {Lv},
  \citenamefont {Weng}, \citenamefont {Fu}, \citenamefont {Wang}, \citenamefont
  {Miao}, \citenamefont {Ma}, \citenamefont {Richard}, \citenamefont {Huang},
  \citenamefont {Zhao}, \citenamefont {Chen}, \citenamefont {Fang},
  \citenamefont {Dai}, \citenamefont {Qian},\ and\ \citenamefont
  {Ding}}]{Ding-PRX}%
  \BibitemOpen
  \bibfield  {author} {\bibinfo {author} {\bibfnamefont {B.~Q.}\ \bibnamefont
  {Lv}}, \bibinfo {author} {\bibfnamefont {H.~M.}\ \bibnamefont {Weng}},
  \bibinfo {author} {\bibfnamefont {B.~B.}\ \bibnamefont {Fu}}, \bibinfo
  {author} {\bibfnamefont {X.~P.}\ \bibnamefont {Wang}}, \bibinfo {author}
  {\bibfnamefont {H.}~\bibnamefont {Miao}}, \bibinfo {author} {\bibfnamefont
  {J.}~\bibnamefont {Ma}}, \bibinfo {author} {\bibfnamefont {P.}~\bibnamefont
  {Richard}}, \bibinfo {author} {\bibfnamefont {X.~C.}\ \bibnamefont {Huang}},
  \bibinfo {author} {\bibfnamefont {L.~X.}\ \bibnamefont {Zhao}}, \bibinfo
  {author} {\bibfnamefont {G.~F.}\ \bibnamefont {Chen}}, \bibinfo {author}
  {\bibfnamefont {Z.}~\bibnamefont {Fang}}, \bibinfo {author} {\bibfnamefont
  {X.}~\bibnamefont {Dai}}, \bibinfo {author} {\bibfnamefont {T.}~\bibnamefont
  {Qian}}, \ and\ \bibinfo {author} {\bibfnamefont {H.}~\bibnamefont {Ding}},\
  }\href {\doibase 10.1103/PhysRevX.5.031013} {\bibfield  {journal} {\bibinfo
  {journal} {Phys. Rev. X}\ }\textbf {\bibinfo {volume} {5}},\ \bibinfo {pages}
  {031013} (\bibinfo {year} {2015})}\BibitemShut {NoStop}%
\bibitem [{\citenamefont {Xu}\ \emph {et~al.}(2015{\natexlab{b}})\citenamefont
  {Xu}, \citenamefont {Alidoust}, \citenamefont {Belopolski}, \citenamefont
  {Yuan}, \citenamefont {Bian}, \citenamefont {Chang}, \citenamefont {Zheng},
  \citenamefont {Strocov}, \citenamefont {Sanchez}, \citenamefont {Chang},
  \citenamefont {Zhang}, \citenamefont {Mou}, \citenamefont {Wu}, \citenamefont
  {Huang}, \citenamefont {Lee}, \citenamefont {Huang}, \citenamefont {Wang},
  \citenamefont {Bansil}, \citenamefont {Jeng}, \citenamefont {Neupert},
  \citenamefont {Kaminski}, \citenamefont {Lin}, \citenamefont {Jia},\ and\
  \citenamefont {{Zahid Hasan}}}]{Hasan-NaturePhysics}%
  \BibitemOpen
  \bibfield  {author} {\bibinfo {author} {\bibfnamefont {S.-Y.}\ \bibnamefont
  {Xu}}, \bibinfo {author} {\bibfnamefont {N.}~\bibnamefont {Alidoust}},
  \bibinfo {author} {\bibfnamefont {I.}~\bibnamefont {Belopolski}}, \bibinfo
  {author} {\bibfnamefont {Z.}~\bibnamefont {Yuan}}, \bibinfo {author}
  {\bibfnamefont {G.}~\bibnamefont {Bian}}, \bibinfo {author} {\bibfnamefont
  {T.-R.}\ \bibnamefont {Chang}}, \bibinfo {author} {\bibfnamefont
  {H.}~\bibnamefont {Zheng}}, \bibinfo {author} {\bibfnamefont {V.~N.}\
  \bibnamefont {Strocov}}, \bibinfo {author} {\bibfnamefont {D.~S.}\
  \bibnamefont {Sanchez}}, \bibinfo {author} {\bibfnamefont {G.}~\bibnamefont
  {Chang}}, \bibinfo {author} {\bibfnamefont {C.}~\bibnamefont {Zhang}},
  \bibinfo {author} {\bibfnamefont {D.}~\bibnamefont {Mou}}, \bibinfo {author}
  {\bibfnamefont {Y.}~\bibnamefont {Wu}}, \bibinfo {author} {\bibfnamefont
  {L.}~\bibnamefont {Huang}}, \bibinfo {author} {\bibfnamefont {C.-C.}\
  \bibnamefont {Lee}}, \bibinfo {author} {\bibfnamefont {S.-M.}\ \bibnamefont
  {Huang}}, \bibinfo {author} {\bibfnamefont {B.}~\bibnamefont {Wang}},
  \bibinfo {author} {\bibfnamefont {A.}~\bibnamefont {Bansil}}, \bibinfo
  {author} {\bibfnamefont {H.-T.}\ \bibnamefont {Jeng}}, \bibinfo {author}
  {\bibfnamefont {T.}~\bibnamefont {Neupert}}, \bibinfo {author} {\bibfnamefont
  {A.}~\bibnamefont {Kaminski}}, \bibinfo {author} {\bibfnamefont
  {H.}~\bibnamefont {Lin}}, \bibinfo {author} {\bibfnamefont {S.}~\bibnamefont
  {Jia}}, \ and\ \bibinfo {author} {\bibfnamefont {M.}~\bibnamefont {{Zahid
  Hasan}}},\ }\href {\doibase 10.1038/nphys3437} {\bibfield  {journal}
  {\bibinfo  {journal} {Nat. Phys.}\ }\textbf {\bibinfo {volume} {11}},\
  \bibinfo {pages} {748} (\bibinfo {year} {2015}{\natexlab{b}})}\BibitemShut
  {NoStop}%
\bibitem [{\citenamefont {Xu}\ \emph {et~al.}(2016)\citenamefont {Xu},
  \citenamefont {Weng}, \citenamefont {Lv}, \citenamefont {Matt}, \citenamefont
  {Park}, \citenamefont {Bisti}, \citenamefont {Strocov}, \citenamefont
  {Gawryluk}, \citenamefont {Pomjakushina}, \citenamefont {Conder},
  \citenamefont {Plumb}, \citenamefont {Radovic}, \citenamefont {Aut{\`{e}}s},
  \citenamefont {Yazyev}, \citenamefont {Fang}, \citenamefont {Dai},
  \citenamefont {Qian}, \citenamefont {Mesot}, \citenamefont {Ding},\ and\
  \citenamefont {Shi}}]{Xu-Shi}%
  \BibitemOpen
  \bibfield  {author} {\bibinfo {author} {\bibfnamefont {N.}~\bibnamefont
  {Xu}}, \bibinfo {author} {\bibfnamefont {H.}~\bibnamefont {Weng}}, \bibinfo
  {author} {\bibfnamefont {B.~Q.}\ \bibnamefont {Lv}}, \bibinfo {author}
  {\bibfnamefont {C.~E.}\ \bibnamefont {Matt}}, \bibinfo {author}
  {\bibfnamefont {J.}~\bibnamefont {Park}}, \bibinfo {author} {\bibfnamefont
  {F.}~\bibnamefont {Bisti}}, \bibinfo {author} {\bibfnamefont {V.~N.}\
  \bibnamefont {Strocov}}, \bibinfo {author} {\bibfnamefont {D.}~\bibnamefont
  {Gawryluk}}, \bibinfo {author} {\bibfnamefont {E.}~\bibnamefont
  {Pomjakushina}}, \bibinfo {author} {\bibfnamefont {K.}~\bibnamefont
  {Conder}}, \bibinfo {author} {\bibfnamefont {N.~C.}\ \bibnamefont {Plumb}},
  \bibinfo {author} {\bibfnamefont {M.}~\bibnamefont {Radovic}}, \bibinfo
  {author} {\bibfnamefont {G.}~\bibnamefont {Aut{\`{e}}s}}, \bibinfo {author}
  {\bibfnamefont {O.~V.}\ \bibnamefont {Yazyev}}, \bibinfo {author}
  {\bibfnamefont {Z.}~\bibnamefont {Fang}}, \bibinfo {author} {\bibfnamefont
  {X.}~\bibnamefont {Dai}}, \bibinfo {author} {\bibfnamefont {T.}~\bibnamefont
  {Qian}}, \bibinfo {author} {\bibfnamefont {J.}~\bibnamefont {Mesot}},
  \bibinfo {author} {\bibfnamefont {H.}~\bibnamefont {Ding}}, \ and\ \bibinfo
  {author} {\bibfnamefont {M.}~\bibnamefont {Shi}},\ }\href {\doibase
  10.1038/ncomms11006} {\bibfield  {journal} {\bibinfo  {journal} {Nat.
  Commun.}\ }\textbf {\bibinfo {volume} {7}},\ \bibinfo {pages} {11006}
  (\bibinfo {year} {2016})}\BibitemShut {NoStop}%
\bibitem [{\citenamefont {Inoue}\ \emph {et~al.}(2016)\citenamefont {Inoue},
  \citenamefont {Gyenis}, \citenamefont {Wang}, \citenamefont {Li},
  \citenamefont {Oh}, \citenamefont {Jiang}, \citenamefont {Ni}, \citenamefont
  {Bernevig},\ and\ \citenamefont {Yazdani}}]{yazdani-1}%
  \BibitemOpen
  \bibfield  {author} {\bibinfo {author} {\bibfnamefont {H.}~\bibnamefont
  {Inoue}}, \bibinfo {author} {\bibfnamefont {A.}~\bibnamefont {Gyenis}},
  \bibinfo {author} {\bibfnamefont {Z.}~\bibnamefont {Wang}}, \bibinfo {author}
  {\bibfnamefont {J.}~\bibnamefont {Li}}, \bibinfo {author} {\bibfnamefont
  {S.~W.}\ \bibnamefont {Oh}}, \bibinfo {author} {\bibfnamefont
  {S.}~\bibnamefont {Jiang}}, \bibinfo {author} {\bibfnamefont
  {N.}~\bibnamefont {Ni}}, \bibinfo {author} {\bibfnamefont {B.~A.}\
  \bibnamefont {Bernevig}}, \ and\ \bibinfo {author} {\bibfnamefont
  {A.}~\bibnamefont {Yazdani}},\ }\href {\doibase 10.1126/science.aad8766}
  {\bibfield  {journal} {\bibinfo  {journal} {Science}\ }\textbf {\bibinfo
  {volume} {351}},\ \bibinfo {pages} {1184} (\bibinfo {year}
  {2016})}\BibitemShut {NoStop}%
\bibitem [{\citenamefont {Kourtis}\ \emph {et~al.}(2016)\citenamefont
  {Kourtis}, \citenamefont {Li}, \citenamefont {Wang}, \citenamefont
  {Yazdani},\ and\ \citenamefont {Bernevig}}]{yazdani-2}%
  \BibitemOpen
  \bibfield  {author} {\bibinfo {author} {\bibfnamefont {S.}~\bibnamefont
  {Kourtis}}, \bibinfo {author} {\bibfnamefont {J.}~\bibnamefont {Li}},
  \bibinfo {author} {\bibfnamefont {Z.}~\bibnamefont {Wang}}, \bibinfo {author}
  {\bibfnamefont {A.}~\bibnamefont {Yazdani}}, \ and\ \bibinfo {author}
  {\bibfnamefont {B.~A.}\ \bibnamefont {Bernevig}},\ }\href {\doibase
  10.1103/PhysRevB.93.041109} {\bibfield  {journal} {\bibinfo  {journal} {Phys.
  Rev. B}\ }\textbf {\bibinfo {volume} {93}},\ \bibinfo {pages} {041109}
  (\bibinfo {year} {2016})}\BibitemShut {NoStop}%
\bibitem [{\citenamefont {Chang}\ \emph {et~al.}(2016)\citenamefont {Chang},
  \citenamefont {Xu}, \citenamefont {Zheng}, \citenamefont {Lee}, \citenamefont
  {Huang}, \citenamefont {Belopolski}, \citenamefont {Sanchez}, \citenamefont
  {Bian}, \citenamefont {Alidoust}, \citenamefont {Chang}, \citenamefont {Hsu},
  \citenamefont {Jeng}, \citenamefont {Bansil}, \citenamefont {Lin},\ and\
  \citenamefont {Hasan}}]{TaAs5}%
  \BibitemOpen
  \bibfield  {author} {\bibinfo {author} {\bibfnamefont {G.}~\bibnamefont
  {Chang}}, \bibinfo {author} {\bibfnamefont {S.-Y.}\ \bibnamefont {Xu}},
  \bibinfo {author} {\bibfnamefont {H.}~\bibnamefont {Zheng}}, \bibinfo
  {author} {\bibfnamefont {C.-C.}\ \bibnamefont {Lee}}, \bibinfo {author}
  {\bibfnamefont {S.-M.}\ \bibnamefont {Huang}}, \bibinfo {author}
  {\bibfnamefont {I.}~\bibnamefont {Belopolski}}, \bibinfo {author}
  {\bibfnamefont {D.~S.}\ \bibnamefont {Sanchez}}, \bibinfo {author}
  {\bibfnamefont {G.}~\bibnamefont {Bian}}, \bibinfo {author} {\bibfnamefont
  {N.}~\bibnamefont {Alidoust}}, \bibinfo {author} {\bibfnamefont {T.-R.}\
  \bibnamefont {Chang}}, \bibinfo {author} {\bibfnamefont {C.-H.}\ \bibnamefont
  {Hsu}}, \bibinfo {author} {\bibfnamefont {H.-T.}\ \bibnamefont {Jeng}},
  \bibinfo {author} {\bibfnamefont {A.}~\bibnamefont {Bansil}}, \bibinfo
  {author} {\bibfnamefont {H.}~\bibnamefont {Lin}}, \ and\ \bibinfo {author}
  {\bibfnamefont {M.~Z.}\ \bibnamefont {Hasan}},\ }\href {\doibase
  10.1103/PhysRevLett.116.066601} {\bibfield  {journal} {\bibinfo  {journal}
  {Phys. Rev. Lett.}\ }\textbf {\bibinfo {volume} {116}},\ \bibinfo {pages}
  {066601} (\bibinfo {year} {2016})}\BibitemShut {NoStop}%
\bibitem [{\citenamefont {Wang}\ \emph
  {et~al.}(2016{\natexlab{a}})\citenamefont {Wang}, \citenamefont {Zheng},
  \citenamefont {Shen}, \citenamefont {Lu}, \citenamefont {Fang}, \citenamefont
  {Sheng}, \citenamefont {Zhou}, \citenamefont {Yang}, \citenamefont {Li},
  \citenamefont {Feng},\ and\ \citenamefont {Xu}}]{Wang-Xu}%
  \BibitemOpen
  \bibfield  {author} {\bibinfo {author} {\bibfnamefont {Z.}~\bibnamefont
  {Wang}}, \bibinfo {author} {\bibfnamefont {Y.}~\bibnamefont {Zheng}},
  \bibinfo {author} {\bibfnamefont {Z.}~\bibnamefont {Shen}}, \bibinfo {author}
  {\bibfnamefont {Y.}~\bibnamefont {Lu}}, \bibinfo {author} {\bibfnamefont
  {H.}~\bibnamefont {Fang}}, \bibinfo {author} {\bibfnamefont {F.}~\bibnamefont
  {Sheng}}, \bibinfo {author} {\bibfnamefont {Y.}~\bibnamefont {Zhou}},
  \bibinfo {author} {\bibfnamefont {X.}~\bibnamefont {Yang}}, \bibinfo {author}
  {\bibfnamefont {Y.}~\bibnamefont {Li}}, \bibinfo {author} {\bibfnamefont
  {C.}~\bibnamefont {Feng}}, \ and\ \bibinfo {author} {\bibfnamefont {Z.-A.}\
  \bibnamefont {Xu}},\ }\href {\doibase 10.1103/PhysRevB.93.121112} {\bibfield
  {journal} {\bibinfo  {journal} {Phys. Rev. B}\ }\textbf {\bibinfo {volume}
  {93}},\ \bibinfo {pages} {121112} (\bibinfo {year}
  {2016}{\natexlab{a}})}\BibitemShut {NoStop}%
\bibitem [{\citenamefont {Burkov}\ and\ \citenamefont
  {Balents}(2011)}]{Burkov1}%
  \BibitemOpen
  \bibfield  {author} {\bibinfo {author} {\bibfnamefont {A.~A.}\ \bibnamefont
  {Burkov}}\ and\ \bibinfo {author} {\bibfnamefont {L.}~\bibnamefont
  {Balents}},\ }\href {\doibase 10.1103/PhysRevLett.107.127205} {\bibfield
  {journal} {\bibinfo  {journal} {Phys. Rev. Lett.}\ }\textbf {\bibinfo
  {volume} {107}},\ \bibinfo {pages} {127205} (\bibinfo {year}
  {2011})}\BibitemShut {NoStop}%
\bibitem [{\citenamefont {Burkov}\ \emph {et~al.}(2011)\citenamefont {Burkov},
  \citenamefont {Hook},\ and\ \citenamefont {Balents}}]{Burkov2}%
  \BibitemOpen
  \bibfield  {author} {\bibinfo {author} {\bibfnamefont {A.~A.}\ \bibnamefont
  {Burkov}}, \bibinfo {author} {\bibfnamefont {M.~D.}\ \bibnamefont {Hook}}, \
  and\ \bibinfo {author} {\bibfnamefont {L.}~\bibnamefont {Balents}},\ }\href
  {\doibase 10.1103/PhysRevB.84.235126} {\bibfield  {journal} {\bibinfo
  {journal} {Phys. Rev. B}\ }\textbf {\bibinfo {volume} {84}},\ \bibinfo
  {pages} {235126} (\bibinfo {year} {2011})}\BibitemShut {NoStop}%
\bibitem [{\citenamefont {Zyuzin}\ \emph {et~al.}(2012)\citenamefont {Zyuzin},
  \citenamefont {Wu},\ and\ \citenamefont {Burkov}}]{Zyuzin1}%
  \BibitemOpen
  \bibfield  {author} {\bibinfo {author} {\bibfnamefont {A.~A.}\ \bibnamefont
  {Zyuzin}}, \bibinfo {author} {\bibfnamefont {S.}~\bibnamefont {Wu}}, \ and\
  \bibinfo {author} {\bibfnamefont {A.~A.}\ \bibnamefont {Burkov}},\ }\href
  {\doibase 10.1103/PhysRevB.85.165110} {\bibfield  {journal} {\bibinfo
  {journal} {Phys. Rev. B}\ }\textbf {\bibinfo {volume} {85}},\ \bibinfo
  {pages} {165110} (\bibinfo {year} {2012})}\BibitemShut {NoStop}%
\bibitem [{\citenamefont {Das}(2013)}]{Das}%
  \BibitemOpen
  \bibfield  {author} {\bibinfo {author} {\bibfnamefont {T.}~\bibnamefont
  {Das}},\ }\href {\doibase 10.1103/PhysRevB.88.035444} {\bibfield  {journal}
  {\bibinfo  {journal} {Phys. Rev. B}\ }\textbf {\bibinfo {volume} {88}},\
  \bibinfo {pages} {035444} (\bibinfo {year} {2013})}\BibitemShut {NoStop}%
\bibitem [{\citenamefont {Gibson}\ \emph {et~al.}(2015)\citenamefont {Gibson},
  \citenamefont {Schoop}, \citenamefont {Muechler}, \citenamefont {Xie},
  \citenamefont {Hirschberger}, \citenamefont {Ong}, \citenamefont {Car},\ and\
  \citenamefont {Cava}}]{Cava2}%
  \BibitemOpen
  \bibfield  {author} {\bibinfo {author} {\bibfnamefont {Q.~D.}\ \bibnamefont
  {Gibson}}, \bibinfo {author} {\bibfnamefont {L.~M.}\ \bibnamefont {Schoop}},
  \bibinfo {author} {\bibfnamefont {L.}~\bibnamefont {Muechler}}, \bibinfo
  {author} {\bibfnamefont {L.~S.}\ \bibnamefont {Xie}}, \bibinfo {author}
  {\bibfnamefont {M.}~\bibnamefont {Hirschberger}}, \bibinfo {author}
  {\bibfnamefont {N.~P.}\ \bibnamefont {Ong}}, \bibinfo {author} {\bibfnamefont
  {R.}~\bibnamefont {Car}}, \ and\ \bibinfo {author} {\bibfnamefont {R.~J.}\
  \bibnamefont {Cava}},\ }\href {\doibase 10.1103/PhysRevB.91.205128}
  {\bibfield  {journal} {\bibinfo  {journal} {Phys. Rev. B}\ }\textbf {\bibinfo
  {volume} {91}},\ \bibinfo {pages} {205128} (\bibinfo {year}
  {2015})}\BibitemShut {NoStop}%
\bibitem [{\citenamefont {Ganeshan}\ and\ \citenamefont
  {Das~Sarma}(2015)}]{babaganesh}%
  \BibitemOpen
  \bibfield  {author} {\bibinfo {author} {\bibfnamefont {S.}~\bibnamefont
  {Ganeshan}}\ and\ \bibinfo {author} {\bibfnamefont {S.}~\bibnamefont
  {Das~Sarma}},\ }\href {\doibase 10.1103/PhysRevB.91.125438} {\bibfield
  {journal} {\bibinfo  {journal} {Phys. Rev. B}\ }\textbf {\bibinfo {volume}
  {91}},\ \bibinfo {pages} {125438} (\bibinfo {year} {2015})}\BibitemShut
  {NoStop}%
\bibitem [{\citenamefont {Wan}\ \emph {et~al.}(2011)\citenamefont {Wan},
  \citenamefont {Turner}, \citenamefont {Vishwanath},\ and\ \citenamefont
  {Savrasov}}]{vishwanath}%
  \BibitemOpen
  \bibfield  {author} {\bibinfo {author} {\bibfnamefont {X.}~\bibnamefont
  {Wan}}, \bibinfo {author} {\bibfnamefont {A.~M.}\ \bibnamefont {Turner}},
  \bibinfo {author} {\bibfnamefont {A.}~\bibnamefont {Vishwanath}}, \ and\
  \bibinfo {author} {\bibfnamefont {S.~Y.}\ \bibnamefont {Savrasov}},\ }\href
  {\doibase 10.1103/PhysRevB.83.205101} {\bibfield  {journal} {\bibinfo
  {journal} {Phys. Rev. B}\ }\textbf {\bibinfo {volume} {83}},\ \bibinfo
  {pages} {205101} (\bibinfo {year} {2011})}\BibitemShut {NoStop}%
\bibitem [{\citenamefont {Moon}\ \emph {et~al.}(2013)\citenamefont {Moon},
  \citenamefont {Xu}, \citenamefont {Kim},\ and\ \citenamefont
  {Balents}}]{balents}%
  \BibitemOpen
  \bibfield  {author} {\bibinfo {author} {\bibfnamefont {E.-G.}\ \bibnamefont
  {Moon}}, \bibinfo {author} {\bibfnamefont {C.}~\bibnamefont {Xu}}, \bibinfo
  {author} {\bibfnamefont {Y.~B.}\ \bibnamefont {Kim}}, \ and\ \bibinfo
  {author} {\bibfnamefont {L.}~\bibnamefont {Balents}},\ }\href {\doibase
  10.1103/PhysRevLett.111.206401} {\bibfield  {journal} {\bibinfo  {journal}
  {Phys. Rev. Lett.}\ }\textbf {\bibinfo {volume} {111}},\ \bibinfo {pages}
  {206401} (\bibinfo {year} {2013})}\BibitemShut {NoStop}%
\bibitem [{\citenamefont {Lee}\ \emph {et~al.}(2013)\citenamefont {Lee},
  \citenamefont {Paramekanti},\ and\ \citenamefont {Kim}}]{sunbin}%
  \BibitemOpen
  \bibfield  {author} {\bibinfo {author} {\bibfnamefont {S.~B.}\ \bibnamefont
  {Lee}}, \bibinfo {author} {\bibfnamefont {A.}~\bibnamefont {Paramekanti}}, \
  and\ \bibinfo {author} {\bibfnamefont {Y.~B.}\ \bibnamefont {Kim}},\ }\href
  {\doibase 10.1103/PhysRevLett.111.196601} {\bibfield  {journal} {\bibinfo
  {journal} {Phys. Rev. Lett.}\ }\textbf {\bibinfo {volume} {111}},\ \bibinfo
  {pages} {196601} (\bibinfo {year} {2013})}\BibitemShut {NoStop}%
\bibitem [{\citenamefont {Goswami}\ \emph {et~al.}()\citenamefont {Goswami},
  \citenamefont {Roy},\ and\ \citenamefont {{Das
  Sarma}}}]{goswami-roy-dassarma}%
  \BibitemOpen
  \bibfield  {author} {\bibinfo {author} {\bibfnamefont {P.}~\bibnamefont
  {Goswami}}, \bibinfo {author} {\bibfnamefont {B.}~\bibnamefont {Roy}}, \ and\
  \bibinfo {author} {\bibfnamefont {S.}~\bibnamefont {{Das Sarma}}},\
  }\href@noop {} {\ }\Eprint {http://arxiv.org/abs/arXiv:1603.02273 (2016)}
  {arXiv:1603.02273 (2016)} \BibitemShut {NoStop}%
\bibitem [{\citenamefont {Herring}(1937)}]{herring}%
  \BibitemOpen
  \bibfield  {author} {\bibinfo {author} {\bibfnamefont {C.}~\bibnamefont
  {Herring}},\ }\href {\doibase 10.1103/PhysRev.52.365} {\bibfield  {journal}
  {\bibinfo  {journal} {Phys. Rev.}\ }\textbf {\bibinfo {volume} {52}},\
  \bibinfo {pages} {365} (\bibinfo {year} {1937})}\BibitemShut {NoStop}%
\bibitem [{\citenamefont {Nielsen}\ and\ \citenamefont
  {Ninomiya}(1981)}]{nielsen-nogo}%
  \BibitemOpen
  \bibfield  {author} {\bibinfo {author} {\bibfnamefont {H.~B.}\ \bibnamefont
  {Nielsen}}\ and\ \bibinfo {author} {\bibfnamefont {M.}~\bibnamefont
  {Ninomiya}},\ }\href {\doibase 10.1016/0370-2693(81)91026-1} {\bibfield
  {journal} {\bibinfo  {journal} {Phys. Lett. B}\ }\textbf {\bibinfo {volume}
  {105}},\ \bibinfo {pages} {219} (\bibinfo {year} {1981})}\BibitemShut
  {NoStop}%
\bibitem [{\citenamefont {Fang}\ \emph {et~al.}(2012)\citenamefont {Fang},
  \citenamefont {Gilbert}, \citenamefont {Dai},\ and\ \citenamefont
  {Bernevig}}]{Fang2012:Weyl}%
  \BibitemOpen
  \bibfield  {author} {\bibinfo {author} {\bibfnamefont {C.}~\bibnamefont
  {Fang}}, \bibinfo {author} {\bibfnamefont {M.~J.}\ \bibnamefont {Gilbert}},
  \bibinfo {author} {\bibfnamefont {X.}~\bibnamefont {Dai}}, \ and\ \bibinfo
  {author} {\bibfnamefont {B.~A.}\ \bibnamefont {Bernevig}},\ }\href {\doibase
  10.1103/PhysRevLett.108.266802} {\bibfield  {journal} {\bibinfo  {journal}
  {Phys. Rev. Lett.}\ }\textbf {\bibinfo {volume} {108}},\ \bibinfo {pages}
  {266802} (\bibinfo {year} {2012})}\BibitemShut {NoStop}%
\bibitem [{\citenamefont {Yang}\ and\ \citenamefont
  {Nagaosa}(2014)}]{nagaosa-yang}%
  \BibitemOpen
  \bibfield  {author} {\bibinfo {author} {\bibfnamefont {B.-J.}\ \bibnamefont
  {Yang}}\ and\ \bibinfo {author} {\bibfnamefont {N.}~\bibnamefont {Nagaosa}},\
  }\href {\doibase 10.1038/ncomms5898} {\bibfield  {journal} {\bibinfo
  {journal} {Nat. Commun.}\ }\textbf {\bibinfo {volume} {5}},\ \bibinfo {pages}
  {4898} (\bibinfo {year} {2014})}\BibitemShut {NoStop}%
\bibitem [{\citenamefont {Huang}\ \emph {et~al.}(2016)\citenamefont {Huang},
  \citenamefont {Xu}, \citenamefont {Belopolski}, \citenamefont {Lee},
  \citenamefont {Chang}, \citenamefont {Chang}, \citenamefont {Wang},
  \citenamefont {Alidoust}, \citenamefont {Bian}, \citenamefont {Neupane},
  \citenamefont {Sanchez}, \citenamefont {Zheng}, \citenamefont {Jeng},
  \citenamefont {Bansil}, \citenamefont {Neupert}, \citenamefont {Lin},\ and\
  \citenamefont {Hasan}}]{hasan-DW}%
  \BibitemOpen
  \bibfield  {author} {\bibinfo {author} {\bibfnamefont {S.-M.}\ \bibnamefont
  {Huang}}, \bibinfo {author} {\bibfnamefont {S.-Y.}\ \bibnamefont {Xu}},
  \bibinfo {author} {\bibfnamefont {I.}~\bibnamefont {Belopolski}}, \bibinfo
  {author} {\bibfnamefont {C.-C.}\ \bibnamefont {Lee}}, \bibinfo {author}
  {\bibfnamefont {G.}~\bibnamefont {Chang}}, \bibinfo {author} {\bibfnamefont
  {T.-R.}\ \bibnamefont {Chang}}, \bibinfo {author} {\bibfnamefont
  {B.}~\bibnamefont {Wang}}, \bibinfo {author} {\bibfnamefont {N.}~\bibnamefont
  {Alidoust}}, \bibinfo {author} {\bibfnamefont {G.}~\bibnamefont {Bian}},
  \bibinfo {author} {\bibfnamefont {M.}~\bibnamefont {Neupane}}, \bibinfo
  {author} {\bibfnamefont {D.}~\bibnamefont {Sanchez}}, \bibinfo {author}
  {\bibfnamefont {H.}~\bibnamefont {Zheng}}, \bibinfo {author} {\bibfnamefont
  {H.-T.}\ \bibnamefont {Jeng}}, \bibinfo {author} {\bibfnamefont
  {A.}~\bibnamefont {Bansil}}, \bibinfo {author} {\bibfnamefont
  {T.}~\bibnamefont {Neupert}}, \bibinfo {author} {\bibfnamefont
  {H.}~\bibnamefont {Lin}}, \ and\ \bibinfo {author} {\bibfnamefont {M.~Z.}\
  \bibnamefont {Hasan}},\ }\href {\doibase 10.1073/pnas.1514581113} {\bibfield
  {journal} {\bibinfo  {journal} {Proc. Natl. Acad. Sci.}\ }\textbf {\bibinfo
  {volume} {113}},\ \bibinfo {pages} {1180} (\bibinfo {year}
  {2016})}\BibitemShut {NoStop}%
\bibitem [{\citenamefont {Potter}\ \emph {et~al.}(2014)\citenamefont {Potter},
  \citenamefont {Kimchi},\ and\ \citenamefont {Vishwanath}}]{Potter2014}%
  \BibitemOpen
  \bibfield  {author} {\bibinfo {author} {\bibfnamefont {A.~C.}\ \bibnamefont
  {Potter}}, \bibinfo {author} {\bibfnamefont {I.}~\bibnamefont {Kimchi}}, \
  and\ \bibinfo {author} {\bibfnamefont {A.}~\bibnamefont {Vishwanath}},\
  }\href {\doibase 10.1038/ncomms6161} {\bibfield  {journal} {\bibinfo
  {journal} {Nat. Commun.}\ }\textbf {\bibinfo {volume} {5}},\ \bibinfo {pages}
  {5161} (\bibinfo {year} {2014})}\BibitemShut {NoStop}%
\bibitem [{\citenamefont {Zhang}\ \emph
  {et~al.}(2016{\natexlab{a}})\citenamefont {Zhang}, \citenamefont {Bulmash},
  \citenamefont {Hosur}, \citenamefont {Potter},\ and\ \citenamefont
  {Vishwanath}}]{Zhang2016}%
  \BibitemOpen
  \bibfield  {author} {\bibinfo {author} {\bibfnamefont {Y.}~\bibnamefont
  {Zhang}}, \bibinfo {author} {\bibfnamefont {D.}~\bibnamefont {Bulmash}},
  \bibinfo {author} {\bibfnamefont {P.}~\bibnamefont {Hosur}}, \bibinfo
  {author} {\bibfnamefont {A.~C.}\ \bibnamefont {Potter}}, \ and\ \bibinfo
  {author} {\bibfnamefont {A.}~\bibnamefont {Vishwanath}},\ }\href {\doibase
  10.1038/srep23741} {\bibfield  {journal} {\bibinfo  {journal} {Sci. Rep.}\
  }\textbf {\bibinfo {volume} {6}},\ \bibinfo {pages} {23741} (\bibinfo {year}
  {2016}{\natexlab{a}})}\BibitemShut {NoStop}%
\bibitem [{\citenamefont {Pippard}(1989)}]{Pippard}%
  \BibitemOpen
  \bibfield  {author} {\bibinfo {author} {\bibfnamefont {A.~B.}\ \bibnamefont
  {Pippard}},\ }\href@noop {} {\emph {\bibinfo {title} {Magnetoresistance in
  metals}}},\ \bibinfo {series} {Cambridge Studies in Low temperature Physics},
  Vol.~\bibinfo {volume} {2}\ (\bibinfo  {publisher} {Cambridge University
  Press},\ \bibinfo {year} {1989})\BibitemShut {NoStop}%
\bibitem [{\citenamefont {Zhu}\ \emph {et~al.}(2011)\citenamefont {Zhu},
  \citenamefont {Fauqu\'e}, \citenamefont {Fuseya},\ and\ \citenamefont
  {Behnia}}]{behnia3}%
  \BibitemOpen
  \bibfield  {author} {\bibinfo {author} {\bibfnamefont {Z.}~\bibnamefont
  {Zhu}}, \bibinfo {author} {\bibfnamefont {B.}~\bibnamefont {Fauqu\'e}},
  \bibinfo {author} {\bibfnamefont {Y.}~\bibnamefont {Fuseya}}, \ and\ \bibinfo
  {author} {\bibfnamefont {K.}~\bibnamefont {Behnia}},\ }\href {\doibase
  10.1103/PhysRevB.84.115137} {\bibfield  {journal} {\bibinfo  {journal} {Phys.
  Rev. B}\ }\textbf {\bibinfo {volume} {84}},\ \bibinfo {pages} {115137}
  (\bibinfo {year} {2011})}\BibitemShut {NoStop}%
\bibitem [{\citenamefont {Peskin}\ and\ \citenamefont
  {Schroeder}(1995)}]{peskin}%
  \BibitemOpen
  \bibfield  {author} {\bibinfo {author} {\bibfnamefont {M.}~\bibnamefont
  {Peskin}}\ and\ \bibinfo {author} {\bibfnamefont {D.}~\bibnamefont
  {Schroeder}},\ }\href@noop {} {\emph {\bibinfo {title} {An Introduction To
  Quantum Field Theory}}}\ (\bibinfo  {publisher} {Westview Press},\ \bibinfo
  {year} {1995})\BibitemShut {NoStop}%
\bibitem [{\citenamefont {Fujikawa}\ and\ \citenamefont
  {Suzuki}(2004)}]{fujikawa}%
  \BibitemOpen
  \bibfield  {author} {\bibinfo {author} {\bibfnamefont {K.}~\bibnamefont
  {Fujikawa}}\ and\ \bibinfo {author} {\bibfnamefont {H.}~\bibnamefont
  {Suzuki}},\ }\href@noop {} {\emph {\bibinfo {title} {Path Integrals and
  Quantum Anomalies}}},\ International Series of Monographs on Physics\
  (\bibinfo  {publisher} {OUP Oxford},\ \bibinfo {year} {2004})\BibitemShut
  {NoStop}%
\bibitem [{\citenamefont {Adler}(1969)}]{adler}%
  \BibitemOpen
  \bibfield  {author} {\bibinfo {author} {\bibfnamefont {S.~L.}\ \bibnamefont
  {Adler}},\ }\href {\doibase 10.1103/PhysRev.177.2426} {\bibfield  {journal}
  {\bibinfo  {journal} {Phys. Rev.}\ }\textbf {\bibinfo {volume} {177}},\
  \bibinfo {pages} {2426} (\bibinfo {year} {1969})}\BibitemShut {NoStop}%
\bibitem [{\citenamefont {Bell}\ and\ \citenamefont {Jackiw}(1969)}]{bell}%
  \BibitemOpen
  \bibfield  {author} {\bibinfo {author} {\bibfnamefont {J.~S.}\ \bibnamefont
  {Bell}}\ and\ \bibinfo {author} {\bibfnamefont {R.}~\bibnamefont {Jackiw}},\
  }\href {\doibase 10.1007/BF02823296} {\bibfield  {journal} {\bibinfo
  {journal} {Il Nuovo Cimento A}\ }\textbf {\bibinfo {volume} {60}},\ \bibinfo
  {pages} {47} (\bibinfo {year} {1969})}\BibitemShut {NoStop}%
\bibitem [{\citenamefont {Osada}(2008)}]{Osada}%
  \BibitemOpen
  \bibfield  {author} {\bibinfo {author} {\bibfnamefont {T.}~\bibnamefont
  {Osada}},\ }\href {\doibase 10.1143/JPSJ.77.084711} {\bibfield  {journal}
  {\bibinfo  {journal} {J. Phys. Soc. Jpn.}\ }\textbf {\bibinfo {volume}
  {77}},\ \bibinfo {pages} {084711} (\bibinfo {year} {2008})}\BibitemShut
  {NoStop}%
\bibitem [{\citenamefont {Fukushima}\ \emph {et~al.}(2008)\citenamefont
  {Fukushima}, \citenamefont {Kharzeev},\ and\ \citenamefont
  {Warringa}}]{fukushima}%
  \BibitemOpen
  \bibfield  {author} {\bibinfo {author} {\bibfnamefont {K.}~\bibnamefont
  {Fukushima}}, \bibinfo {author} {\bibfnamefont {D.~E.}\ \bibnamefont
  {Kharzeev}}, \ and\ \bibinfo {author} {\bibfnamefont {H.~J.}\ \bibnamefont
  {Warringa}},\ }\href {\doibase 10.1103/PhysRevD.78.074033} {\bibfield
  {journal} {\bibinfo  {journal} {Phys. Rev. D}\ }\textbf {\bibinfo {volume}
  {78}},\ \bibinfo {pages} {074033} (\bibinfo {year} {2008})}\BibitemShut
  {NoStop}%
\bibitem [{\citenamefont {Xu}\ \emph {et~al.}(2011)\citenamefont {Xu},
  \citenamefont {Weng}, \citenamefont {Wang}, \citenamefont {Dai},\ and\
  \citenamefont {Fang}}]{xu}%
  \BibitemOpen
  \bibfield  {author} {\bibinfo {author} {\bibfnamefont {G.}~\bibnamefont
  {Xu}}, \bibinfo {author} {\bibfnamefont {H.}~\bibnamefont {Weng}}, \bibinfo
  {author} {\bibfnamefont {Z.}~\bibnamefont {Wang}}, \bibinfo {author}
  {\bibfnamefont {X.}~\bibnamefont {Dai}}, \ and\ \bibinfo {author}
  {\bibfnamefont {Z.}~\bibnamefont {Fang}},\ }\href {\doibase
  10.1103/PhysRevLett.107.186806} {\bibfield  {journal} {\bibinfo  {journal}
  {Phys. Rev. Lett.}\ }\textbf {\bibinfo {volume} {107}},\ \bibinfo {pages}
  {186806} (\bibinfo {year} {2011})}\BibitemShut {NoStop}%
\bibitem [{\citenamefont {Cho}()}]{Cho}%
  \BibitemOpen
  \bibfield  {author} {\bibinfo {author} {\bibfnamefont {G.~Y.}\ \bibnamefont
  {Cho}},\ }\href@noop {} {\ }\Eprint {http://arxiv.org/abs/arXiv:1110.1939
  (2011)} {arXiv:1110.1939 (2011)} \BibitemShut {NoStop}%
\bibitem [{\citenamefont {Grushin}(2012)}]{Grushin}%
  \BibitemOpen
  \bibfield  {author} {\bibinfo {author} {\bibfnamefont {A.~G.}\ \bibnamefont
  {Grushin}},\ }\href {\doibase 10.1103/PhysRevD.86.045001} {\bibfield
  {journal} {\bibinfo  {journal} {Phys. Rev. D}\ }\textbf {\bibinfo {volume}
  {86}},\ \bibinfo {pages} {045001} (\bibinfo {year} {2012})}\BibitemShut
  {NoStop}%
\bibitem [{\citenamefont {Aji}(2012)}]{aji}%
  \BibitemOpen
  \bibfield  {author} {\bibinfo {author} {\bibfnamefont {V.}~\bibnamefont
  {Aji}},\ }\href {\doibase 10.1103/PhysRevB.85.241101} {\bibfield  {journal}
  {\bibinfo  {journal} {Phys. Rev. B}\ }\textbf {\bibinfo {volume} {85}},\
  \bibinfo {pages} {241101} (\bibinfo {year} {2012})}\BibitemShut {NoStop}%
\bibitem [{\citenamefont {Zyuzin}\ and\ \citenamefont
  {Burkov}(2012)}]{Zyuzin3}%
  \BibitemOpen
  \bibfield  {author} {\bibinfo {author} {\bibfnamefont {A.~A.}\ \bibnamefont
  {Zyuzin}}\ and\ \bibinfo {author} {\bibfnamefont {A.~A.}\ \bibnamefont
  {Burkov}},\ }\href {\doibase 10.1103/PhysRevB.86.115133} {\bibfield
  {journal} {\bibinfo  {journal} {Phys. Rev. B}\ }\textbf {\bibinfo {volume}
  {86}},\ \bibinfo {pages} {115133} (\bibinfo {year} {2012})}\BibitemShut
  {NoStop}%
\bibitem [{\citenamefont {Goswami}\ and\ \citenamefont
  {Tewari}(2013)}]{GoswamiTewari}%
  \BibitemOpen
  \bibfield  {author} {\bibinfo {author} {\bibfnamefont {P.}~\bibnamefont
  {Goswami}}\ and\ \bibinfo {author} {\bibfnamefont {S.}~\bibnamefont
  {Tewari}},\ }\href {\doibase 10.1103/PhysRevB.88.245107} {\bibfield
  {journal} {\bibinfo  {journal} {Phys. Rev. B}\ }\textbf {\bibinfo {volume}
  {88}},\ \bibinfo {pages} {245107} (\bibinfo {year} {2013})}\BibitemShut
  {NoStop}%
\bibitem [{\citenamefont {Chen}\ \emph {et~al.}(2013)\citenamefont {Chen},
  \citenamefont {Wu},\ and\ \citenamefont {Burkov}}]{burkovsu}%
  \BibitemOpen
  \bibfield  {author} {\bibinfo {author} {\bibfnamefont {Y.}~\bibnamefont
  {Chen}}, \bibinfo {author} {\bibfnamefont {S.}~\bibnamefont {Wu}}, \ and\
  \bibinfo {author} {\bibfnamefont {A.~A.}\ \bibnamefont {Burkov}},\ }\href
  {\doibase 10.1103/PhysRevB.88.125105} {\bibfield  {journal} {\bibinfo
  {journal} {Phys. Rev. B}\ }\textbf {\bibinfo {volume} {88}},\ \bibinfo
  {pages} {125105} (\bibinfo {year} {2013})}\BibitemShut {NoStop}%
\bibitem [{\citenamefont {Son}\ and\ \citenamefont {Spivak}(2013)}]{son}%
  \BibitemOpen
  \bibfield  {author} {\bibinfo {author} {\bibfnamefont {D.~T.}\ \bibnamefont
  {Son}}\ and\ \bibinfo {author} {\bibfnamefont {B.~Z.}\ \bibnamefont
  {Spivak}},\ }\href {\doibase 10.1103/PhysRevB.88.104412} {\bibfield
  {journal} {\bibinfo  {journal} {Phys. Rev. B}\ }\textbf {\bibinfo {volume}
  {88}},\ \bibinfo {pages} {104412} (\bibinfo {year} {2013})}\BibitemShut
  {NoStop}%
\bibitem [{\citenamefont {Vazifeh}\ and\ \citenamefont
  {Franz}(2013)}]{vazifeh}%
  \BibitemOpen
  \bibfield  {author} {\bibinfo {author} {\bibfnamefont {M.~M.}\ \bibnamefont
  {Vazifeh}}\ and\ \bibinfo {author} {\bibfnamefont {M.}~\bibnamefont
  {Franz}},\ }\href {\doibase 10.1103/PhysRevLett.111.027201} {\bibfield
  {journal} {\bibinfo  {journal} {Phys. Rev. Lett.}\ }\textbf {\bibinfo
  {volume} {111}},\ \bibinfo {pages} {027201} (\bibinfo {year}
  {2013})}\BibitemShut {NoStop}%
\bibitem [{\citenamefont {Parameswaran}\ \emph {et~al.}(2014)\citenamefont
  {Parameswaran}, \citenamefont {Grover}, \citenamefont {Abanin}, \citenamefont
  {Pesin},\ and\ \citenamefont {Vishwanath}}]{vishwanathsid}%
  \BibitemOpen
  \bibfield  {author} {\bibinfo {author} {\bibfnamefont {S.~A.}\ \bibnamefont
  {Parameswaran}}, \bibinfo {author} {\bibfnamefont {T.}~\bibnamefont
  {Grover}}, \bibinfo {author} {\bibfnamefont {D.~A.}\ \bibnamefont {Abanin}},
  \bibinfo {author} {\bibfnamefont {D.~A.}\ \bibnamefont {Pesin}}, \ and\
  \bibinfo {author} {\bibfnamefont {A.}~\bibnamefont {Vishwanath}},\ }\href
  {\doibase 10.1103/PhysRevX.4.031035} {\bibfield  {journal} {\bibinfo
  {journal} {Phys. Rev. X}\ }\textbf {\bibinfo {volume} {4}},\ \bibinfo {pages}
  {031035} (\bibinfo {year} {2014})}\BibitemShut {NoStop}%
\bibitem [{\citenamefont {Gorbar}\ \emph {et~al.}(2014)\citenamefont {Gorbar},
  \citenamefont {Miransky},\ and\ \citenamefont {Shovkovy}}]{shovkovy}%
  \BibitemOpen
  \bibfield  {author} {\bibinfo {author} {\bibfnamefont {E.~V.}\ \bibnamefont
  {Gorbar}}, \bibinfo {author} {\bibfnamefont {V.~A.}\ \bibnamefont
  {Miransky}}, \ and\ \bibinfo {author} {\bibfnamefont {I.~A.}\ \bibnamefont
  {Shovkovy}},\ }\href {\doibase 10.1103/PhysRevB.89.085126} {\bibfield
  {journal} {\bibinfo  {journal} {Phys. Rev. B}\ }\textbf {\bibinfo {volume}
  {89}},\ \bibinfo {pages} {085126} (\bibinfo {year} {2014})}\BibitemShut
  {NoStop}%
\bibitem [{\citenamefont {Burkov}(2014)}]{burkov-prl}%
  \BibitemOpen
  \bibfield  {author} {\bibinfo {author} {\bibfnamefont {A.~A.}\ \bibnamefont
  {Burkov}},\ }\href {\doibase 10.1103/PhysRevLett.113.247203} {\bibfield
  {journal} {\bibinfo  {journal} {Phys. Rev. Lett.}\ }\textbf {\bibinfo
  {volume} {113}},\ \bibinfo {pages} {247203} (\bibinfo {year}
  {2014})}\BibitemShut {NoStop}%
\bibitem [{\citenamefont {Pesin}\ \emph {et~al.}(2015)\citenamefont {Pesin},
  \citenamefont {Mishchenko},\ and\ \citenamefont {Levchenko}}]{Pesin}%
  \BibitemOpen
  \bibfield  {author} {\bibinfo {author} {\bibfnamefont {D.~A.}\ \bibnamefont
  {Pesin}}, \bibinfo {author} {\bibfnamefont {E.~G.}\ \bibnamefont
  {Mishchenko}}, \ and\ \bibinfo {author} {\bibfnamefont {A.}~\bibnamefont
  {Levchenko}},\ }\href {\doibase 10.1103/PhysRevB.92.174202} {\bibfield
  {journal} {\bibinfo  {journal} {Phys. Rev. B}\ }\textbf {\bibinfo {volume}
  {92}},\ \bibinfo {pages} {174202} (\bibinfo {year} {2015})}\BibitemShut
  {NoStop}%
\bibitem [{\citenamefont {Jimenez-Alba}\ \emph {et~al.}(2015)\citenamefont
  {Jimenez-Alba}, \citenamefont {Landsteiner}, \citenamefont {Liu},\ and\
  \citenamefont {Sun}}]{Yawen}%
  \BibitemOpen
  \bibfield  {author} {\bibinfo {author} {\bibfnamefont {A.}~\bibnamefont
  {Jimenez-Alba}}, \bibinfo {author} {\bibfnamefont {K.}~\bibnamefont
  {Landsteiner}}, \bibinfo {author} {\bibfnamefont {Y.}~\bibnamefont {Liu}}, \
  and\ \bibinfo {author} {\bibfnamefont {Y.-W.}\ \bibnamefont {Sun}},\ }\href
  {\doibase 10.1007/JHEP07(2015)117} {\bibfield  {journal} {\bibinfo  {journal}
  {J. High Energy Phys.}\ }\textbf {\bibinfo {volume} {2015}},\ \bibinfo
  {pages} {117} (\bibinfo {year} {2015})}\BibitemShut {NoStop}%
\bibitem [{\citenamefont {Burkov}(2015{\natexlab{b}})}]{Burkov}%
  \BibitemOpen
  \bibfield  {author} {\bibinfo {author} {\bibfnamefont {A.~A.}\ \bibnamefont
  {Burkov}},\ }\href {\doibase 10.1103/PhysRevB.91.245157} {\bibfield
  {journal} {\bibinfo  {journal} {Phys. Rev. B}\ }\textbf {\bibinfo {volume}
  {91}},\ \bibinfo {pages} {245157} (\bibinfo {year}
  {2015}{\natexlab{b}})}\BibitemShut {NoStop}%
\bibitem [{\citenamefont {Shen}\ \emph {et~al.}(2016)\citenamefont {Shen},
  \citenamefont {Deng}, \citenamefont {Kotliar},\ and\ \citenamefont
  {Ni}}]{NiNi}%
  \BibitemOpen
  \bibfield  {author} {\bibinfo {author} {\bibfnamefont {B.}~\bibnamefont
  {Shen}}, \bibinfo {author} {\bibfnamefont {X.}~\bibnamefont {Deng}}, \bibinfo
  {author} {\bibfnamefont {G.}~\bibnamefont {Kotliar}}, \ and\ \bibinfo
  {author} {\bibfnamefont {N.}~\bibnamefont {Ni}},\ }\href {\doibase
  10.1103/PhysRevB.93.195119} {\bibfield  {journal} {\bibinfo  {journal} {Phys.
  Rev. B}\ }\textbf {\bibinfo {volume} {93}},\ \bibinfo {pages} {195119}
  (\bibinfo {year} {2016})}\BibitemShut {NoStop}%
\bibitem [{\citenamefont {Zhang}\ \emph
  {et~al.}(2016{\natexlab{b}})\citenamefont {Zhang}, \citenamefont {Lu},\ and\
  \citenamefont {Shen}}]{Shun-Qing}%
  \BibitemOpen
  \bibfield  {author} {\bibinfo {author} {\bibfnamefont {S.-B.}\ \bibnamefont
  {Zhang}}, \bibinfo {author} {\bibfnamefont {H.-Z.}\ \bibnamefont {Lu}}, \
  and\ \bibinfo {author} {\bibfnamefont {S.-Q.}\ \bibnamefont {Shen}},\ }\href
  {\doibase 10.1088/1367-2630/18/5/053039} {\bibfield  {journal} {\bibinfo
  {journal} {New J. Phys.}\ }\textbf {\bibinfo {volume} {18}},\ \bibinfo
  {pages} {053039} (\bibinfo {year} {2016}{\natexlab{b}})}\BibitemShut
  {NoStop}%
\bibitem [{\citenamefont {Chen}\ \emph {et~al.}(2016)\citenamefont {Chen},
  \citenamefont {Liu}, \citenamefont {Jiang},\ and\ \citenamefont {Xie}}]{Xie}%
  \BibitemOpen
  \bibfield  {author} {\bibinfo {author} {\bibfnamefont {C.-Z.}\ \bibnamefont
  {Chen}}, \bibinfo {author} {\bibfnamefont {H.}~\bibnamefont {Liu}}, \bibinfo
  {author} {\bibfnamefont {H.}~\bibnamefont {Jiang}}, \ and\ \bibinfo {author}
  {\bibfnamefont {X.~C.}\ \bibnamefont {Xie}},\ }\href {\doibase
  10.1103/PhysRevB.93.165420} {\bibfield  {journal} {\bibinfo  {journal} {Phys.
  Rev. B}\ }\textbf {\bibinfo {volume} {93}},\ \bibinfo {pages} {165420}
  (\bibinfo {year} {2016})}\BibitemShut {NoStop}%
\bibitem [{\citenamefont {Sun}\ and\ \citenamefont {Yang}(2016)}]{QingYang}%
  \BibitemOpen
  \bibfield  {author} {\bibinfo {author} {\bibfnamefont {Y.-W.}\ \bibnamefont
  {Sun}}\ and\ \bibinfo {author} {\bibfnamefont {Q.}~\bibnamefont {Yang}},\
  }\href {\doibase 10.1007/JHEP07(2015)117} {\bibfield  {journal} {\bibinfo
  {journal} {J. High Energy Phys.}\ }\textbf {\bibinfo {volume} {2016}},\
  \bibinfo {pages} {122} (\bibinfo {year} {2016})}\BibitemShut {NoStop}%
\bibitem [{\citenamefont {Spivak}\ and\ \citenamefont
  {Andreev}(2016)}]{Andreev}%
  \BibitemOpen
  \bibfield  {author} {\bibinfo {author} {\bibfnamefont {B.~Z.}\ \bibnamefont
  {Spivak}}\ and\ \bibinfo {author} {\bibfnamefont {A.~V.}\ \bibnamefont
  {Andreev}},\ }\href {\doibase 10.1103/PhysRevB.93.085107} {\bibfield
  {journal} {\bibinfo  {journal} {Phys. Rev. B}\ }\textbf {\bibinfo {volume}
  {93}},\ \bibinfo {pages} {085107} (\bibinfo {year} {2016})}\BibitemShut
  {NoStop}%
\bibitem [{\citenamefont {Zyuzin}()}]{Zyuzin}%
  \BibitemOpen
  \bibfield  {author} {\bibinfo {author} {\bibfnamefont {V.~A.}\ \bibnamefont
  {Zyuzin}},\ }\href@noop {} {\ }\Eprint {http://arxiv.org/abs/arXiv:1608.01286
  (2016)} {arXiv:1608.01286 (2016)} \BibitemShut {NoStop}%
\bibitem [{\citenamefont {Tajima}\ \emph {et~al.}(2009)\citenamefont {Tajima},
  \citenamefont {Sugawara}, \citenamefont {Kato}, \citenamefont {Nishio},\ and\
  \citenamefont {Kajita}}]{Tajima2}%
  \BibitemOpen
  \bibfield  {author} {\bibinfo {author} {\bibfnamefont {N.}~\bibnamefont
  {Tajima}}, \bibinfo {author} {\bibfnamefont {S.}~\bibnamefont {Sugawara}},
  \bibinfo {author} {\bibfnamefont {R.}~\bibnamefont {Kato}}, \bibinfo {author}
  {\bibfnamefont {Y.}~\bibnamefont {Nishio}}, \ and\ \bibinfo {author}
  {\bibfnamefont {K.}~\bibnamefont {Kajita}},\ }\href {\doibase
  10.1103/PhysRevLett.102.176403} {\bibfield  {journal} {\bibinfo  {journal}
  {Phys. Rev. Lett.}\ }\textbf {\bibinfo {volume} {102}},\ \bibinfo {pages}
  {176403} (\bibinfo {year} {2009})}\BibitemShut {NoStop}%
\bibitem [{\citenamefont {Kim}\ \emph {et~al.}(2013)\citenamefont {Kim},
  \citenamefont {Kim}, \citenamefont {Wang}, \citenamefont {Sasaki},
  \citenamefont {Satoh}, \citenamefont {Ohnishi}, \citenamefont {Kitaura},
  \citenamefont {Yang},\ and\ \citenamefont {Li}}]{kim1}%
  \BibitemOpen
  \bibfield  {author} {\bibinfo {author} {\bibfnamefont {H.-J.}\ \bibnamefont
  {Kim}}, \bibinfo {author} {\bibfnamefont {K.-S.}\ \bibnamefont {Kim}},
  \bibinfo {author} {\bibfnamefont {J.-F.}\ \bibnamefont {Wang}}, \bibinfo
  {author} {\bibfnamefont {M.}~\bibnamefont {Sasaki}}, \bibinfo {author}
  {\bibfnamefont {N.}~\bibnamefont {Satoh}}, \bibinfo {author} {\bibfnamefont
  {A.}~\bibnamefont {Ohnishi}}, \bibinfo {author} {\bibfnamefont
  {M.}~\bibnamefont {Kitaura}}, \bibinfo {author} {\bibfnamefont
  {M.}~\bibnamefont {Yang}}, \ and\ \bibinfo {author} {\bibfnamefont
  {L.}~\bibnamefont {Li}},\ }\href {\doibase 10.1103/PhysRevLett.111.246603}
  {\bibfield  {journal} {\bibinfo  {journal} {Phys. Rev. Lett.}\ }\textbf
  {\bibinfo {volume} {111}},\ \bibinfo {pages} {246603} (\bibinfo {year}
  {2013})}\BibitemShut {NoStop}%
\bibitem [{\citenamefont {Zhao}\ \emph {et~al.}(2015)\citenamefont {Zhao},
  \citenamefont {Liu}, \citenamefont {Zhang}, \citenamefont {Wang},
  \citenamefont {Wang}, \citenamefont {Lin}, \citenamefont {Xing},
  \citenamefont {Lu}, \citenamefont {Liu}, \citenamefont {Wang}, \citenamefont
  {Brombosz}, \citenamefont {Xiao}, \citenamefont {Jia}, \citenamefont {Xie},\
  and\ \citenamefont {Wang}}]{zhao}%
  \BibitemOpen
  \bibfield  {author} {\bibinfo {author} {\bibfnamefont {Y.}~\bibnamefont
  {Zhao}}, \bibinfo {author} {\bibfnamefont {H.}~\bibnamefont {Liu}}, \bibinfo
  {author} {\bibfnamefont {C.}~\bibnamefont {Zhang}}, \bibinfo {author}
  {\bibfnamefont {H.}~\bibnamefont {Wang}}, \bibinfo {author} {\bibfnamefont
  {J.}~\bibnamefont {Wang}}, \bibinfo {author} {\bibfnamefont {Z.}~\bibnamefont
  {Lin}}, \bibinfo {author} {\bibfnamefont {Y.}~\bibnamefont {Xing}}, \bibinfo
  {author} {\bibfnamefont {H.}~\bibnamefont {Lu}}, \bibinfo {author}
  {\bibfnamefont {J.}~\bibnamefont {Liu}}, \bibinfo {author} {\bibfnamefont
  {Y.}~\bibnamefont {Wang}}, \bibinfo {author} {\bibfnamefont {S.~M.}\
  \bibnamefont {Brombosz}}, \bibinfo {author} {\bibfnamefont {Z.}~\bibnamefont
  {Xiao}}, \bibinfo {author} {\bibfnamefont {S.}~\bibnamefont {Jia}}, \bibinfo
  {author} {\bibfnamefont {X.~C.}\ \bibnamefont {Xie}}, \ and\ \bibinfo
  {author} {\bibfnamefont {J.}~\bibnamefont {Wang}},\ }\href {\doibase
  10.1103/PhysRevX.5.031037} {\bibfield  {journal} {\bibinfo  {journal} {Phys.
  Rev. X}\ }\textbf {\bibinfo {volume} {5}},\ \bibinfo {pages} {031037}
  (\bibinfo {year} {2015})}\BibitemShut {NoStop}%
\bibitem [{\citenamefont {Xiong}\ \emph {et~al.}()\citenamefont {Xiong},
  \citenamefont {Kushwaha}, \citenamefont {Liang}, \citenamefont {Krizan},
  \citenamefont {Wang}, \citenamefont {Cava},\ and\ \citenamefont
  {Ong}}]{XiongOng}%
  \BibitemOpen
  \bibfield  {author} {\bibinfo {author} {\bibfnamefont {J.}~\bibnamefont
  {Xiong}}, \bibinfo {author} {\bibfnamefont {S.~K.}\ \bibnamefont {Kushwaha}},
  \bibinfo {author} {\bibfnamefont {T.}~\bibnamefont {Liang}}, \bibinfo
  {author} {\bibfnamefont {J.~W.}\ \bibnamefont {Krizan}}, \bibinfo {author}
  {\bibfnamefont {W.}~\bibnamefont {Wang}}, \bibinfo {author} {\bibfnamefont
  {R.~J.}\ \bibnamefont {Cava}}, \ and\ \bibinfo {author} {\bibfnamefont
  {N.~P.}\ \bibnamefont {Ong}},\ }\href@noop {} {\ }\Eprint
  {http://arxiv.org/abs/arXiv:1503.08179 (2015)} {arXiv:1503.08179 (2015)}
  \BibitemShut {NoStop}%
\bibitem [{\citenamefont {Liang}\ \emph {et~al.}(2015)\citenamefont {Liang},
  \citenamefont {Gibson}, \citenamefont {Ali}, \citenamefont {Liu},
  \citenamefont {Cava},\ and\ \citenamefont {Ong}}]{ong}%
  \BibitemOpen
  \bibfield  {author} {\bibinfo {author} {\bibfnamefont {T.}~\bibnamefont
  {Liang}}, \bibinfo {author} {\bibfnamefont {Q.}~\bibnamefont {Gibson}},
  \bibinfo {author} {\bibfnamefont {M.~N.}\ \bibnamefont {Ali}}, \bibinfo
  {author} {\bibfnamefont {M.}~\bibnamefont {Liu}}, \bibinfo {author}
  {\bibfnamefont {R.~J.}\ \bibnamefont {Cava}}, \ and\ \bibinfo {author}
  {\bibfnamefont {N.~P.}\ \bibnamefont {Ong}},\ }\href {\doibase
  10.1038/nmat4143} {\bibfield  {journal} {\bibinfo  {journal} {Nat. Mater.}\
  }\textbf {\bibinfo {volume} {14}},\ \bibinfo {pages} {280} (\bibinfo {year}
  {2015})}\BibitemShut {NoStop}%
\bibitem [{\citenamefont {Huang}\ \emph
  {et~al.}(2015{\natexlab{b}})\citenamefont {Huang}, \citenamefont {Zhao},
  \citenamefont {Long}, \citenamefont {Wang}, \citenamefont {Chen},
  \citenamefont {Yang}, \citenamefont {Liang}, \citenamefont {Xue},
  \citenamefont {Weng}, \citenamefont {Fang}, \citenamefont {Dai},\ and\
  \citenamefont {Chen}}]{XHuang}%
  \BibitemOpen
  \bibfield  {author} {\bibinfo {author} {\bibfnamefont {X.}~\bibnamefont
  {Huang}}, \bibinfo {author} {\bibfnamefont {L.}~\bibnamefont {Zhao}},
  \bibinfo {author} {\bibfnamefont {Y.}~\bibnamefont {Long}}, \bibinfo {author}
  {\bibfnamefont {P.}~\bibnamefont {Wang}}, \bibinfo {author} {\bibfnamefont
  {D.}~\bibnamefont {Chen}}, \bibinfo {author} {\bibfnamefont {Z.}~\bibnamefont
  {Yang}}, \bibinfo {author} {\bibfnamefont {H.}~\bibnamefont {Liang}},
  \bibinfo {author} {\bibfnamefont {M.}~\bibnamefont {Xue}}, \bibinfo {author}
  {\bibfnamefont {H.}~\bibnamefont {Weng}}, \bibinfo {author} {\bibfnamefont
  {Z.}~\bibnamefont {Fang}}, \bibinfo {author} {\bibfnamefont {X.}~\bibnamefont
  {Dai}}, \ and\ \bibinfo {author} {\bibfnamefont {G.}~\bibnamefont {Chen}},\
  }\href {\doibase 10.1103/PhysRevX.5.031023} {\bibfield  {journal} {\bibinfo
  {journal} {Phys. Rev. X}\ }\textbf {\bibinfo {volume} {5}},\ \bibinfo {pages}
  {031023} (\bibinfo {year} {2015}{\natexlab{b}})}\BibitemShut {NoStop}%
\bibitem [{\citenamefont {Zhang}\ \emph
  {et~al.}(2016{\natexlab{c}})\citenamefont {Zhang}, \citenamefont {Xu},
  \citenamefont {Belopolski}, \citenamefont {Yuan}, \citenamefont {Lin},
  \citenamefont {Tong}, \citenamefont {Bian}, \citenamefont {Alidoust},
  \citenamefont {Lee}, \citenamefont {Huang}, \citenamefont {Chang},
  \citenamefont {Chang}, \citenamefont {Hsu}, \citenamefont {Jeng},
  \citenamefont {Neupane}, \citenamefont {Sanchez}, \citenamefont {Zheng},
  \citenamefont {Wang}, \citenamefont {Lin}, \citenamefont {Zhang},
  \citenamefont {Lu}, \citenamefont {Shen}, \citenamefont {Neupert},
  \citenamefont {{Zahid Hasan}},\ and\ \citenamefont {Jia}}]{CZhang}%
  \BibitemOpen
  \bibfield  {author} {\bibinfo {author} {\bibfnamefont {C.-L.}\ \bibnamefont
  {Zhang}}, \bibinfo {author} {\bibfnamefont {S.-Y.}\ \bibnamefont {Xu}},
  \bibinfo {author} {\bibfnamefont {I.}~\bibnamefont {Belopolski}}, \bibinfo
  {author} {\bibfnamefont {Z.}~\bibnamefont {Yuan}}, \bibinfo {author}
  {\bibfnamefont {Z.}~\bibnamefont {Lin}}, \bibinfo {author} {\bibfnamefont
  {B.}~\bibnamefont {Tong}}, \bibinfo {author} {\bibfnamefont {G.}~\bibnamefont
  {Bian}}, \bibinfo {author} {\bibfnamefont {N.}~\bibnamefont {Alidoust}},
  \bibinfo {author} {\bibfnamefont {C.-C.}\ \bibnamefont {Lee}}, \bibinfo
  {author} {\bibfnamefont {S.-M.}\ \bibnamefont {Huang}}, \bibinfo {author}
  {\bibfnamefont {T.-R.}\ \bibnamefont {Chang}}, \bibinfo {author}
  {\bibfnamefont {G.}~\bibnamefont {Chang}}, \bibinfo {author} {\bibfnamefont
  {C.-H.}\ \bibnamefont {Hsu}}, \bibinfo {author} {\bibfnamefont {H.-T.}\
  \bibnamefont {Jeng}}, \bibinfo {author} {\bibfnamefont {M.}~\bibnamefont
  {Neupane}}, \bibinfo {author} {\bibfnamefont {D.~S.}\ \bibnamefont
  {Sanchez}}, \bibinfo {author} {\bibfnamefont {H.}~\bibnamefont {Zheng}},
  \bibinfo {author} {\bibfnamefont {J.}~\bibnamefont {Wang}}, \bibinfo {author}
  {\bibfnamefont {H.}~\bibnamefont {Lin}}, \bibinfo {author} {\bibfnamefont
  {C.}~\bibnamefont {Zhang}}, \bibinfo {author} {\bibfnamefont {H.-Z.}\
  \bibnamefont {Lu}}, \bibinfo {author} {\bibfnamefont {S.-Q.}\ \bibnamefont
  {Shen}}, \bibinfo {author} {\bibfnamefont {T.}~\bibnamefont {Neupert}},
  \bibinfo {author} {\bibfnamefont {M.}~\bibnamefont {{Zahid Hasan}}}, \ and\
  \bibinfo {author} {\bibfnamefont {S.}~\bibnamefont {Jia}},\ }\href {\doibase
  10.1038/ncomms10735} {\bibfield  {journal} {\bibinfo  {journal} {Nat.
  Commun.}\ }\textbf {\bibinfo {volume} {7}},\ \bibinfo {pages} {10735}
  (\bibinfo {year} {2016}{\natexlab{c}})}\BibitemShut {NoStop}%
\bibitem [{\citenamefont {Li}\ \emph {et~al.}(2016{\natexlab{a}})\citenamefont
  {Li}, \citenamefont {Kharzeev}, \citenamefont {Zhang}, \citenamefont {Huang},
  \citenamefont {Pletikosi{\'c}}, \citenamefont {Fedorov}, \citenamefont
  {Zhong}, \citenamefont {Schneeloch}, \citenamefont {Gu},\ and\ \citenamefont
  {Valla}}]{kharzeev}%
  \BibitemOpen
  \bibfield  {author} {\bibinfo {author} {\bibfnamefont {Q.}~\bibnamefont
  {Li}}, \bibinfo {author} {\bibfnamefont {D.~E.}\ \bibnamefont {Kharzeev}},
  \bibinfo {author} {\bibfnamefont {C.}~\bibnamefont {Zhang}}, \bibinfo
  {author} {\bibfnamefont {Y.}~\bibnamefont {Huang}}, \bibinfo {author}
  {\bibfnamefont {I.}~\bibnamefont {Pletikosi{\'c}}}, \bibinfo {author}
  {\bibfnamefont {A.~V.}\ \bibnamefont {Fedorov}}, \bibinfo {author}
  {\bibfnamefont {R.~D.}\ \bibnamefont {Zhong}}, \bibinfo {author}
  {\bibfnamefont {J.~A.}\ \bibnamefont {Schneeloch}}, \bibinfo {author}
  {\bibfnamefont {G.~D.}\ \bibnamefont {Gu}}, \ and\ \bibinfo {author}
  {\bibfnamefont {T.}~\bibnamefont {Valla}},\ }\href {\doibase
  10.1038/nphys3648} {\bibfield  {journal} {\bibinfo  {journal} {Nat. Phys.}\
  }\textbf {\bibinfo {volume} {12}},\ \bibinfo {pages} {550} (\bibinfo {year}
  {2016}{\natexlab{a}})}\BibitemShut {NoStop}%
\bibitem [{\citenamefont {Kikugawa}\ \emph {et~al.}(2016)\citenamefont
  {Kikugawa}, \citenamefont {Goswami}, \citenamefont {Kiswandhi}, \citenamefont
  {Choi}, \citenamefont {Graf}, \citenamefont {Baumbach}, \citenamefont
  {Brooks}, \citenamefont {Sugii}, \citenamefont {Iida}, \citenamefont
  {Nishio}, \citenamefont {Uji}, \citenamefont {Terashima}, \citenamefont
  {Rourke}, \citenamefont {Hussey}, \citenamefont {Takatsu}, \citenamefont
  {Yonezawa}, \citenamefont {Maeno},\ and\ \citenamefont {Balicas}}]{balicas}%
  \BibitemOpen
  \bibfield  {author} {\bibinfo {author} {\bibfnamefont {N.}~\bibnamefont
  {Kikugawa}}, \bibinfo {author} {\bibfnamefont {P.}~\bibnamefont {Goswami}},
  \bibinfo {author} {\bibfnamefont {A.}~\bibnamefont {Kiswandhi}}, \bibinfo
  {author} {\bibfnamefont {E.~S.}\ \bibnamefont {Choi}}, \bibinfo {author}
  {\bibfnamefont {D.}~\bibnamefont {Graf}}, \bibinfo {author} {\bibfnamefont
  {R.~E.}\ \bibnamefont {Baumbach}}, \bibinfo {author} {\bibfnamefont {J.~S.}\
  \bibnamefont {Brooks}}, \bibinfo {author} {\bibfnamefont {K.}~\bibnamefont
  {Sugii}}, \bibinfo {author} {\bibfnamefont {Y.}~\bibnamefont {Iida}},
  \bibinfo {author} {\bibfnamefont {M.}~\bibnamefont {Nishio}}, \bibinfo
  {author} {\bibfnamefont {S.}~\bibnamefont {Uji}}, \bibinfo {author}
  {\bibfnamefont {T.}~\bibnamefont {Terashima}}, \bibinfo {author}
  {\bibfnamefont {P.~M.~C.}\ \bibnamefont {Rourke}}, \bibinfo {author}
  {\bibfnamefont {N.~E.}\ \bibnamefont {Hussey}}, \bibinfo {author}
  {\bibfnamefont {H.}~\bibnamefont {Takatsu}}, \bibinfo {author} {\bibfnamefont
  {S.}~\bibnamefont {Yonezawa}}, \bibinfo {author} {\bibfnamefont
  {Y.}~\bibnamefont {Maeno}}, \ and\ \bibinfo {author} {\bibfnamefont
  {L.}~\bibnamefont {Balicas}},\ }\href {\doibase 10.1038/ncomms10903}
  {\bibfield  {journal} {\bibinfo  {journal} {Nat. Commun.}\ }\textbf {\bibinfo
  {volume} {7}},\ \bibinfo {pages} {10903} (\bibinfo {year}
  {2016})}\BibitemShut {NoStop}%
\bibitem [{\citenamefont {Arnold}\ \emph {et~al.}(2016)\citenamefont {Arnold},
  \citenamefont {Shekhar}, \citenamefont {Wu}, \citenamefont {Sun},
  \citenamefont {dos Reis}, \citenamefont {Kumar}, \citenamefont {Naumann},
  \citenamefont {Ajeesh}, \citenamefont {Schmidt}, \citenamefont {Grushin},
  \citenamefont {Bardarson}, \citenamefont {Baenitz}, \citenamefont {Sokolov},
  \citenamefont {Borrmann}, \citenamefont {Nicklas}, \citenamefont {Felser},
  \citenamefont {Hassinger},\ and\ \citenamefont {Yan}}]{Shekhar-2}%
  \BibitemOpen
  \bibfield  {author} {\bibinfo {author} {\bibfnamefont {F.}~\bibnamefont
  {Arnold}}, \bibinfo {author} {\bibfnamefont {C.}~\bibnamefont {Shekhar}},
  \bibinfo {author} {\bibfnamefont {S.-C.}\ \bibnamefont {Wu}}, \bibinfo
  {author} {\bibfnamefont {Y.}~\bibnamefont {Sun}}, \bibinfo {author}
  {\bibfnamefont {R.~D.}\ \bibnamefont {dos Reis}}, \bibinfo {author}
  {\bibfnamefont {N.}~\bibnamefont {Kumar}}, \bibinfo {author} {\bibfnamefont
  {M.}~\bibnamefont {Naumann}}, \bibinfo {author} {\bibfnamefont {M.~O.}\
  \bibnamefont {Ajeesh}}, \bibinfo {author} {\bibfnamefont {M.}~\bibnamefont
  {Schmidt}}, \bibinfo {author} {\bibfnamefont {A.~G.}\ \bibnamefont
  {Grushin}}, \bibinfo {author} {\bibfnamefont {J.~H.}\ \bibnamefont
  {Bardarson}}, \bibinfo {author} {\bibfnamefont {M.}~\bibnamefont {Baenitz}},
  \bibinfo {author} {\bibfnamefont {D.}~\bibnamefont {Sokolov}}, \bibinfo
  {author} {\bibfnamefont {H.}~\bibnamefont {Borrmann}}, \bibinfo {author}
  {\bibfnamefont {M.}~\bibnamefont {Nicklas}}, \bibinfo {author} {\bibfnamefont
  {C.}~\bibnamefont {Felser}}, \bibinfo {author} {\bibfnamefont
  {E.}~\bibnamefont {Hassinger}}, \ and\ \bibinfo {author} {\bibfnamefont
  {B.}~\bibnamefont {Yan}},\ }\href {\doibase 10.1038/ncomms11615} {\bibfield
  {journal} {\bibinfo  {journal} {Nat. Commun.}\ }\textbf {\bibinfo {volume}
  {7}},\ \bibinfo {pages} {11615} (\bibinfo {year} {2016})}\BibitemShut
  {NoStop}%
\bibitem [{\citenamefont {Wang}\ \emph
  {et~al.}(2016{\natexlab{b}})\citenamefont {Wang}, \citenamefont {Yu},
  \citenamefont {Guo}, \citenamefont {Liu},\ and\ \citenamefont
  {Xia}}]{Wang-Xia}%
  \BibitemOpen
  \bibfield  {author} {\bibinfo {author} {\bibfnamefont {Y.-Y.}\ \bibnamefont
  {Wang}}, \bibinfo {author} {\bibfnamefont {Q.-H.}\ \bibnamefont {Yu}},
  \bibinfo {author} {\bibfnamefont {P.-J.}\ \bibnamefont {Guo}}, \bibinfo
  {author} {\bibfnamefont {K.}~\bibnamefont {Liu}}, \ and\ \bibinfo {author}
  {\bibfnamefont {T.-L.}\ \bibnamefont {Xia}},\ }\href {\doibase
  10.1103/PhysRevB.94.041103} {\bibfield  {journal} {\bibinfo  {journal} {Phys.
  Rev. B}\ }\textbf {\bibinfo {volume} {94}},\ \bibinfo {pages} {041103}
  (\bibinfo {year} {2016}{\natexlab{b}})}\BibitemShut {NoStop}%
\bibitem [{\citenamefont {Wang}\ \emph
  {et~al.}(2016{\natexlab{c}})\citenamefont {Wang}, \citenamefont {Li},
  \citenamefont {Liu}, \citenamefont {Yan}, \citenamefont {Wang}, \citenamefont
  {Liu}, \citenamefont {Lin}, \citenamefont {Li}, \citenamefont {Wang},
  \citenamefont {Li}, \citenamefont {Mandrus}, \citenamefont {Xie},
  \citenamefont {Feng},\ and\ \citenamefont {Wang}}]{Wang-Wang}%
  \BibitemOpen
  \bibfield  {author} {\bibinfo {author} {\bibfnamefont {H.}~\bibnamefont
  {Wang}}, \bibinfo {author} {\bibfnamefont {C.-K.}\ \bibnamefont {Li}},
  \bibinfo {author} {\bibfnamefont {H.}~\bibnamefont {Liu}}, \bibinfo {author}
  {\bibfnamefont {J.}~\bibnamefont {Yan}}, \bibinfo {author} {\bibfnamefont
  {J.}~\bibnamefont {Wang}}, \bibinfo {author} {\bibfnamefont {J.}~\bibnamefont
  {Liu}}, \bibinfo {author} {\bibfnamefont {Z.}~\bibnamefont {Lin}}, \bibinfo
  {author} {\bibfnamefont {Y.}~\bibnamefont {Li}}, \bibinfo {author}
  {\bibfnamefont {Y.}~\bibnamefont {Wang}}, \bibinfo {author} {\bibfnamefont
  {L.}~\bibnamefont {Li}}, \bibinfo {author} {\bibfnamefont {D.}~\bibnamefont
  {Mandrus}}, \bibinfo {author} {\bibfnamefont {X.~C.}\ \bibnamefont {Xie}},
  \bibinfo {author} {\bibfnamefont {J.}~\bibnamefont {Feng}}, \ and\ \bibinfo
  {author} {\bibfnamefont {J.}~\bibnamefont {Wang}},\ }\href {\doibase
  10.1103/PhysRevB.93.165127} {\bibfield  {journal} {\bibinfo  {journal} {Phys.
  Rev. B}\ }\textbf {\bibinfo {volume} {93}},\ \bibinfo {pages} {165127}
  (\bibinfo {year} {2016}{\natexlab{c}})}\BibitemShut {NoStop}%
\bibitem [{\citenamefont {Zheng}\ \emph {et~al.}(2016)\citenamefont {Zheng},
  \citenamefont {Lu}, \citenamefont {Zhu}, \citenamefont {Ning}, \citenamefont
  {Han}, \citenamefont {Zhang}, \citenamefont {Zhang}, \citenamefont {Xi},
  \citenamefont {Yang}, \citenamefont {Du}, \citenamefont {Yang}, \citenamefont
  {Zhang},\ and\ \citenamefont {Tian}}]{Zheng-Tian}%
  \BibitemOpen
  \bibfield  {author} {\bibinfo {author} {\bibfnamefont {G.}~\bibnamefont
  {Zheng}}, \bibinfo {author} {\bibfnamefont {J.}~\bibnamefont {Lu}}, \bibinfo
  {author} {\bibfnamefont {X.}~\bibnamefont {Zhu}}, \bibinfo {author}
  {\bibfnamefont {W.}~\bibnamefont {Ning}}, \bibinfo {author} {\bibfnamefont
  {Y.}~\bibnamefont {Han}}, \bibinfo {author} {\bibfnamefont {H.}~\bibnamefont
  {Zhang}}, \bibinfo {author} {\bibfnamefont {J.}~\bibnamefont {Zhang}},
  \bibinfo {author} {\bibfnamefont {C.}~\bibnamefont {Xi}}, \bibinfo {author}
  {\bibfnamefont {J.}~\bibnamefont {Yang}}, \bibinfo {author} {\bibfnamefont
  {H.}~\bibnamefont {Du}}, \bibinfo {author} {\bibfnamefont {K.}~\bibnamefont
  {Yang}}, \bibinfo {author} {\bibfnamefont {Y.}~\bibnamefont {Zhang}}, \ and\
  \bibinfo {author} {\bibfnamefont {M.}~\bibnamefont {Tian}},\ }\href {\doibase
  10.1103/PhysRevB.93.115414} {\bibfield  {journal} {\bibinfo  {journal} {Phys.
  Rev. B}\ }\textbf {\bibinfo {volume} {93}},\ \bibinfo {pages} {115414}
  (\bibinfo {year} {2016})}\BibitemShut {NoStop}%
\bibitem [{\citenamefont {Wiedmann}\ \emph {et~al.}(2016)\citenamefont
  {Wiedmann}, \citenamefont {Jost}, \citenamefont {Fauqu\'e}, \citenamefont
  {van Dijk}, \citenamefont {Meijer}, \citenamefont {Khouri}, \citenamefont
  {Pezzini}, \citenamefont {Grauer}, \citenamefont {Schreyeck}, \citenamefont
  {Br\"une}, \citenamefont {Buhmann}, \citenamefont {Molenkamp},\ and\
  \citenamefont {Hussey}}]{Wiedmann}%
  \BibitemOpen
  \bibfield  {author} {\bibinfo {author} {\bibfnamefont {S.}~\bibnamefont
  {Wiedmann}}, \bibinfo {author} {\bibfnamefont {A.}~\bibnamefont {Jost}},
  \bibinfo {author} {\bibfnamefont {B.}~\bibnamefont {Fauqu\'e}}, \bibinfo
  {author} {\bibfnamefont {J.}~\bibnamefont {van Dijk}}, \bibinfo {author}
  {\bibfnamefont {M.~J.}\ \bibnamefont {Meijer}}, \bibinfo {author}
  {\bibfnamefont {T.}~\bibnamefont {Khouri}}, \bibinfo {author} {\bibfnamefont
  {S.}~\bibnamefont {Pezzini}}, \bibinfo {author} {\bibfnamefont
  {S.}~\bibnamefont {Grauer}}, \bibinfo {author} {\bibfnamefont
  {S.}~\bibnamefont {Schreyeck}}, \bibinfo {author} {\bibfnamefont
  {C.}~\bibnamefont {Br\"une}}, \bibinfo {author} {\bibfnamefont
  {H.}~\bibnamefont {Buhmann}}, \bibinfo {author} {\bibfnamefont {L.~W.}\
  \bibnamefont {Molenkamp}}, \ and\ \bibinfo {author} {\bibfnamefont {N.~E.}\
  \bibnamefont {Hussey}},\ }\href {\doibase 10.1103/PhysRevB.94.081302}
  {\bibfield  {journal} {\bibinfo  {journal} {Phys. Rev. B}\ }\textbf {\bibinfo
  {volume} {94}},\ \bibinfo {pages} {081302(R)} (\bibinfo {year}
  {2016})}\BibitemShut {NoStop}%
\bibitem [{\citenamefont {Nielsen}\ and\ \citenamefont
  {Ninomiya}(1983)}]{nielsen}%
  \BibitemOpen
  \bibfield  {author} {\bibinfo {author} {\bibfnamefont {H.~B.}\ \bibnamefont
  {Nielsen}}\ and\ \bibinfo {author} {\bibfnamefont {M.}~\bibnamefont
  {Ninomiya}},\ }\href {\doibase 10.1016/0370-2693(83)91529-0} {\bibfield
  {journal} {\bibinfo  {journal} {Phys. Lett. B}\ }\textbf {\bibinfo {volume}
  {130}},\ \bibinfo {pages} {389} (\bibinfo {year} {1983})}\BibitemShut
  {NoStop}%
\bibitem [{\citenamefont {Argyres}\ and\ \citenamefont
  {Adams}(1956)}]{Argyres1956:Transport}%
  \BibitemOpen
  \bibfield  {author} {\bibinfo {author} {\bibfnamefont {P.~N.}\ \bibnamefont
  {Argyres}}\ and\ \bibinfo {author} {\bibfnamefont {E.~N.}\ \bibnamefont
  {Adams}},\ }\href {\doibase 10.1103/PhysRev.104.900} {\bibfield  {journal}
  {\bibinfo  {journal} {Phys. Rev.}\ }\textbf {\bibinfo {volume} {104}},\
  \bibinfo {pages} {900} (\bibinfo {year} {1956})}\BibitemShut {NoStop}%
\bibitem [{\citenamefont {Goswami}\ \emph {et~al.}(2015)\citenamefont
  {Goswami}, \citenamefont {Pixley},\ and\ \citenamefont
  {Das~Sarma}}]{Goswami2015}%
  \BibitemOpen
  \bibfield  {author} {\bibinfo {author} {\bibfnamefont {P.}~\bibnamefont
  {Goswami}}, \bibinfo {author} {\bibfnamefont {J.~H.}\ \bibnamefont {Pixley}},
  \ and\ \bibinfo {author} {\bibfnamefont {S.}~\bibnamefont {Das~Sarma}},\
  }\href {\doibase 10.1103/PhysRevB.92.075205} {\bibfield  {journal} {\bibinfo
  {journal} {Phys. Rev. B}\ }\textbf {\bibinfo {volume} {92}},\ \bibinfo
  {pages} {075205} (\bibinfo {year} {2015})}\BibitemShut {NoStop}%
\bibitem [{\citenamefont {Ashcroft}\ and\ \citenamefont
  {Mermin}(1976)}]{ashcroft}%
  \BibitemOpen
  \bibfield  {author} {\bibinfo {author} {\bibfnamefont {N.}~\bibnamefont
  {Ashcroft}}\ and\ \bibinfo {author} {\bibfnamefont {N.}~\bibnamefont
  {Mermin}},\ }\href@noop {} {\emph {\bibinfo {title} {Solid State Physics}}},\
  HRW international editions\ (\bibinfo  {publisher} {Holt, Rinehart and
  Winston},\ \bibinfo {year} {1976})\BibitemShut {NoStop}%
\bibitem [{\citenamefont {Hikami}\ \emph {et~al.}(1980)\citenamefont {Hikami},
  \citenamefont {Larkin},\ and\ \citenamefont {Nagaoka}}]{hikami}%
  \BibitemOpen
  \bibfield  {author} {\bibinfo {author} {\bibfnamefont {S.}~\bibnamefont
  {Hikami}}, \bibinfo {author} {\bibfnamefont {A.~I.}\ \bibnamefont {Larkin}},
  \ and\ \bibinfo {author} {\bibfnamefont {Y.}~\bibnamefont {Nagaoka}},\ }\href
  {\doibase 10.1143/PTP.63.707} {\bibfield  {journal} {\bibinfo  {journal}
  {Prog. Theor. Phys.}\ }\textbf {\bibinfo {volume} {63}},\ \bibinfo {pages}
  {707} (\bibinfo {year} {1980})}\BibitemShut {NoStop}%
\bibitem [{\citenamefont {Bergmann}(1984)}]{bergmann1984weak}%
  \BibitemOpen
  \bibfield  {author} {\bibinfo {author} {\bibfnamefont {G.}~\bibnamefont
  {Bergmann}},\ }\href {\doibase 10.1016/0370-1573(84)90103-0} {\bibfield
  {journal} {\bibinfo  {journal} {Phys. Rep.}\ }\textbf {\bibinfo {volume}
  {107}},\ \bibinfo {pages} {1} (\bibinfo {year} {1984})}\BibitemShut {NoStop}%
\bibitem [{\citenamefont {Imry}\ and\ \citenamefont {Ma}(1975)}]{Imry}%
  \BibitemOpen
  \bibfield  {author} {\bibinfo {author} {\bibfnamefont {Y.}~\bibnamefont
  {Imry}}\ and\ \bibinfo {author} {\bibfnamefont {S.-k.}\ \bibnamefont {Ma}},\
  }\href {\doibase 10.1103/PhysRevLett.35.1399} {\bibfield  {journal} {\bibinfo
   {journal} {Phys. Rev. Lett.}\ }\textbf {\bibinfo {volume} {35}},\ \bibinfo
  {pages} {1399} (\bibinfo {year} {1975})}\BibitemShut {NoStop}%
\bibitem [{\citenamefont {Bera}\ \emph {et~al.}(2016)\citenamefont {Bera},
  \citenamefont {Sau},\ and\ \citenamefont {Roy}}]{roy-bera}%
  \BibitemOpen
  \bibfield  {author} {\bibinfo {author} {\bibfnamefont {S.}~\bibnamefont
  {Bera}}, \bibinfo {author} {\bibfnamefont {J.~D.}\ \bibnamefont {Sau}}, \
  and\ \bibinfo {author} {\bibfnamefont {B.}~\bibnamefont {Roy}},\ }\href
  {\doibase 10.1103/PhysRevB.93.201302} {\bibfield  {journal} {\bibinfo
  {journal} {Phys. Rev. B}\ }\textbf {\bibinfo {volume} {93}},\ \bibinfo
  {pages} {201302} (\bibinfo {year} {2016})}\BibitemShut {NoStop}%
\bibitem [{\citenamefont {Goswami}\ and\ \citenamefont
  {Nevidomskyy}(2015)}]{goswami-topology}%
  \BibitemOpen
  \bibfield  {author} {\bibinfo {author} {\bibfnamefont {P.}~\bibnamefont
  {Goswami}}\ and\ \bibinfo {author} {\bibfnamefont {A.~H.}\ \bibnamefont
  {Nevidomskyy}},\ }\href {\doibase 10.1103/PhysRevB.92.214504} {\bibfield
  {journal} {\bibinfo  {journal} {Phys. Rev. B}\ }\textbf {\bibinfo {volume}
  {92}},\ \bibinfo {pages} {214504} (\bibinfo {year} {2015})}\BibitemShut
  {NoStop}%
\bibitem [{\citenamefont {Roy}\ and\ \citenamefont {Sau}(2015)}]{roy-sau}%
  \BibitemOpen
  \bibfield  {author} {\bibinfo {author} {\bibfnamefont {B.}~\bibnamefont
  {Roy}}\ and\ \bibinfo {author} {\bibfnamefont {J.~D.}\ \bibnamefont {Sau}},\
  }\href {\doibase 10.1103/PhysRevB.92.125141} {\bibfield  {journal} {\bibinfo
  {journal} {Phys. Rev. B}\ }\textbf {\bibinfo {volume} {92}},\ \bibinfo
  {pages} {125141} (\bibinfo {year} {2015})}\BibitemShut {NoStop}%
\bibitem [{\citenamefont {Chen}\ and\ \citenamefont {Fiete}(2016)}]{fiete}%
  \BibitemOpen
  \bibfield  {author} {\bibinfo {author} {\bibfnamefont {Q.}~\bibnamefont
  {Chen}}\ and\ \bibinfo {author} {\bibfnamefont {G.~A.}\ \bibnamefont
  {Fiete}},\ }\href {\doibase 10.1103/PhysRevB.93.155125} {\bibfield  {journal}
  {\bibinfo  {journal} {Phys. Rev. B}\ }\textbf {\bibinfo {volume} {93}},\
  \bibinfo {pages} {155125} (\bibinfo {year} {2016})}\BibitemShut {NoStop}%
\bibitem [{\citenamefont {Yoshioka}\ and\ \citenamefont
  {Fukuyama}(1981)}]{fukuyama}%
  \BibitemOpen
  \bibfield  {author} {\bibinfo {author} {\bibfnamefont {D.}~\bibnamefont
  {Yoshioka}}\ and\ \bibinfo {author} {\bibfnamefont {H.}~\bibnamefont
  {Fukuyama}},\ }\href {\doibase 10.1143/JPSJ.50.725} {\bibfield  {journal}
  {\bibinfo  {journal} {J. Phys. Soc. Jpn.}\ }\textbf {\bibinfo {volume}
  {50}},\ \bibinfo {pages} {725} (\bibinfo {year} {1981})}\BibitemShut
  {NoStop}%
\bibitem [{\citenamefont {Bardasis}\ and\ \citenamefont
  {Das~Sarma}(1984)}]{Sankar1}%
  \BibitemOpen
  \bibfield  {author} {\bibinfo {author} {\bibfnamefont {A.}~\bibnamefont
  {Bardasis}}\ and\ \bibinfo {author} {\bibfnamefont {S.}~\bibnamefont
  {Das~Sarma}},\ }\href {\doibase 10.1103/PhysRevB.29.780} {\bibfield
  {journal} {\bibinfo  {journal} {Phys. Rev. B}\ }\textbf {\bibinfo {volume}
  {29}},\ \bibinfo {pages} {780} (\bibinfo {year} {1984})}\BibitemShut
  {NoStop}%
\bibitem [{\citenamefont {MacDonald}\ and\ \citenamefont
  {Bryant}(1987)}]{bryant}%
  \BibitemOpen
  \bibfield  {author} {\bibinfo {author} {\bibfnamefont {A.~H.}\ \bibnamefont
  {MacDonald}}\ and\ \bibinfo {author} {\bibfnamefont {G.~W.}\ \bibnamefont
  {Bryant}},\ }\href {\doibase 10.1103/PhysRevLett.58.515} {\bibfield
  {journal} {\bibinfo  {journal} {Phys. Rev. Lett.}\ }\textbf {\bibinfo
  {volume} {58}},\ \bibinfo {pages} {515} (\bibinfo {year} {1987})}\BibitemShut
  {NoStop}%
\bibitem [{\citenamefont {Li}\ \emph {et~al.}(2015)\citenamefont {Li},
  \citenamefont {Roy},\ and\ \citenamefont {Das~Sarma}}]{Li-Roy-DasSarma}%
  \BibitemOpen
  \bibfield  {author} {\bibinfo {author} {\bibfnamefont {X.}~\bibnamefont
  {Li}}, \bibinfo {author} {\bibfnamefont {B.}~\bibnamefont {Roy}}, \ and\
  \bibinfo {author} {\bibfnamefont {S.}~\bibnamefont {Das~Sarma}},\ }\href
  {\doibase 10.1103/PhysRevB.92.235144} {\bibfield  {journal} {\bibinfo
  {journal} {Phys. Rev. B}\ }\textbf {\bibinfo {volume} {92}},\ \bibinfo
  {pages} {235144} (\bibinfo {year} {2015})}\BibitemShut {NoStop}%
\bibitem [{\citenamefont {Li}\ \emph {et~al.}(2016{\natexlab{b}})\citenamefont
  {Li}, \citenamefont {Zhang},\ and\ \citenamefont {MacDonald}}]{Xiao-SnTe}%
  \BibitemOpen
  \bibfield  {author} {\bibinfo {author} {\bibfnamefont {X.}~\bibnamefont
  {Li}}, \bibinfo {author} {\bibfnamefont {F.}~\bibnamefont {Zhang}}, \ and\
  \bibinfo {author} {\bibfnamefont {A.~H.}\ \bibnamefont {MacDonald}},\ }\href
  {\doibase 10.1103/PhysRevLett.116.026803} {\bibfield  {journal} {\bibinfo
  {journal} {Phys. Rev. Lett.}\ }\textbf {\bibinfo {volume} {116}},\ \bibinfo
  {pages} {026803} (\bibinfo {year} {2016}{\natexlab{b}})}\BibitemShut
  {NoStop}%
\bibitem [{\citenamefont {Li}\ \emph {et~al.}(2014)\citenamefont {Li},
  \citenamefont {Zhang}, \citenamefont {Niu},\ and\ \citenamefont
  {MacDonald}}]{Xiao-DW}%
  \BibitemOpen
  \bibfield  {author} {\bibinfo {author} {\bibfnamefont {X.}~\bibnamefont
  {Li}}, \bibinfo {author} {\bibfnamefont {F.}~\bibnamefont {Zhang}}, \bibinfo
  {author} {\bibfnamefont {Q.}~\bibnamefont {Niu}}, \ and\ \bibinfo {author}
  {\bibfnamefont {A.~H.}\ \bibnamefont {MacDonald}},\ }\href {\doibase
  10.1103/PhysRevLett.113.116803} {\bibfield  {journal} {\bibinfo  {journal}
  {Phys. Rev. Lett.}\ }\textbf {\bibinfo {volume} {113}},\ \bibinfo {pages}
  {116803} (\bibinfo {year} {2014})}\BibitemShut {NoStop}%
\bibitem [{\citenamefont {Roy}\ and\ \citenamefont
  {Herbut}(2010)}]{roy-herbut}%
  \BibitemOpen
  \bibfield  {author} {\bibinfo {author} {\bibfnamefont {B.}~\bibnamefont
  {Roy}}\ and\ \bibinfo {author} {\bibfnamefont {I.~F.}\ \bibnamefont
  {Herbut}},\ }\href {\doibase 10.1103/PhysRevB.82.035429} {\bibfield
  {journal} {\bibinfo  {journal} {Phys. Rev. B}\ }\textbf {\bibinfo {volume}
  {82}},\ \bibinfo {pages} {035429} (\bibinfo {year} {2010})}\BibitemShut
  {NoStop}%
\bibitem [{\citenamefont {Miransky}\ and\ \citenamefont
  {Shovkovy}(2015)}]{miransky}%
  \BibitemOpen
  \bibfield  {author} {\bibinfo {author} {\bibfnamefont {V.~A.}\ \bibnamefont
  {Miransky}}\ and\ \bibinfo {author} {\bibfnamefont {I.~A.}\ \bibnamefont
  {Shovkovy}},\ }\href {\doibase 10.1016/j.physrep.2015.02.003} {\bibfield
  {journal} {\bibinfo  {journal} {Phys. Rep.}\ }\textbf {\bibinfo {volume}
  {576}},\ \bibinfo {pages} {1} (\bibinfo {year} {2015})}\BibitemShut {NoStop}%
\bibitem [{\citenamefont {Roy}\ \emph {et~al.}(2014)\citenamefont {Roy},
  \citenamefont {Kennett},\ and\ \citenamefont {Das~Sarma}}]{kennett}%
  \BibitemOpen
  \bibfield  {author} {\bibinfo {author} {\bibfnamefont {B.}~\bibnamefont
  {Roy}}, \bibinfo {author} {\bibfnamefont {M.~P.}\ \bibnamefont {Kennett}}, \
  and\ \bibinfo {author} {\bibfnamefont {S.}~\bibnamefont {Das~Sarma}},\ }\href
  {\doibase 10.1103/PhysRevB.90.201409} {\bibfield  {journal} {\bibinfo
  {journal} {Phys. Rev. B}\ }\textbf {\bibinfo {volume} {90}},\ \bibinfo
  {pages} {201409} (\bibinfo {year} {2014})}\BibitemShut {NoStop}%
\bibitem [{\citenamefont {Hwang}\ \emph {et~al.}(2007)\citenamefont {Hwang},
  \citenamefont {Adam},\ and\ \citenamefont {{Das Sarma}}}]{EHHwang1}%
  \BibitemOpen
  \bibfield  {author} {\bibinfo {author} {\bibfnamefont {E.~H.}\ \bibnamefont
  {Hwang}}, \bibinfo {author} {\bibfnamefont {S.}~\bibnamefont {Adam}}, \ and\
  \bibinfo {author} {\bibfnamefont {S.}~\bibnamefont {{Das Sarma}}},\ }\href
  {\doibase 10.1103/PhysRevLett.98.186806} {\bibfield  {journal} {\bibinfo
  {journal} {Phys. Rev. Lett.}\ }\textbf {\bibinfo {volume} {98}},\ \bibinfo
  {pages} {186806} (\bibinfo {year} {2007})}\BibitemShut {NoStop}%
\bibitem [{\citenamefont {Tan}\ \emph {et~al.}(2007)\citenamefont {Tan},
  \citenamefont {Zhang}, \citenamefont {Bolotin}, \citenamefont {Zhao},
  \citenamefont {Adam}, \citenamefont {Hwang}, \citenamefont {Das~Sarma},
  \citenamefont {Stormer},\ and\ \citenamefont {Kim}}]{EHHwang2}%
  \BibitemOpen
  \bibfield  {author} {\bibinfo {author} {\bibfnamefont {Y.-W.}\ \bibnamefont
  {Tan}}, \bibinfo {author} {\bibfnamefont {Y.}~\bibnamefont {Zhang}}, \bibinfo
  {author} {\bibfnamefont {K.}~\bibnamefont {Bolotin}}, \bibinfo {author}
  {\bibfnamefont {Y.}~\bibnamefont {Zhao}}, \bibinfo {author} {\bibfnamefont
  {S.}~\bibnamefont {Adam}}, \bibinfo {author} {\bibfnamefont {E.~H.}\
  \bibnamefont {Hwang}}, \bibinfo {author} {\bibfnamefont {S.}~\bibnamefont
  {Das~Sarma}}, \bibinfo {author} {\bibfnamefont {H.~L.}\ \bibnamefont
  {Stormer}}, \ and\ \bibinfo {author} {\bibfnamefont {P.}~\bibnamefont
  {Kim}},\ }\href {\doibase 10.1103/PhysRevLett.99.246803} {\bibfield
  {journal} {\bibinfo  {journal} {Phys. Rev. Lett.}\ }\textbf {\bibinfo
  {volume} {99}},\ \bibinfo {pages} {246803} (\bibinfo {year}
  {2007})}\BibitemShut {NoStop}%
\bibitem [{\citenamefont {Adam}\ \emph {et~al.}(2007)\citenamefont {Adam},
  \citenamefont {Hwang}, \citenamefont {Galitski},\ and\ \citenamefont {{Das
  Sarma}}}]{adam2007self}%
  \BibitemOpen
  \bibfield  {author} {\bibinfo {author} {\bibfnamefont {S.}~\bibnamefont
  {Adam}}, \bibinfo {author} {\bibfnamefont {E.~H.}\ \bibnamefont {Hwang}},
  \bibinfo {author} {\bibfnamefont {V.~M.}\ \bibnamefont {Galitski}}, \ and\
  \bibinfo {author} {\bibfnamefont {S.}~\bibnamefont {{Das Sarma}}},\ }\href
  {\doibase 10.1073/pnas.0704772104} {\bibfield  {journal} {\bibinfo  {journal}
  {Proc. Natl. Acad. Sci. USA}\ }\textbf {\bibinfo {volume} {104}},\ \bibinfo
  {pages} {18392} (\bibinfo {year} {2007})}\BibitemShut {NoStop}%
\bibitem [{\citenamefont {Chen}\ \emph {et~al.}(2008)\citenamefont {Chen},
  \citenamefont {Jang}, \citenamefont {Adam}, \citenamefont {Fuhrer},
  \citenamefont {Williams},\ and\ \citenamefont {Ishigami}}]{Adam2008charged}%
  \BibitemOpen
  \bibfield  {author} {\bibinfo {author} {\bibfnamefont {J.-H.}\ \bibnamefont
  {Chen}}, \bibinfo {author} {\bibfnamefont {C.}~\bibnamefont {Jang}}, \bibinfo
  {author} {\bibfnamefont {S.}~\bibnamefont {Adam}}, \bibinfo {author}
  {\bibfnamefont {M.~S.}\ \bibnamefont {Fuhrer}}, \bibinfo {author}
  {\bibfnamefont {E.~D.}\ \bibnamefont {Williams}}, \ and\ \bibinfo {author}
  {\bibfnamefont {M.}~\bibnamefont {Ishigami}},\ }\href {\doibase
  10.1038/nphys935} {\bibfield  {journal} {\bibinfo  {journal} {Nat. Phys.}\
  }\textbf {\bibinfo {volume} {4}},\ \bibinfo {pages} {377} (\bibinfo {year}
  {2008})}\BibitemShut {NoStop}%
\end{thebibliography}%

\end{document}